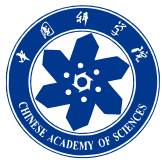

**University of Chinese Academy of Sciences**

# 博士学位论文

扭结-介子非弹性散射

| | |
|---|---|
| 作者姓名： | 刘晖 |
| 指导教师： | Jarah Evslin 研究员 中国科学院近代物理研究所 |
| 学位类别： | 理学博士 |
| 学科专业： | 理论物理 |
| 培养单位： | 中国科学院理论物理研究所 |

2023 年 6 月

# Kink-Meson Inelastic Scattering

A dissertation submitted to

University of Chinese Academy of Sciences

in partial fulfillment of the requirement

for the degree of

Doctor of Philosophy

in Theoretical Physics

By

Hui Liu

Supervisor: Professor Jarah Evslin

Institute of Theoretical Physics, Chinese Academy of Sciences

June, 2023

## 中国科学院大学
## 学位论文原创性声明

本人郑重声明：所呈交的学位论文是本人在导师的指导下独立进行研究工作所取得的成果。尽我所知，除文中已经注明引用的内容外，本论文不包含任何其他个人或集体已经发表或撰写过的研究成果。对论文所涉及的研究工作做出贡献的其他个人和集体，均已在文中以明确方式标明或致谢。

作者签名：

日　　期：

## 中国科学院大学
## 学位论文授权使用声明

本人完全了解并同意遵守中国科学院有关保存和使用学位论文的规定，即中国科学院有权保留送交学位论文的副本，允许该论文被查阅，可以按照学术研究公开原则和保护知识产权的原则公布该论文的全部或部分内容，可以采用影印、缩印或其他复制手段保存、汇编本学位论文。

涉密及延迟公开的学位论文在解密或延迟期后适用本声明。

作者签名：　　　　　　　　　导师签名：

日　　期：　　　　　　　　　日　　期：

# 摘　要

经典扭结-（反）扭结散射一直是一个热门研究课题，相比之下，更为简单的扭结-介子散射反而被研究得相对较少。扭结-介子散射大致可分为两种情形：经典情形和量子情形。在经典扭结-介子散射中允许两种过程：介子融合和介子反射。在量子情形，扭结-介子散射则有相比于经典情形更丰富的现象，其中包括弹性散射和非弹性散射。本文的研究对象正是一个介子和一个扭结的非弹性散射的量子理论。

本文中，我们首先回顾了近年来发展的线性化的孤子微扰理论，其形式在单扭结空间（one-kink sector）尤为简单，借助于它，求解扭结-介子非弹性散射过程的振幅和概率可以简化为扭结标架中的微扰问题。尽管人们通常感兴趣的 Sine-Gordon 孤子和 $\phi^4$ 扭结都是非反射性扭结，为了考虑更一般的情况，我们研究了量子反射性扭结，并发现介子波包在传播过程中不同位置的振幅和量子力学中粒子通过对称势垒或势阱的散射的反射系数和透射系数相对应。我们计算了扭结态的约化内积，以解决不可归一化态的红外发散问题。接着我们考虑扭结和介子在 1+1 维标量量子场论中的非弹性散射。领头阶存在三个非弹性散射过程：（1）介子倍增（末态为两个介子和一个扭结）；（2）斯托克斯散射（末态为介子和激发态扭结）；（3）反斯托克斯散射（初态中扭结处于激发态，末态中扭结处于退激发态）。我们首次计算了这三种过程的领头阶概率以及对于不同动量的末态介子的微分概率。我们首先对任意标量扭结得出一般性的结果，然后将它们应用于 $\phi^4$ 双势阱模型的扭结以得出具体的解析和数值结果。最后，我们相信我们的方法可以被推广到更高维，例如磁单极子情形。

**关键词：** 孤子，扭结，介子，约化内积，非弹性散射

I

# Abstract

Classical kink-(anti)kink scattering has always been a popular research topic, whereas the simpler kink-meson scattering has been relatively less studied. Kink-meson scattering can be roughly divided into two situations: classical and quantum. In classical kink-meson scattering, two processes are allowed: meson fusion and meson reflection. In the quantum case, kink-meson scattering exhibits richer phenomena than in the classical case, including elastic and inelastic scattering. The research subject of this thesis is the quantum theory of inelastic scattering of a meson with a kink.

In this thesis, we first review the linearized soliton perturbation theory developed in recent years, which is particularly simple in the one-kink sector. Using it, the amplitude and probability of kink-meson inelastic scattering can be simplified into a perturbative problem in the kink frame. Although the Sine-Gordon soliton and $\phi^4$ kink, which people are usually interested in, are both reflectionless kinks, in order to consider more general cases, we study quantum reflective kinks and find that the amplitude of meson wave packets at different positions during propagation corresponds to the reflection and transmission coefficients of particles scattered by symmetric potential barriers or potential wells in quantum mechanics. We calculate the reduced inner product of the kink states to solve the infrared divergence problem of non-normalizable states. Then we consider the inelastic scattering of a meson off of a kink in a (1+1)-dimensional scalar quantum field theory. At leading order there are three inelastic scattering processes: (1) meson multiplication (the final state is two mesons and a kink); (2) Stokes scattering (the final state is a meson and an excited kink); (3) anti-Stokes scattering (the initial kink is excited and the final kink is de-excited). For the first time, we calculate the leading-order probabilities of these three processes and the differential probabilities for final-state mesons with different momenta. We first obtain general results for arbitrary scalar kinks and then apply them to the kinks of the $\phi^4$ double well model to obtain analytical and numerical results. Finally, we believe that our method can be generalized to higher dimensions, such as the case of monopoles.

...

# 目 录

# 图形列表

x

# 记号列表

| 算符 | 描述 |
|---|---|
| $\phi(x), \pi(x)$ | 实标量场和它的共轭动量 |
| $B_k^\ddagger, B_k$ | 以正规模为基的产生和湮灭算符 |
| $\phi_0, \pi_0$ | 以正规模为基的 $\phi(x)$ 和 $\pi(x)$ 的零模 |

| 哈密顿量 | 描述 |
|---|---|
| $H$ | 初始哈密顿量 |
| $H'$ | 初始哈密顿量 $H$ 中 $\phi(x)$ 被扭结解 $f(x)$ 平移后的新哈密顿量 |
| $H'_n$ | $H'$ 中的 $\phi^n$ 项 |

| 符号 | 描述 |
|---|---|
| $m$ | $\phi$ 场的质量项（介子质量） |
| $\beta$ | $m/2$ 的简化记号（介子质量的一半） |
| $f(x)$ | 扭结的经典解 |
| $\mathcal{D}_f$ | 将 $\phi(x)$ 平移一个经典扭结解的算符 |
| $\mathfrak{g}_B$ | 正规模中的零模 |
| $\mathfrak{g}_S$ | 正规模中的形模 |
| $\mathfrak{g}_k$ | 正规模中的连续模 |
| $\gamma_i^{mn}$ | $\phi_0^m B^{\ddagger n}$ 的第 $i$ 阶系数 |
| $V_{ijk}$ | 势的导数和各个模函数的缩并 |
| $\mathcal{I}(x)$ | 根据维克定理得到的缩并因子 |
| $k_i$ | 正规模中的动量类比 |
| $\omega_k$ | $k$ 对应的频率 |
| $\tilde{\mathfrak{g}}$ | 正规模 $\mathfrak{g}$ 的傅立叶反变换 |
| $Q_n$ | 扭结能量（质量）的 $n$ 圈修正 |

# 第 1 章　绪论

## 1.1　研究动机

自从人们在文献 [1] 中的 $\phi^4$ 双势阱模型中发现共振窗口的分形模式以来，经典扭结-（反）扭结散射一直是一个热门研究课题。例如，最近的一系列工作试图了解共振到底是由单个扭结的形模（shape mode）引起的，还是由结合的扭结-反扭结系统的集体束缚模引起的 [2–4]。

相比之下，人们对更简单的扭结-介子散射过程反而关注得相对较少 [5–10]。在经典的 $\phi^4$ 场论中，文献 [11, 12] 发现了介子会对扭结施加一个负压。产生这种负压的原因很简单，在扭结内部，$n$ 个频率为 $\omega$ 的介子可以融合成一个频率为 $n\omega$ 的介子，由于质壳条件，它的动量比原始的 $n$ 个介子的动量总和更多。动量守恒要求这个额外的动量来自于扭结，这迫使扭结朝相反方向移动。

这种介子融合似乎是经典 $\phi^4$ 模型中介子-扭结散射过程唯一可能发生的现象。但在反射性扭结的情形下，介子也可能发生反射，导致正压力 [13]。因此，一般来说，在经典扭结-介子散射中允许两种过程：介子融合和介子反射。

在量子理论中，我们期望出现更丰富的现象。其一，我们预期会出现与多形模激发相对应的无限个不稳定激发态。这些应当作为扭结-介子弹性散射中的窄共振出现。其次我们也预期存在扭结-介子非弹性散射。例如，拉曼光谱的出现将成为可能（单色介子与扭结散射，激发内部形模）。扭结的激发谱可以从观察到的散射介子的能量减少中读出。在逆过程中，形模的退激发也是可能的。最后，在量子理论中，我们预期不仅存在介子融合现象，还存在介子裂变现象。

## 1.2　研究意义

我们认为研究扭结-介子散射的量子理论具有它本身的意义，此外它还有助于我们理解扭结与其环境的相互作用。在线性框架下，我们期望这种相互作用仅由三种过程主导。另一方面，离线性框架不太远的地方，将出现辐射振幅更高阶的过程，例如介子融合 [11–13]。随着我们朝经典框架过渡，这些变得更加相关。在不久的将来，我们希望理解量子理论中此类高阶过程。尽管本文我们是在 1+1 维情况下讨论，我们相信我们的方法可以被推广到更高维，例如磁单极子的情形，这将极大地帮助我们理解质量隙的产生机制，事实上这也正是我们进行这一系列工作的动机。本文中，我们的研究对象是扭结-介子非弹性散射的

量子理论，因其计算过程相比于扭结-介子弹性散射更为简单，图 1-1 展示了我们期望的三种扭结-介子非弹性散射过程（为简单起见，只展示了向前散射的情形）。我们将在未来的工作中研究扭结-介子弹性散射。

## 1.3 线性化的孤子微扰理论

在文献 [14] 中引入的扭结哈密顿量中，所有这些过程都有望在树图级别发生。然而，在允许多圈计算的传统方法中，例如文献 [15, 16] 中的集体坐标方法，它们的计算难度非常大。近年来，一种全新的方法线性化的孤子微扰理论被提出，该方法首次在文献 [17] 中在一圈水平被提出，在随后的文献 [18] 中发展到了更高圈。这种新的形式在单扭结空间（one-kink sector）尤为简单[a]（单扭结空间包含单个扭结和任意数量的介子和杂质）。线性化的孤子微扰理论的优势在于，相比于传统方法极大地降低了计算难度，并且使得对一些用传统方法本质上无法计算的物理量的计算成为可能。

线性化的孤子微扰理论有着极其丰富的应用场景。比如文献 [19] 中使用截断而非正规序来正规化发散，解释了为何直接对扭结哈密顿量进行能量截断会导致错误的一圈扭结质量结果；文献 [20] 计算了一个不稳定的多形模激发态的寿命；文献 [21] 计算了 $\phi^4$ 扭结的形状因子；文献 [22] 中讨论了一般情况下的扭结-介子散射的形状因子；谱阱是位置空间，超过它形模会合并到连续模，其在经典理论中有非连续性，而文献 [23] 表明在量子理论中它们是连续的；文献 [24] 发现在平移算符中增加一个扭结构型的量子修正后，蝌蚪图可以被消除，从而极大地简化计算；文献 [25, 26] 研究了 $\phi^4$ 扭结的两圈质量修正；文献 [27] 研究了形模激发态能量的领头阶修正，并预期在有限的耦合下，形模会融合到连续模；文献 [28] 发现了激发态扭结对应的量子态；文献 [29] 概述了对两圈态的研究；文献 [30] 给出了扭结态的薛定谔波函数泛函。

同样地，线性化的孤子微扰理论也非常适用于本文将要研究的课题——介子和非相对论性扭结的散射。作为全文的铺垫，我们首先在本节回顾线性化的孤子微扰理论，以及将在下一节简要地研究量子反射性扭结 [31]。

文献 [17, 18] 中引入了一种新的哈密顿形式，用于 1+1 维标量量子场论的扭结空间（kink sector）中的计算。扭结空间是由有限数量的基本介子和单个量子扭结组成的福克空间。我们把不含扭结的介子福克空间称为真空空间（vacuum sector）。

---

[a]注意我们需要在模空间中选择基点。如果扭结距离基点太远，比如扭结-（反）扭结散射情形会发生的那样，那么必须把在不同基点处计算出的演化加以组合。

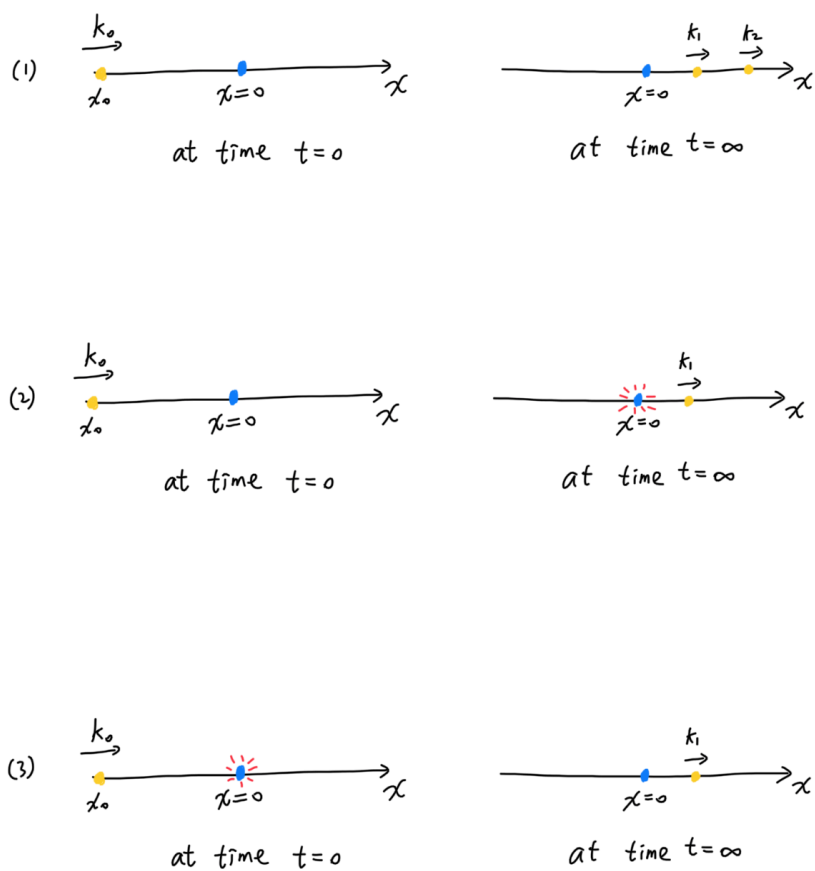

**图 1-1 扭结-介子非弹性散射的三种期望过程。** 黄点表示介子，蓝点表示扭结。初始 $t=0$ 时刻，介子均位于 $x=x_0$ 处且动量为 $k_0$，扭结位于 $x=0$ 处（由于我们在质心系中讨论，且扭结质量远大于介子质量，可近似认为散射前后扭结位置始终在 $x=0$）。（**1**）散射前后扭结均为基态，介子由一个分裂为两个，我们称此过程为介子倍增；（**2**）散射前后介子数目未发生变化，扭结从基态被激发到激发态，我们称此过程为斯托克斯散射；**2**）散射前后介子数目未发生变化，扭结从激发态退激到基态，我们称此过程为反斯托克斯散射。

**Figure 1-1 Three expected processes of kink-meson inelastic scattering.** The yellow dots represent mesons, and the blue dots represent kinks. At the initial time $t = 0$, the mesons are located at $x = x_0$ and have a momentum of $k_0$, while the kinks are located at $x = 0$ (since we are discussing in the center-of-mass frame and the kink mass is much larger than the meson mass, we can assume that the kink position remains at $x = 0$ before and after scattering). The three processes are: (1) before and after scattering, the kink is in the ground state, and one meson splits into two, which we call meson multiplication; (2) the number of meson remains unchanged before and after scattering, and the kink is excited from the ground state to the excited state, which we call Stokes scattering; (3) the number of meson remains unchanged before and after scattering, and the kink is de-excited from the excited state to the ground state, which we call anti-Stokes scattering.

真空空间中的态可以在微扰理论中构建。人们可以以一组平面波的基分解场，构造生成和湮灭算符。真空被定义为被所有的湮灭算符湮灭的态，真空空间是由有限数量的生成算符作用于真空而产生的。我们可以通过求解哈密顿本征值问题微扰地得出哈密顿本征态。

然而这种微扰方法对扭结空间失效。这在经典理论中十分明显，因为场的高阶矩不会趋于零。扭结空间对应于接近经典扭结解 $\phi(x,t) = f(x)$ 的经典场构型。而 $\phi(x,t) - f(x)$ 的高阶矩很小，因此人们期望通过对 $\phi(x,t) - f(x)$ 进行微扰处理能够得到扭结空间中的态。

线性化的孤子微扰理论是在量子场论中实现上述思路的一种形式。我们可以在薛定谔绘景中构造如下的幺正平移算符 $\mathcal{D}_f$

$$\mathcal{D}_f = \mathrm{Exp}\left[-i\int dx f(x)\pi(x)\right], \qquad \mathcal{D}_f^\dagger \phi(x)\mathcal{D}_f = \phi(x) + f(x). \tag{1-1}$$

这里我们将 $\mathcal{D}_f$ 视为被动变换，即重新命名希尔伯特空间的坐标系并对作用于其上的算符进行变换。更准确地说，我们将 扭结标架 定义为希尔伯特空间上的坐标系，其中右矢 $|\psi\rangle$ 表示在通常的"定义标架"中定义的态 $\mathcal{D}_f|\psi\rangle$。这个平移算符 $\mathcal{D}_f$ 在量子理论中的作用与经典理论中的平移相同

$$: F[\phi(x), \pi(x)] :_a \mathcal{D}_f = \mathcal{D}_f : F[\phi(x) + f(x), \pi(x)] :_a. \tag{1-2}$$

小结一下，我们上面做法的本质是构建了一个扭结哈密顿量 $H'$，它与原始哈密顿量 $H$（我们也称之为定义哈密顿量，因其定义了理论）具有相同的能谱。这个原始哈密顿量具有一般的形式。当定义哈密顿量在希尔伯特空间的定义标架中生成时间演化时，幺正变换 $\mathcal{D}_f^\dagger$ 将此定义标架变换到扭结标架。我们可以使用 $H'$ 在扭结标架中测量质量并进行时间演化。等谱特性意味着我们可以正确地用下面的方法来计算扭结质量

$$H|K\rangle = E|K\rangle \Rightarrow H'\mathcal{D}_f^\dagger|K\rangle = E\mathcal{D}_f^\dagger|K\rangle. \tag{1-3}$$

而时间演化算符和空间平移算符分别由扭结哈密顿量 $H'$ 和扭结动量 $P'$ 给出

$$H' = \mathcal{D}_f^\dagger H \mathcal{D}_f, \qquad P' = \mathcal{D}_f^\dagger P \mathcal{D}_f = P + \sqrt{Q_0}\pi_0. \tag{1-4}$$

来看一下如此做给我们带来的好处。在扭结标架中，扭结空间是使用生成算符微扰构造而成的。因此，在存在一个扭结的情况下，哈密顿本征态、形状因子、甚至各种过程的振幅和概率的构建都被简化为扭结标架中的微扰问题。

然而在获得好处的同时我们也有所失去。我们必须选择一个特定的扭结解 $f(x)$。在满足平移不变性的理论中，对每一个实数 $x_0$ 选择的 $f(x-x_0)$ 都对应一个模空间。因此我们失去了显然的平移不变性。我们必须在靠近模空间中的某个基点局域地研究物理。然而，如果我们对具有平移不变性的态（更准确地说是动量本征态）感兴趣（这正是本文的情况）[b]，那么理解模空间中的任意一个特定区域就足以保证我们能够理解所有区域。所以这将不是问题，因此我们只是以 $x_0$ 为基点进行微扰，并且可以随时使用平移不变性以简化表达式。对于运动的扭结的情形，文献 [32] 中有着详细的研究，其中对静止扭结进行伪转动（boost）来定义量子运动扭结，并且定义了扭结波包。

### 1.3.1 细节实现

尽管我们相信线性化的孤子微扰理论这种形式能够非常普遍地应用于我们感兴趣的更多的现象学，到目前为止，我们只是将它应用到具有以下形式的薛定谔绘景中的哈密顿量

$$H = \int dx : \mathcal{H}(x) :_a, \quad \mathcal{H}(x) = \frac{\pi^2(x)}{2} + \frac{(\partial_x \phi(x))^2}{2} + \frac{V(\sqrt{\lambda}\phi(x))}{\lambda}. \quad (1\text{-}5)$$

这里 $V$ 是简并势，$\phi(x,t) = f(x)$ 是经典运动方程的解，它在一维空间 $x$ 的正负无穷远处的取值对应于势能取两个简并最小值的场值。

符号 $::_a$ 表示平面波的产生和湮灭算符的正规序。它定义于质量 $m$，其有两个定义

$$m^2 = V^{(2)}(\sqrt{\lambda}f(\pm\infty)), \quad V^{(n)}(\sqrt{\lambda}\phi(x)) = \frac{\partial^n V(\sqrt{\lambda}\phi(x))}{(\partial\sqrt{\lambda}\phi(x))^n}, \quad (1\text{-}6)$$

分别对应于无穷远处的势的两个最小值处的标量质量。如果它们不一致，那么量子修正会使简并破缺，扭结会变成一个加速的假真空气泡壁 [33]，本文中我们将不考虑这种情况。

我们在耦合常数为 $\lambda$ 的情形下得出了扭结哈密顿量 $H'$ 的本征态。为此，我们以 $\lambda$ 的幂次来分解所有物理量。例如，扭结的基态能量 $Q$ 被分解为 $\sum_i Q_i$，其中每个 $Q_i$ 的阶数为 $O(\lambda^{i-1})$。此处 $Q_0$ 是经典扭结解的经典能量。

正规序中，扭结哈密顿量本身分解为 $H'_i$ 之和，它由 $i$ 个基本场构成。它们包括

$$H'_0 = Q_0, \quad H'_1 = 0, \quad H'_{n>2} = \lambda^{\frac{n}{2}-1} \int dx \frac{V^{(n)}(\sqrt{\lambda}f(x))}{n!} : \phi^n(x) :_a. \quad (1\text{-}7)$$

---

[b]本文始终使用扭结和介子的质心系，因此所有态都是总动量算符的本征态，且本征值为零。波包将由具有不同动量的介子构成，注意扭结的动量总是与介子动量之和大小相等而方向相反。平移不变性是指将扭结和介子作为一个系统，整体同时平移。

其中最重要的是 $H_2'$，因为它的特征向量在扭结空间的态的微扰展开的第一步就会被用到。假如用 $x$ 表示它则可以看出一些不同寻常之处，它的形式很接近自由哈密顿量，但质量项却与位置有关。

为了把它表达得更清楚，我们引入扭结的正规模 $\mathfrak{g}(x)$，它定义为扭结附近的小的经典扰动，同时是下面的施图姆-刘维尔方程的解

$$V^{(2)}(\sqrt{\lambda}f(x))\mathfrak{g}(x) = \omega^2\mathfrak{g}(x) + \mathfrak{g}''(x), \qquad \phi(x,t) = e^{-i\omega t}\mathfrak{g}(x). \tag{1-8}$$

它们按各自的频率 $\omega$ 分类。我们把 $\omega_B = 0$ 的实数解 $\mathfrak{g}_B(x)$ 称为零模。任何满足 $0 < \omega_S < m$ 的实数解 $\mathfrak{g}_S(x)$ 称为形模。在这之上是连续模 $\mathfrak{g}_k(x)$，其中 $\omega_k = \sqrt{m^2 + k^2}$。我们进行如下约定

$$\omega_k = \sqrt{m^2 + k^2}, \qquad \mathfrak{g}_k^*(x) = \mathfrak{g}_{-k}(x) \tag{1-9}$$

$$\int dx |\mathfrak{g}_B(x)|^2 = 1, \quad \int dx \mathfrak{g}_{k_1}(x)\mathfrak{g}_{k_2}^*(x) = 2\pi\delta(k_1 - k_2), \quad \int dx \mathfrak{g}_{S_1}(x)\mathfrak{g}_{S_2}^*(x) = \delta_{S_1 S_2}.$$

我们通过下式来约定 $\mathfrak{g}_B$ 的符号

$$\mathfrak{g}_B(x) = -\frac{f'(x)}{\sqrt{Q_0}}. \tag{1-10}$$

正规模可以生成所有的有界函数，因此，我们可以用它们代替平面波来对场进行分解 [34]

$$\phi(x) = \phi_0 \mathfrak{g}_B(x) + \sum\!\!\!\!\!\!\int \frac{dk}{2\pi}\left(B_k^\ddagger + \frac{B_{-k}}{2\omega_k}\right)\mathfrak{g}_k(x), \qquad B_k^\ddagger = \frac{B_k^\dagger}{(2\omega_k)}, \qquad B_S^\ddagger = \frac{B_S^\dagger}{(2\omega_S)} \tag{1-11}$$

$$\pi(x) = \pi_0 \mathfrak{g}_B(x) + i\sum\!\!\!\!\!\!\int \frac{dk}{2\pi}\left(\omega_k B_k^\ddagger - \frac{B_{-k}}{2}\right)\mathfrak{g}_k(x), \qquad B_{-S} = B_S, \qquad \sum\!\!\!\!\!\!\int \frac{dk}{2\pi} = \int \frac{dk}{2\pi} + \sum_S.$$

可以看到我们用算符 $\phi_0$, $\pi_0$, $B$ 和 $B^\ddagger$ 对场进行了分解。这为我们接下来的算符代数计算提供了一组新的基，并且任何算符都可以用这些基本算符来表示。$\phi(x)$ 和 $\pi(x)$ 满足的正则对易关系意味着我们有如下对易关系

$$\left[\phi_0, \pi_0\right] = i, \quad \left[B_{S_1}, B_{S_2}^\ddagger\right] = \delta_{S_1 S_2}, \quad \left[B_{k_1}, B_{k_2}^\ddagger\right] = 2\pi\delta\left(k_1 - k_2\right). \tag{1-12}$$

现在我们可以用这些算符表示 $H_2'$，它的形式如下 [34]

$$H_2' = Q_1 + H_{\text{free}}, \qquad H_{\text{free}} = \frac{\pi_0^2}{2} + \sum_S \omega_S B_S^\ddagger B_S + \int \frac{dk}{2\pi}\omega_k B_k^\ddagger B_k. \tag{1-13}$$

这里 $Q_1$ 是扭结质量的单圈修正

$$Q_1 = -\frac{1}{4}\sum\!\!\!\!\!\!\!\int \frac{dk}{2\pi}\int\frac{dp}{2\pi}\frac{(\omega_p-\omega_k)^2}{\omega_p}\tilde{\mathfrak{g}}_k^2(p) - \frac{1}{4}\int\frac{dp}{2\pi}\omega_p\tilde{\mathfrak{g}}_B(p)\tilde{\mathfrak{g}}_B(p), \qquad \tilde{\mathfrak{g}}(p) = \int dx\mathfrak{g}(x)e^{ipx}. \tag{1-14}$$

$\pi_0^2$ 项是质量为 $Q_0$ 且位置算符为 $\phi_0/\sqrt{Q_0}$ 的自由量子力学粒子的动能。该粒子是扭结的质心。其他项是形模 $S$ 和连续谱模 $k$ 的量子谐振子。$B_S^\ddagger$ 激发一个形模，而 $B_k^\ddagger$ 激发一个连续模。

所有的哈密顿量本征态可以用下面的半经典展开的方式分解

$$|\psi\rangle = \sum_{i=0}^\infty |\psi\rangle_i, \tag{1-15}$$

其中 $|\psi\rangle_i$ 的阶数为 $O(\lambda^{i/2})$。所有本征态 $|\psi\rangle$ 的领头阶分量 $|\psi\rangle_0$ 是 $H'$ 的领头阶本征值方程的解，因此被定义为 $H_2'$ 的本征态。

在扭结标架中，扭结哈密顿量 $H'$ 的扭结基态 $|0\rangle$ 可以按阶数 $O(\lambda^{i/2})$ 分解为 $|0\rangle_i$。我们的半经典展开中的第一项 $|0\rangle_0$ 是 $H_2'$ 的真空。它是 (1-13) 中每一项的基态，因此完全由下面的这些条件表征

$$\pi_0|0\rangle_0 = B_k|0\rangle_0 = B_S|0\rangle_0 = 0. \tag{1-16}$$

类似地，我们可以在领头阶定义包含一个扭结和一个介子的态，包含一个扭结和两个介子的态（将用于介子倍增情形），以及包含一个激发态扭结和一个介子的态（将用于斯托克斯散射和反斯托克斯散射情形）

$$|k\rangle_0 = B_k^\ddagger|0\rangle_0, \qquad |kk'\rangle_0 = B_k^\ddagger B_{k'}^\ddagger|0\rangle_0, \qquad |Sk\rangle_0 = B_S^\ddagger B_k^\ddagger|0\rangle_0. \tag{1-17}$$

## 1.4 量子反射性扭结

现在我们开始简要地研究量子反射性扭结。正如本章开头提到的那样，我们将要在树图级别研究扭结-介子非弹性散射这个新现象。然而到目前为止，形式的应用受到两个限制。首先，所有显式计算都只考虑了由无反射势描述的扭结[c]；其次，到目前为止所有计算都是和时间无关的。

因此在本节中，我们将使用自由扭结哈密顿量来处理扭结-介子散射。如此一来我们解决了上述的两个限制。特别地，我们不仅处理了无反射性扭结，还处理了反射性扭结。我们对系统进行了时间演化，然后我们观察到我们在自由理论中对反射性扭结的处理再现了和相对论量子力学中计算的本质相同的反射和透射系数。

---

[c] 文献 [23] 是个例外，然而对反射性正规模的特殊处理方式体现在执行数值积分的代码中，并且仅在该文献的附录中有粗略描述，本文的处理方式与上述文献中使用的数值代码中的处理方式有关。

### 1.4.1 正规模

考虑具有如下渐近行为的正规模

$$\mathfrak{g}_k(x) = \begin{cases} \mathcal{B}_k e^{-ikx} + \mathcal{C}_k e^{ikx} & \text{若} \quad x \ll -1/m \\ \mathcal{D}_k e^{-ikx} + \mathcal{E}_k e^{ikx} & \text{若} \quad x \gg 1/m \end{cases} \tag{1-18}$$

$$\mathcal{B}_k^* = \mathcal{B}_{-k}, \qquad \mathcal{C}_k^* = \mathcal{C}_{-k}, \qquad \mathcal{D}_k^* = \mathcal{D}_{-k}, \qquad \mathcal{E}_k^* = \mathcal{E}_{-k}.$$

系数 $\mathcal{B}$、$\mathcal{C}$、$\mathcal{D}$、$\mathcal{E}$ 受完备性关系

$$\int dx |\mathfrak{g}_B(x)|^2 = 1, \quad \int dx \mathfrak{g}_{k_1}(x) \mathfrak{g}_{k_2}^*(x) = 2\pi \delta(k_1 - k_2), \quad \int dx \mathfrak{g}_{S_1}(x) \mathfrak{g}_{S_2}^*(x) = \delta_{S_1 S_2} \tag{1-19}$$

以及它在 $x$ 空间中的表达式

$$\mathfrak{g}_B(x)\mathfrak{g}_B(y) + \sum_S \int \frac{dk}{2\pi} \mathfrak{g}_k(x)\mathfrak{g}_k^*(y) = \delta(x-y), \qquad \sum_S \int \frac{dk}{2\pi} = \int \frac{dk}{2\pi} + \sum_S \tag{1-20}$$

的约束。$|k|$ 的不同值的正规模将自动正交，因为它们满足具有不同特征值的相同施图姆-刘维尔方程。然而，我们需要注意确保 $\mathfrak{g}_k(x)$ 和 $\mathfrak{g}_{-k}(x)$ 正交并且正确归一化。由于这里的积分无穷大，我们只需考虑大 $|x|$ 的区域。

我们给模 $\mathfrak{g}_k$ 应用 $k$-空间的完备性关系 (1-19)。由于我们只对大的 $|x|$ 感兴趣，所以计算下式已经足够

$$\lim_{L \to \infty} \left[ \int_{x=-2L}^{x=-L} + \int_{x=L}^{x=2L} \right] dx \, \mathfrak{g}_{k_1}(x) \mathfrak{g}_{k_2}(x) = \begin{cases} 2L & \text{若} \quad k_1 = -k_2 \\ 0 & \text{若} \quad k_1 \neq -k_2. \end{cases} \tag{1-21}$$

易得

$$\lim_{L \to \infty} \left[ \int_{x=-2L}^{x=-L} + \int_{x=L}^{x=2L} \right] \frac{dx}{L} \mathfrak{g}_{k_1}(x) \mathfrak{g}_{k_2}(x)$$

$$= \begin{cases} |\mathcal{B}_k|^2 + |\mathcal{C}_k|^2 + |\mathcal{D}_k|^2 + |\mathcal{E}_k|^2 & \text{若} \quad k = k_1 = -k_2 \\ 2\mathcal{B}_k \mathcal{C}_k + 2\mathcal{D}_k \mathcal{E}_k & \text{若} \quad k = k_1 = k_2 \\ 0 & \text{若} \quad k_1 \neq \pm k_2. \end{cases} \tag{1-22}$$

然后我们有

$$|\mathcal{B}_k|^2 + |\mathcal{C}_k|^2 + |\mathcal{D}_k|^2 + |\mathcal{E}_k|^2 = 2, \qquad \mathcal{B}_k \mathcal{C}_k + \mathcal{D}_k \mathcal{E}_k = 0. \tag{1-23}$$

接下来我们在位置空间 (1-20) 中应用完备性关系。当 $|x|, |y| \gg 1/m$ 时，只有连续模才有贡献，于是只剩下

$$\int \frac{dk}{2\pi} \mathfrak{g}_k(x) \mathfrak{g}_{-k}(y) = \delta(x-y). \tag{1-24}$$

计算等式左边我们得到

$$\int \frac{dk}{2\pi} \mathfrak{g}_k(x)\mathfrak{g}_{-k}(y) = \int \frac{dk}{2\pi} \begin{cases} \left(|\mathcal{B}_k|^2 + |\mathcal{C}_k|^2\right) e^{ik(x-y)} & \text{若} \quad x, y \ll -1/m \\ \left(|\mathcal{D}_k|^2 + |\mathcal{E}_k|^2\right) e^{ik(x-y)} & \text{若} \quad x, y \gg 1/m \\ \left(\mathcal{B}_k \mathcal{E}_k^* + \mathcal{C}_k^* \mathcal{D}_k\right) e^{-ik(x+y)} & \text{若} \quad x \ll -1/m, \ 1/m \ll y. \end{cases} \tag{1-25}$$

我们对 $|x|$ 和 $|y|$ 非常大的情况感兴趣，在这种情况下 $\mathcal{B}_k$、$\mathcal{C}_k$、$\mathcal{D}_k$ 以及 $\mathcal{E}_k$ 相对于 $k$ 的变化比平面波的因子慢得多。为了避免在 $x = -y$ 处产生不必要的贡献，我们提出如下限制

$$\mathcal{B}_k \mathcal{E}_k^* + \mathcal{C}_k^* \mathcal{D}_k = 0. \tag{1-26}$$

此外，为了在大 $|x|$ 和 $|y|$ 的限制下正确地归一化 delta 函数，我们得到了如下约束

$$|\mathcal{B}_k|^2 + |\mathcal{C}_k|^2 = |\mathcal{D}_k|^2 + |\mathcal{E}_k|^2 = 1. \tag{1-27}$$

我们可以用下面的幺正矩阵来总结所有这些约束

$$U = \begin{pmatrix} \mathcal{B}_k & \mathcal{C}_k^* \\ \mathcal{E}_k & \mathcal{D}_k^* \end{pmatrix}, \qquad U^\dagger U = 1 \tag{1-28}$$

### 1.4.2 波包的传播

考虑一个移动的波包

$$\Phi(x) = \text{Exp}\left[-\frac{(x-x_0)^2}{4\sigma^2} + ixk_0\right], \qquad x_0 \ll -1/m, \qquad k_0 \gg \frac{1}{\sigma}, \qquad \sigma \ll |x_0|. \tag{1-29}$$

注意我们始终使用质心系，由于扭结相对于介子的极大质量，可以近似地看作静止在位置坐标的原点 $x = 0$。上式对应于一个介子初始位于扭结的左边，在 $x = x_0$ 处（因而 $x_0 < 0$）以动量 $k_0$ 向右移动。我们将假设它移动得足够快以至于可以忽略色散。

正规模的完备性意味着，在 $|x| \gg 1/m$，任何波包都可以分解为

$$\Phi(x) = \int \frac{dk}{2\pi} \alpha_k \mathfrak{g}_{-k}(x), \qquad \alpha_k = \int dx \Phi(x) \mathfrak{g}_k(x). \tag{1-30}$$

波包表达式 (1-29) 在 $x \ll -1/m$ 处成立，因此我们可以代入在 $x \ll -1/m$ 处有效的 $\mathfrak{g}_k(x)$ 的渐近表达式。然后我们有

$$\alpha_k = 2\sigma\sqrt{\pi} \left(\mathcal{B}_k e^{-i(k-k_0)x_0} e^{-(k-k_0)^2 \sigma^2} + \mathcal{C}_k e^{i(k+k_0)x_0} e^{-(k+k_0)^2 \sigma^2}\right). \tag{1-31}$$

扭结-介子非弹性散射使用狄拉克符号，在扭结标架中，波包对应于单介子态

$$|\Phi\rangle_0 = \int dx \Phi(x)|x\rangle_0 = \int \frac{dk}{2\pi} \alpha_k |k\rangle_0, \qquad |k\rangle_0 = B_k^{\ddagger}|0\rangle_0, \qquad |x\rangle_0 = \int \frac{dk}{2\pi} \mathfrak{g}_k(x)|k\rangle_0. \tag{1-32}$$

它是单扭结空间的介子福克空间分解的一部分。因此，我们将初始位于 $x = x_0$ 处的介子向 $x \sim 0$ 处的扭结移动。我们提醒读者，方程式 (1-16) 中定义的 $|0\rangle_0$ 是单个扭结基态的最低阶近似值，并且是用希尔伯特空间的扭结标架所表示。

在时刻 $t$ 波包演化为

$$|\Phi(t)\rangle_0 = \int dx \Phi(x,t)|x\rangle_0, \qquad \Phi(x,t) = \int \frac{dk}{2\pi} e^{-i\omega_k t} \alpha_k \mathfrak{g}_{-k}(x). \tag{1-33}$$

当 $\sigma$ 足够大时，$k$ 总是接近于 $k_0$ 或 $-k_0$。因此，我们将以 $(k \pm k_0)$ 展开到一阶。波包色散仅从第二阶开始出现，因此在这样的近似中不必考虑。在第一阶有

$$\omega_k = \omega_{k_0} + (\pm k - k_0)\frac{k_0}{\omega_{k_0}}. \tag{1-34}$$

将 (1-31) 和 (1-34) 代入等式 (1-33) 即得到波包的演化。

在 $x \ll 0$ 处我们得到波包

$$\begin{aligned}\Phi(x,t) &= 2\sigma\sqrt{\pi}e^{-i\omega_{k_0}t} \int \frac{dk}{2\pi} \Bigg( \mathcal{B}_k e^{-i(k-k_0)\left(x_0 + \frac{k_0}{\omega_{k_0}}t\right)} e^{-(k-k_0)^2\sigma^2} \\ &\quad + C_k e^{i(k+k_0)\left(x_0 + \frac{k_0}{\omega_{k_0}}t\right)} e^{-(k+k_0)^2\sigma^2} \Bigg) \left(\mathcal{B}_k^* e^{ikx} + C_k^* e^{-ikx}\right). \end{aligned} \tag{1-35}$$

对于足够大的 $\sigma$，第一个高斯分布集中在 $k = k_0$ 处，第二个高斯分布集中在 $k = -k_0$ 处，因此我们可以用它们在 $k = \pm k_0$ 处的值来简化系数 $\mathcal{B}_k$ 和 $C_k$，得到

$$\begin{aligned}\Phi(x,t) &= e^{-i\omega_{k_0}t}\Bigg(\left(\left|\mathcal{B}_{k_0}\right|^2 + \left|C_{k_0}\right|^2\right)\mathrm{Exp}\left[-\frac{\left(-x + x_0 + \frac{k_0}{\omega_{k_0}}t\right)^2}{4\sigma^2} + ik_0 x\right] \\ &\quad + 2\mathcal{B}_{k_0}C_{k_0}^*\mathrm{Exp}\left[-\frac{\left(x + x_0 + \frac{k_0}{\omega_{k_0}}t\right)^2}{4\sigma^2} - ik_0 x\right]\Bigg). \end{aligned} \tag{1-36}$$

两个高斯因子分别对应于位置 $x$ 等于 $\pm x_t$

$$x_t = x_0 + \frac{k_0}{\omega_{k_0}}t. \tag{1-37}$$

这里我们发现 $k_0/\omega_{k_0}$ 是波包的群速度，$t$ 是传播时间。我们现在考虑 $x \ll 0$ 的情况，因此第一个高斯项在 $x_t \ll 0$ 处，第二个高斯项在 $x_t \gg 0$ 处。

在较早时刻 $t \ll \omega_{k_0}|x_0|/k_0$，只有第一个高斯项可能有贡献，我们得到

$$\Phi(x,t) = e^{-i\omega_{k_0}t+ik_0 x}\text{Exp}\left[-\frac{\left(-x+x_0+\frac{k_0}{\omega_{k_0}}t\right)^2}{4\sigma^2}\right] = e^{-i\left(m^2/\omega_{k_0}\right)t}\Phi\left(x - \frac{k_0}{\omega_{k_0}}t\right). \tag{1-38}$$

这里我们用到了 (1-27)。解释是，在 $(k - k_0)$ 展开的这一阶，波包在到达扭结之前只是简单地向右运动。在 $t \gg \omega_{k_0}|x_0|/k_0$ 时，在介子和扭结发生相互作用之后，只有第二个高斯项可能有贡献。因此，在 $x \ll -1/m$ 处，波函数变为

$$\Phi(x,t) = 2e^{-i\omega_{k_0}t-ik_0 x}\mathcal{B}_{k_0}\mathcal{C}^*_{k_0}\text{Exp}\left[-\frac{\left(x+x_0+\frac{k_0}{\omega_{k_0}}t\right)^2}{4\sigma^2}\right]. \tag{1-39}$$

我们看到扭结的动量改变了正负号，现在 $x < 0$ 处的波包部分正朝相反的方向传播。另外这个位置是 $x = -x_t$，它已经从扭结处发生了反射。不过整个波包并没有发生反射。这里我们只计算了 $x < 0$ 处对波包的贡献，振幅已被减弱了因子 $2\mathcal{B}\mathcal{C}^*$。我们将这个因子解释为反射系数。

我们来看扭结的右侧发生的现象。只需将等式 (1-18) 中 $x \gg 0$ 处的 $\mathfrak{g}_{-k}(x)$ 代入等式 (1-33)，即可得到

$$\begin{aligned}\Phi(x,t) = & e^{-i\omega_{k_0}t}\Bigg(\left(\mathcal{B}_{k_0}\mathcal{D}^*_{k_0}+\mathcal{C}^*_{k_0}\mathcal{E}_{k_0}\right)\text{Exp}\left[-\frac{\left(-x+x_0+\frac{k_0}{\omega_{k_0}}t\right)^2}{4\sigma^2}+ik_0 x\right] \\ & +\left(\mathcal{B}_{k_0}\mathcal{E}^*_{k_0}+\mathcal{C}^*_{k_0}\mathcal{D}_{k_0}\right)\text{Exp}\left[-\frac{\left(x+x_0+\frac{k_0}{\omega_{k_0}}t\right)^2}{4\sigma^2}-ik_0 x\right]\Bigg).\end{aligned} \tag{1-40}$$

由于等式 (1-26) 的约束，第二行为零。第一行的高斯项在 $x = x_t$ 处。

由于我们考虑的是 $x > 0$ 的情况，上面的情况仅在 $x_t > 0$ 时发生，因此 $t \gg \omega_{k_0}|x_0|/k_0$，即介子与扭结发生相互作用之后。这个结果是合理的，因为介子在相互作用之前不可能到达扭结的右边。在散射之后的时刻 $t$，位置是 $x_t$，这意味着在这一阶，介子继续以其初始运动速度经过扭结。我们将 $(\mathcal{B}\mathcal{D}^* + \mathcal{C}^*\mathcal{E})$ 因子解释为透射系数。

### 1.4.3 和量子力学的对比

上述计算和相对论量子力学中的计算是等价的。这是因为量子场论的扭结标架将量子场论中的扭结-介子散射的领头阶问题简化为了一个简单的量子力学练习。当然，上述计算的优点是它可以推广到更高阶，例如可以用来研究介子对扭结的反作用，而这在量子力学中是看不到的，因为在那里扭结被势阱所取代。

但是反射系数果真是 $2\mathcal{BC}^*$ 吗？通常在量子力学中，人们不去使用一组正规模的基来解决这样的问题，因为假使那样做，推导将变得复杂，并且物理图像会变得模糊。而我们之所以在这里使用这一套方法是因为这样做可以与等式 (1-32) 中的扭结标架的介子福克空间简单地联系起来。

为简单起见，我们考虑具有对称势的量子力学。这对应于 $H_2$ 中的对称质量项，它源于量子场论中对称势中的反对称扭结。除了这种情况，一般来说，扭结两侧真空能量的量子修正并不一致，因此扭结会加速 [35]，并且不会对应于哈密顿本征态。虽然这些问题在此处考虑的低阶微扰理论中不会出现，并且它们也不会阻碍我们应用我们的形式，但它们解释了为何我们可以只对对称势感兴趣。

在量子力学中，这一阶的扭结被势阱取代，散射由对称且幺正的 $S$ 矩阵描述

$$S = \begin{pmatrix} t & r \\ r & t \end{pmatrix}, \tag{1-41}$$

其中 $t$ 和 $r$ 是复的透射和反射系数，由于 $S$ 的幺正性，

$$|t|^2 + |r|^2 = 1, \qquad \text{Arg}(r) = \text{Arg}(t) \pm \frac{\pi}{2}. \tag{1-42}$$

考虑 $k > 0$ 的情形。对于单色波，我们的波包 (1-30) 约化为 $\mathfrak{g}_{-k}(x)$。那么 $\mathcal{B}_k^*$ 和 $\mathcal{E}_k^*$ 分别可以解释为从左边和右边入射到势阱中的粒子，而 $\mathcal{C}_k^*$ 和 $\mathcal{D}_k^*$ 对应的是粒子向左和向右远离势阱。$S$ 矩阵将这些系数联系起来

$$\begin{pmatrix} C_k^* \\ \mathcal{D}_k^* \end{pmatrix} = S \begin{pmatrix} \mathcal{E}_k^* \\ \mathcal{B}_k^* \end{pmatrix}. \tag{1-43}$$

矢量 $(C_k, \mathcal{D}_k)$ 和 $(\mathcal{B}_k, \mathcal{E}_k)$ 是正交的（见式 (1-23)）。这可以在相差一个相位 $\phi$ 下固定 $(C_k, \mathcal{D}_k)$

$$S \begin{pmatrix} \mathcal{E}_k^* \\ \mathcal{B}_k^* \end{pmatrix} = \begin{pmatrix} C_k^* \\ \mathcal{D}_k^* \end{pmatrix} = e^{i\phi} \begin{pmatrix} -\mathcal{E}_k^* \\ \mathcal{B}_k^* \end{pmatrix}. \tag{1-44}$$

这一组线性方程的解是

$$\mathcal{B}_k^* = -\frac{t + e^{i\phi}}{r} \mathcal{E}_k^*, \qquad \mathcal{E}_k^* = \frac{e^{i\phi} - t}{r} \mathcal{B}_k^*. \tag{1-45}$$

利用

$$t = |t|e^{i\theta}, \qquad r = i\epsilon_1|r|e^{i\theta}, \qquad \epsilon_1 = \pm 1 \tag{1-46}$$

并结合这两个方程 (1-45)，我们发现

$$\mathcal{E}_k^* = \frac{e^{2i\phi} - |t|^2 e^{2i\theta}}{|r|^2 e^{2i\theta}} \mathcal{E}_k^*. \tag{1-47}$$

这说明

$$e^{i\phi} = \epsilon_2 e^{i\theta}, \qquad \epsilon_2 = \pm 1. \tag{1-48}$$

因此有

$$\mathcal{B}_k = -i\epsilon_1\epsilon_2 \frac{1+\epsilon_2|t|}{|r|} \mathcal{E}_k = -i\epsilon_1\epsilon_2 \sqrt{\frac{1+\epsilon_2|t|}{1-\epsilon_2|t|}} \mathcal{E}_k. \tag{1-49}$$

然后由条件 $|\mathcal{B}_k|^2 + |\mathcal{E}_k|^2 = 1$ 可以导出

$$|\mathcal{B}_k| = \sqrt{\frac{1+\epsilon_2|t|}{2}}, \qquad |\mathcal{E}_k| = \sqrt{\frac{1-\epsilon_2|t|}{2}}. \tag{1-50}$$

因此，使用某个相位 $\lambda$ 表示，

$$\mathcal{B}_k = e^{-i\lambda}\sqrt{\frac{1+\epsilon_2|t|}{2}}, \qquad \mathcal{E}_k = ie^{-i\lambda}\epsilon_1\epsilon_2\sqrt{\frac{1-\epsilon_2|t|}{2}}. \tag{1-51}$$

现在使用最初的方程 (1-44)，我们得到

$$C_k = -e^{-i\phi}\mathcal{E}_k = -ie^{-i(\lambda+\phi)}\epsilon_1\epsilon_2\sqrt{\frac{1-\epsilon_2|t|}{2}}, \qquad \mathcal{D}_k = e^{-i\phi}\mathcal{B}_k = e^{-i(\lambda+\phi)}\sqrt{\frac{1+\epsilon_2|t|}{2}}. \tag{1-52}$$

因此

$$2\mathcal{B}_k C_k^* = 2ie^{i\phi}\epsilon_1\epsilon_2\sqrt{\frac{1+\epsilon_2|t|}{2}}\sqrt{\frac{1-\epsilon_2|t|}{2}} = ie^{i\phi}\epsilon_1\epsilon_2|r| = r. \tag{1-53}$$

因此，量子场论的反射系数 $2\mathcal{BC}^*$ 与量子力学的结果完全一致。类似地

$$\mathcal{B}_k\mathcal{D}_k^* + C_k^*\mathcal{E}_k = e^{i\phi}\left(|\mathcal{B}_k|^2 - |\mathcal{E}_k|^2\right) = e^{i\phi}\epsilon_2|t| = t. \tag{1-54}$$

所以透射系数也与量子力学的结果一致。

### 1.4.4 对量子反射性扭结的小结

本节的计算非常简单，并且这些计算也在过去一个世纪以来的相对论量子力学中进行过无数次，对应于粒子通过对称势垒或势阱的散射。然而我们提出了对计算结果的新颖解释，因为粒子扮演着基本介子场 $\phi(x)$ 的角色，而势阱是对量子扭结的领头阶近似。扭结标架的奇妙之处在于它将非微扰量子场论计算转

化为量子力学中的一个古老的练习。我们现在可以期待扭结-介子散射可以在扭结哈密顿量的领头阶使用量子力学中的技术进行处理，这也将是本文后面的重要部分。

那么我们重复这个古老计算有何收获？我们扩展了文献 [17, 18] 的形式以包括时间演化和反射势。反射势显然很有趣，因为它们描述了大多数的扭结。然而更重要的是，我们认为即使对于自由理论中无反射性的扭结，在树图级别也有反射发生，因此我们认为，即使对于诸如 $\phi^4$ 理论和 Sine-Gordon 模型这些无反射性扭结进行微扰处理，我们在本节开发出的框架也是必要的。

更重要的是，我们终于将时间演化引入了线性化的孤子微扰理论的形式，该形式始终使用薛定谔绘景，其中的所有算符都是时间无关的。此外，除文献 [23] 外，该形式限制了我们对哈密顿本征态的关注，所以即使是态也不会演化。即使在文献 [23] 中，也只是考虑了 $t=0$ 时刻，尽管在那里计算了瞬时加速度。另一方面，本节首次描述了如何将这种形式应用于有限时间的时间演化。

量子场论中的领头阶的态已在每个时刻 $t$ 被求出，即使是在相互作用期间也是如此，尽管 $\mathfrak{g}_k(x)$ 的渐近表达式在那里不能应用。$\mathfrak{g}_k(x)$ 的解析表达式在 Sine-Gordon 和 $\phi^4$ 模型中是已知的，并且在许多其他模型中也发现了非常好的数值近似。因此，等式 (1-31) 和 (1-33) 确实在演化过程中的任何时刻给出了半经典展开式中领头阶的完全态。这相比于欧几里德时间方法是一个进步，欧几里德时间方法通常应用于 $S$ 矩阵的计算，因此本质上仅限于无穷大时间的演化。每一时刻的这些态的领头阶表达式，以及这里引入的扭结标架时间演化的一般形式，当然是计算微扰修正所必需的，其中也包括了我们将要在后几章首次研究的物理现象。

## 1.5 本文余下部分

接下来在第 2 章我们研究扭结态之间的约化内积 [36]，证明了第 3、4 章中经过简化的计算的有效性。然后，我们在第 3、4 章开始计算三种扭结-介子非弹性散射过程发生的概率，其中我们在第 3 章研究介子倍增 [37]，在第 4 章研究斯托克斯散射和反斯托克斯散射 [38]。最后我们在第 5 章做出全文的总结。

# 第 2 章  扭结态之间的约化内积

## 2.1  引言

### 2.1.1  扭结态约化内积的研究动机

文献 [15] 的集体坐标方法允许涉及量子扭结的任意计算[a]。扭结本身的位置被量子化，并且场在这个时间依赖的位置被展开。然而，扭结位置和场展开之间的相互作用是复杂的。为了将算符转化为正则形式以允许量子化，需要经典理论中已有的非线性正则变换。这种变换不会使量子路径积分保持不变，因此在量子理论中，需要在哈密顿量 [16] 中加入无限级数的项。这些复杂性使得解决除最简单的问题之外的所有问题都变得不切实际。例如，扭结质量的两圈修正仅在由于可积性 [40, 41] 或超对称 [42] 已知结果的情况下才能被计算。此外，扭结-介子散射仅限于对有效汤川耦合的领头阶贡献进行计算 [43, 44]。然而，最近这种方法在计算形状因子 [45–47] 方面取得了有前景的发展。

文献 [17, 18] 中提出了一种新的更简单的方法，即线性化的扭结微扰理论，这已经在第 1 章被回顾。在这里扭结场被展开，扭结如同处在一个固定的基点。因此，这些场从一开始就是正则的。代价则是扭结与基点的距离作为耦合中的半经典展开被微扰地处理。因此，如果扭结波包延展超出展开的收敛半径，则结果是不可靠的。在渐近级数的意义上，收敛半径大于扭结的德布罗意波长，但小于其经典直径。这就给出了适用于局部扭结波包的方法[b]，或者更普遍地可以适用于局域孤子波包，如同在许多应用中出现的那样，如孤子暗物质 [48, 49]，固定的 Abrikosov 涡旋和扭结-杂质相互作用 [23, 50]。

然而，有时人们对相反的框架感兴趣，其中扭结处于平移不变态，例如它的基态，或者扭结和有限数目的介子组成的系统的基态。平移不变态是一种量子叠加，对扭结和介子的所有可能的同时且相等的平移求和，这必然使它们的相对距离保持不变，例如文献 [51] 中使用 Skyrme 模型 [52, 53] 处理质子-介子散射的情形，在那里必须同时考虑距基点任意远位置处的多个扭结，因而上面朴素的微扰方法将失效。

这个问题的解决方案是使用平移不变性[c]。关于平移不变的扭结态的所有信

---

[a]从文献 [14] 开始，在一圈水平，人们已经发展了许多强大而有效的处理孤子的方法。近期它们在文献 [39] 中被回顾。

[b]注意应当区分扭结-介子系统质心的波包（它的尺寸被微扰地处理，因此是有界的）和描述与扭结间有着相对位置的一个介子的波包（它可以被精确处理，并且它的单色极限不会使问题复杂化）。

[c]文献 [54] 中提出了一种不需要平移不变性的替代方法。然而到目前为止零模仍没有被包括在内。

息都包含在涉及基点处扭结的构型中，因此只要对该情况进行研究并考虑平移不变性，我们就能得出所有的量。在计算扭结质量的情形，这是通过将态微扰地投影到动量算符的核上，然后求解关于基点的扭结位置的幂级数展开的薛定谔方程来实现的。如果我们对关于基点的展开中的所有系数可以得到解，那么由于平移不变性，任意位置处都可以得解。事实上，该方法已被证明与先前的形状因子计算以及两圈质量修正一致，并且由于其简单性，它提供了对不可积、非超对称扭结质量 [25] 甚至激发态扭结 [28] 以及不可积模型中的扭结形状因子 [21] 的新颖计算。

然而，当人们试图计算动力学量时，这个问题会变得更加严重，因为这里需要计算态的内积。这些平移不变态是不可归一化的，因而它们的内积不存在。通常可以通过以波包的形式对态进行正规化并对波包取大尺寸极限来避免这个问题。不幸的是，在线性化的微扰理论的情况下，这无法完成，因为我们无法处理大于收敛半径的有限波包。人们也许会试图通过压缩空间来避免这个问题。然而，这种紧致化需要特定的边界条件，并且不能保证在紧致化半径取无穷大时有限的贡献不会残留。因此，我们将尽量避免紧致化。

到目前为止，在动力学问题中，我们已经避开了这个复杂处理。在受激扭结衰变 [20] 和本文第 3、4 章将要研究的介子倍增和（反）斯托克斯散射的情况下，出现的内积总是在分子和分母中，其中相同的内积同时出现在概率表达式的分子和分母中，因此我们抵消了它们。然而，在更高阶时，分子和分母中将会出现不能简单抵消的态。

### 2.1.2  本章概述

本章我们提出了一种新的策略来处理平移不变态的内积，它不需要对无穷大进行正规化。由于我们感兴趣的内积总是出现在可观测表达式的分子和分母中，我们用无限体积平移群对两者求商（并注意保持相关的雅可比因子）。这种策略类似于通过全局平移对称性来取规范（同时平移扭结和介子）。直观上，这也与半径为零的紧致化有关，只是扭结和介子之间的距离恒定，因而可以认为它们实际上处于一个无限空间之中。

我们将注意力集中在扭结空间，这是存在一个量子扭结时任意有限数目介子组成的福克空间。然而对其他拓扑空间的推广是显然的。

本章的主要结果是任意两个平移不变态之间的约化内积公式，即式 (2-70)。直观上，普通的内积可以写成集体坐标 $x$ 上的积分，约化内积则通过插入 $\delta(x)$ 来定义。集体坐标通过寻常的刚性平移进行平移变换。在线性化的微扰理论中，我

们不使用集体坐标 $x$，而是使用线性化的坐标 $y$，其平移下的变换相当复杂。我们的公式 (2-70) 将 $\delta(x)$ 替换为 $y'(x)\delta(y)$，其中雅可比因子 $y'(x)$ 是一个算符。令人惊讶的是，由于 $\delta(y)$ 的存在，这个公式比通常的内积公式更简单，因为它不需要用到态对零模（本征值为 $y$）的依赖性。这并不是显然不一致的，因为在平移不变态的情况下，对零模的依赖性完全由平移不变性决定 [18]。这种简化的代价是加入了 $y'(x)$，它在朴素的内积之上包含两个有限的量子修正（这两个量子修正混合了介子数相差一的空间。这些修正反映了扭结的零模在平移时与正规模混合的事实。

本文将涉及到约化内积的三个应用场景。首先，这使我们能够在更坚实的基础上进行形式操作，这些模在第 3 章和第 4 章中被简单地抵消了。此外，它还允许我们处理出现更复杂内积的情况，例如零模的矩阵元。事实上，这在介子倍增振幅和（反）斯托克斯散射振幅的 $O(\sqrt{\lambda})$ 阶贡献已经发生了，这是由于初态或末态有 $O(\sqrt{\lambda})$ 阶修正，而相互作用没有 $O(\sqrt{\lambda})$ 阶修正。我们将用我们发展的形式来计算这些修正，并证明它们在这一阶没有贡献。

最后，我们用它来计算由双介子态组成的单介子态的领头阶修正。它已经在文献 [28] 中进行了计算，但那里的结果涉及哈密顿量筒并位置处的极点，因此处理极点的不同方法会产生合法但不等价的哈密顿本征态。本章中我们将发现对极点的一个特定的处理方法可以对扭结-介子散射得到具有物理动机的初始条件，其中初态永远不会包含两个介子。

在第 2.2 节中，我们展示了我们在量子力学情况下对约化内积的构造。该构造适用于第 2.3 节中量子场论的扭结空间。在第 2.4 节中，我们提供了一些约化内积的例子。最后，在第 2.5 节中，我们应用这种形式来使用平移不变态计算介子倍增的振幅，得到了与第 3 章中相同的结果，那里的主体计算（第 3.3 节及之后）忽略了对初态和末态的量子修正。并且我们直接把这个等价性推广到（反）斯托克斯散射情形。此外我们还计算了与该实验相关的对初态的量子修正，其对应于文献 [28] 中处理极点的方法。

## 2.2 量子力学中的约化内积

我们的主要结果将是单扭结空间中的平移不变态的有限约化内积的表达式。它是通过平移群对普通内积求商而得出的（并注意仔细处理雅可比项）。在本节中，我们将通过在量子力学中定义一个类似的约化内积来作为我们的动机。

雅可比项是非平庸的，因为平移算符 $P'$ 非线性地作用于线性化的坐标 $y$, 这

里 $y$ 定义为 $\phi_0$ 的本征值。然而，平移算符 $P'$ 作为一个简单的平移在集体坐标 $x$ 上线性作用。我们将通过首先计算以集体坐标 $x$ 表示的态的相当简单的约化内积来得出我们的结果，然后我们将推导出集体坐标和线性化的坐标 $y$ 之间的对应条件，这允许我们定义以线性化的坐标表示的态的约化内积。

### 2.2.1 集体坐标：定义

我们首先定义集体坐标以描述量子力学希尔伯特空间中的态。

令 $|e^n\rangle$ 是在平移算符 $P'$ 下不变的态的一组正交基。其中每一个都可以分解为集体坐标算符 $\hat{x}$ 的本征态

$$|e^n\rangle = \int dx |nx\rangle_x, \qquad \hat{x}|nx\rangle_x = x|nx\rangle_x, \qquad [\hat{x}, P'] = i, \tag{2-1}$$

其中 $n$ 是整数量子数，并且

$$P' \int dx F(x)|nx\rangle_x = -i \int dx F'(x)|nx\rangle_x, \qquad {}_x\langle n_1 x_1 | n_2 x_2\rangle_x = \delta_{n_1 n_2} \delta(x_1 - x_2). \tag{2-2}$$

注意，(2-2) 中的第一个关系式意味着 $P'$ 对此基的作用类似于量子力学中的动量算符

$$[\hat{x}, e^{-ix_2 P'}] = x_2 e^{-ix_2 P'} \Rightarrow e^{-ix_2 P'}|nx_1\rangle_x = |n, x_1 + x_2\rangle_x. \tag{2-3}$$

虽然 $|e^n\rangle$ 不可归一化，但我们可以将约化内积定义为

$$\langle e^m | e^n \rangle_{\text{red}} = \delta_{mn}. \tag{2-4}$$

任何平移不变的态 $|\psi\rangle$ 都可以被展开为

$$|\psi\rangle = \sum_n \psi_n |e^n\rangle. \tag{2-5}$$

因此任意的约化内积是

$$\langle \phi | \psi \rangle_{\text{red}} = \sum_n \phi_n^* \psi_n. \tag{2-6}$$

### 2.2.2 线性化的坐标：内积

考虑态 $|ny\rangle_y$ 的另一组基，其中 $\phi_0$ 和 $\pi_0$ 是哈密顿量中的算符，它们满足

$$\phi_0 |ny\rangle_y = y|ny\rangle_y, \qquad \pi_0 \int dy F(y)|ny\rangle_y = -i \int dy F'(y)|ny\rangle_y. \tag{2-7}$$

定义整数量子数 $n$ 使得

$${}_y\langle n_1 y_1 | n_2 y_2\rangle_y = \delta_{n_1 n_2} G_{n_1}(y_1, y_2), \tag{2-8}$$

其中 $G_n$ 是一类我们将要继续确定的函数。

由于 $\phi_0$ 是厄米的，我们有

$$0 = {}_y\langle ny_1|(\phi_0 - \phi_0)|ny_2\rangle_y = (y_1 - y_2){}_y\langle ny_1|ny_2\rangle_y = (y_1 - y_2)G_n(y_1, y_2). \qquad (2\text{-}9)$$

所以只有当 $y_1 = y_2$ 时 $G_n(y_1, y_2)$ 才不为零。因此我们把它简单地写成 $\delta(y_1 - y_2)G_n(y_1)$ 并且

$${}_y\langle n_1 y_1|n_2 y_2\rangle_y = \delta_{n_1 n_2}\delta(y_1 - y_2)G_{n_1}(y_1). \qquad (2\text{-}10)$$

由于 $\pi_0$ 是厄米的，对任意的紧支撑函数 $F_i(y)$ 有

$$\begin{aligned}
0 &= \int dy_1 \int dy_2 \, {}_y\langle ny_1|F_1^*(y_1)(\pi_0 - \pi_0)F_2(y_2)|ny_2\rangle_y & (2\text{-}11)\\
&= \int dy_1 \int dy_2 \, {}_y\langle ny_1|\left[iF_1'^*(y_1)F_2(y_2) + iF_1^*(y_1)F_2'(y_2)\right]|ny_2\rangle_y \\
&= i \int dy_1 \int dy_2 \left[F_1'^*(y_1)F_2(y_2) + F_1^*(y_1)F_2'(y_2)\right]G_n(y_1)\delta(y_1 - y_2) \\
&= i \int dy \partial_y(F_1^*(y)F_2(y))G_n(y) = -i \int dy F_1^*(y)F_2(y)\partial_y G_n(y). & (2\text{-}12)
\end{aligned}$$

由于它对任意的紧支撑函数均成立，我们得到

$$\partial_y G_n(y) = 0 \qquad (2\text{-}13)$$

因此我们将 $G_n(y)$ 替换为 $G_n$，然后可以写出下式

$${}_y\langle n_1 y_1|n_2 y_2\rangle_y = \delta_{n_1 n_2}\delta(y_1 - y_2)G_{n_1}. \qquad (2\text{-}14)$$

最后，我们可以将态 $|ny\rangle_y$ 以因子 $1/\sqrt{G_n}$ 重整化，所以有

$${}_y\langle n_1 y_1|n_2 y_2\rangle_y = \delta_{n_1 n_2}\delta(y_1 - y_2). \qquad (2\text{-}15)$$

### 2.2.3　线性化的坐标：平移

考虑如下平移不变态

$$|\psi\rangle = \sum_n \int dy \hat{\psi}_n(y)|ny\rangle_y \qquad (2\text{-}16)$$

并假设平移算符有如下形式

$$P' = A\pi_0 + B + C\phi_0, \qquad (2\text{-}17)$$

其中 $A$、$B$ 和 $C$ 是与 $\pi_0$ 和 $\phi_0$ 对易的矩阵。注意，^符号并不意味着 $\hat{\psi}_n(y)$ 是算符，它仅仅是 $y$ 基中的系数。

将平移算符作用于平移不变态，我们得到

$$P'|\psi\rangle = \sum_{mn}\int dy\left[-iA_{mn}\hat{\psi}'_n(y) + B_{mn}\hat{\psi}_n(y) + C_{mn}y\hat{\psi}_n(y)\right]|my\rangle_y \quad (2\text{-}18)$$

所以对于平移不变态有

$$A\hat{\psi}'(y) + i(B + yC)\hat{\psi}(y) = 0. \quad (2\text{-}19)$$

特别地，对于小 $\epsilon$ 值有

$$\hat{\psi}(\epsilon) = \hat{\psi}(0) + \epsilon\hat{\psi}'(0) = \hat{\psi}(0) - i\epsilon A^{-1}B\hat{\psi}(0). \quad (2\text{-}20)$$

令 $\hat{v}^j$ 是 $A_{mn}$ 的本征矢，使得

$$\sum_n A_{mn}\hat{v}^j_n = \lambda_j\hat{v}^j_m. \quad (2\text{-}21)$$

考虑由下式定义的平移不变态 $|v^j\rangle$

$$|v^j\rangle = \sum_n\int dy\hat{v}^j_n(y)|ny\rangle_y, \qquad \hat{v}^j_n(0) = \hat{v}^j_n. \quad (2\text{-}22)$$

对其使用平移算符可以得到

$$P'|v^j\rangle = \sum_{mn}\int dy\left[-iA_{mn}\hat{v}^{j\prime}_n(y) + B_{mn}\hat{v}^j_n(y) + C_{mn}y\hat{v}^j_n(y)\right]|my\rangle_y = 0. \quad (2\text{-}23)$$

故

$$\sum_n\left[-iA_{mn}\hat{v}^{j\prime}_n(y) + B_{mn}\hat{v}^j_n(y) + C_{mn}y\hat{v}^j_n(y)\right] = 0. \quad (2\text{-}24)$$

特别地，对于小 $\epsilon$ 值有

$$\hat{v}^j(\epsilon) = (1 - i\epsilon A^{-1}B)\hat{v}^j. \quad (2\text{-}25)$$

现在让我们只考虑固定在 $y$ 处的分量

$$|v^j, y\rangle_y = \sum_n\hat{v}^j_n(y)|ny\rangle_y, \qquad |v^j\rangle = \int dy|v^j, y\rangle_y. \quad (2\text{-}26)$$

它不是平移不变的

$$P'|v^j, 0\rangle_y$$

$$= P'\sum_n\hat{v}^j_n|n0\rangle_y = \sum_n\hat{v}^j_n P'\int dy\delta(y)|ny\rangle_y = \sum_n\hat{v}^j_n\lim_{\sigma\to 0}\frac{1}{\sigma\sqrt{2\pi}}P'\int dye^{-\frac{y^2}{2\sigma^2}}|ny\rangle_y$$

$$= \sum_{mn}\hat{v}^j_n\lim_{\sigma\to 0}\frac{1}{\sigma\sqrt{2\pi}}\int dye^{-\frac{y^2}{2\sigma^2}}\left[iA_{mn}\frac{y}{\sigma^2} + B_{mn} + C_{mn}y\right]|my\rangle_y$$

$$= \sum_{mn}\hat{v}^j_n\lim_{\sigma\to 0}\frac{1}{\sigma\sqrt{2\pi}}\int dye^{-\frac{y^2}{2\sigma^2}}\left[iA_{mn}\frac{y}{\sigma^2} + B_{mn}\right]|my\rangle_y. \quad (2\text{-}27)$$

在最后一个等式中，我们使用了如下事实，即当 $\sigma \to 0$ 并且 $y/\sigma$ 固定时，也有 $y \to 0$。$C$ 项的系数为 $y$，因此 $C$ 项也趋于零。

现在让我们考虑一个有限距离 $\epsilon$ 的变换。我们将通过极限内部的一阶变换来近似它，当我们取 $\sigma \to 0$ 时，我们同时也取 $\epsilon/\sigma \to 0$，因此这是合法的。变换是

$$
\begin{aligned}
& e^{-i\epsilon P'}|v^j,0\rangle_y \\
&= \sum_n \hat{v}_n^j \lim_{\sigma \to 0} \frac{1}{\sigma\sqrt{2\pi}} e^{-i\epsilon P'} \int dy\, e^{-\frac{y^2}{2\sigma^2}} |ny\rangle_y \\
&= \sum_n \hat{v}_n^j \lim_{\sigma \to 0} \frac{1}{\sigma\sqrt{2\pi}} (1 - i\epsilon P') \int dy\, e^{-\frac{y^2}{2\sigma^2}} |ny\rangle_y \\
&= \sum_{mn} \hat{v}_n^j \lim_{\sigma \to 0} \frac{1}{\sigma\sqrt{2\pi}} \int dy\, e^{-\frac{y^2}{2\sigma^2}} \left[\delta_{mn} + \epsilon A_{mn}\frac{y}{\sigma^2} - i\epsilon B_{mn}\right] |my\rangle_y \\
&= \sum_{mn} \hat{v}_n^j \lim_{\sigma \to 0} \frac{1}{\sigma\sqrt{2\pi}} \int dy\, e^{-\frac{y^2}{2\sigma^2}} \left[\delta_{mn} + \epsilon \delta_{mn}\lambda_j\frac{y}{\sigma^2} - i\epsilon\lambda_j(A^{-1}B)_{mn}\right] |my\rangle_y.
\end{aligned}
$$
(2-28)

使用展开 (2-25)

$$
e^{-\frac{(y-\lambda_j\epsilon)^2}{2\sigma^2}} \hat{v}^j(\lambda_j\epsilon) = e^{-\frac{y^2}{2\sigma^2}} \left[1 + \frac{\lambda_j\epsilon y}{\sigma^2} - i\lambda_j\epsilon A^{-1}B\right] \hat{v}^j.
$$
(2-29)

于是我们最后得到

$$
\begin{aligned}
e^{-i\epsilon P'}|v^j,0\rangle_y &= \sum_n \hat{v}_n^j(\lambda_j\epsilon) \lim_{\sigma \to 0} \frac{1}{\sigma\sqrt{2\pi}} \int dy\, e^{-\frac{(y-\lambda_j\epsilon)^2}{2\sigma^2}} |ny\rangle_y \\
&= \sum_n \hat{v}_n^j(\lambda_j\epsilon)|n,\lambda_j\epsilon\rangle_y = |v^j,\lambda_j\epsilon\rangle_y.
\end{aligned}
$$
(2-30)

我们看到线性化的 $y$ 坐标类似于集体 $x$ 坐标，除了由 $\epsilon$ 引起的平移令 $x$ 增加 $\epsilon$，而令 $y$ 增加 $\lambda_j\epsilon$。特别地，这个比率取决于被变换的态的指标 $j$。

### 2.2.4 线性化的坐标：模

要计算线性化的 $y$ 基中的约化模，我们需要将 $y$ 基和 $x$ 基联系在一起。为此，我们需要将它们的普通归一化匹配起来，我们将通过匹配它们的模来实现这一点。$|v^j\rangle$ 和 $|v^j,0\rangle$ 两者都有无穷模。这促使我们定义

$$
|v^j;\epsilon\rangle_y = \int_0^\epsilon dz\, e^{-izP'}|v^j,0\rangle_y.
$$
(2-31)

对于小 $\epsilon$ 的情况，它的模很容易计算

$$\begin{aligned}
\left|\left|v^j;\epsilon\right\rangle_y\right|^2 &= \int_0^\epsilon dz_1 \int_0^\epsilon dz_2 \,_y\langle v^j,0|e^{-i(z_2-z_1)P'}|v^j,0\rangle_y \quad (2\text{-}32)\\
&= \sum_{n_1 n_2} \int_0^\epsilon dz_1 \int_0^\epsilon dz_2 \hat{v}_{n_1}^{j*}(\lambda_j z_1)\hat{v}_{n_2}^{j}(\lambda_j z_2)\,_y\langle n_1,\lambda_j z_1|n_2,\lambda_j z_2\rangle_y\\
&= \sum_{n_1 n_2} \int_0^\epsilon dz_1 \int_0^\epsilon dz_2 \hat{v}_{n_1}^{j*}(\lambda_j z_1)\hat{v}_{n_2}^{j}(\lambda_j z_2)\delta_{n_1 n_2}\delta(\lambda_j(z_1-z_2))\\
&= \frac{1}{\lambda_j} \sum_n \int_0^\epsilon dz \hat{v}_n^{j*}(\lambda_j z)\hat{v}_n^{j}(\lambda_j z).
\end{aligned}$$

现在，直到 $O(\epsilon)$ 阶的修正，我们可以使用近似 $\hat{v}(\lambda_j z) = \hat{v}$。然后我们得到

$$\left|\left|v^j;\epsilon\right\rangle_y\right|^2 = \frac{1}{\lambda_j}\sum_n \hat{v}_n^{j*}\hat{v}_n^{j}\int_0^\epsilon dz = \frac{\epsilon \sum_n \hat{v}_n^{j*}\hat{v}_n^{j}}{\lambda_j} = \frac{\epsilon|\hat{v}^j|^2}{\lambda_j}. \quad (2\text{-}33)$$

### 2.2.5 集体坐标：模

让我们在集体坐标基下写出相同的态 $|v^j\rangle$

$$|v^j\rangle = \sum_n v_n^j |e^n\rangle = \sum_n v_n^j \int dx |nx\rangle_x. \quad (2\text{-}34)$$

再一次地，我们可以通过集体坐标来分解这个态

$$|v^j,x\rangle_x = \sum_n v_n^j |nx\rangle_x. \quad (2\text{-}35)$$

这里由 $\epsilon$ 引起的平移表现为

$$e^{-i\epsilon P'}|v^j,0\rangle_x = |v^j,\epsilon\rangle_x. \quad (2\text{-}36)$$

为了获得有限大小的模，我们再次定义

$$|v^j;\epsilon\rangle_x = \int_0^\epsilon dx|v^j,x\rangle_x. \quad (2\text{-}37)$$

和上面的计算一样，它的模是

$$\left|\left|v^j;\epsilon\right\rangle_x\right|^2 = \epsilon \sum_n v_n^{j*}v_n^j = \epsilon|v^j|^2. \quad (2\text{-}38)$$

### 2.2.6 集体坐标和线性化的坐标的匹配

现在我们来看在希尔伯特空间中使用 $x$ 基和 $y$ 基的匹配。显然，这种匹配必须保持模不变，所以我们选择

$$|v^j;\epsilon\rangle_y = \frac{1}{\sqrt{\lambda_j}}\frac{|\hat{v}^j|}{|v^j|}|v^j;\epsilon\rangle_x. \quad (2\text{-}39)$$

除以 $\epsilon$ 并取极限 $\epsilon \to 0$，这变为

$$\sum_n \hat{v}_n^j |n0\rangle_y = |v^j, 0\rangle_y = \frac{1}{\sqrt{\lambda_j}} \frac{|\hat{v}^j|}{|v^j|} |v^j, 0\rangle_x = \frac{1}{\sqrt{\lambda_j}} \frac{|\hat{v}^j|}{|v^j|} \sum_n v_n^j |n0\rangle_x, \quad (2\text{-}40)$$

因而有

$$\sum_n \frac{\hat{v}_n^j}{|\hat{v}^j|} |n0\rangle_y = \frac{1}{\sqrt{\lambda_j}} \sum_n \frac{v_n^j}{|v^j|} |n0\rangle_x. \quad (2\text{-}41)$$

在 $\epsilon$ 的领头阶，由平移变换可以得出

$$\sum_n \frac{\hat{v}_n^j}{|\hat{v}^j|} |n, \lambda_j \epsilon\rangle_y = \frac{1}{\sqrt{\lambda_j}} \sum_n \frac{v_n^j}{|v^j|} |n\epsilon\rangle_x. \quad (2\text{-}42)$$

### 2.2.7 检查 $v^j$ 的正交性

回忆等式 (2-2) 中，$|n0\rangle_x$ 基是正交的，从等式 (2-15) 也可以看出，$|n0\rangle_y$ 基是正交的。由于 $\hat{v}^j$ 是矩阵 $A$ 的本征矢（我们假设 $A$ 是厄米矩阵），因此 $\hat{v}^j$ 之间也是正交的。

接下来看 $v^j$。它们只是用 $|n0\rangle_x$ 基表示而不是用 $|n0\rangle_y$ 基表示的 $\hat{v}^j$（相差一个 $\lambda_j$ 的缩放因子）。由于两组基分别都是正交的，我们希望 $v^j$ 仍然保持正交。我们来检查一下情况是否确实如此。

让我们将等式 (2-39) 除以 $\sqrt{\epsilon}$ 接着在两个不同的本征值 $j_1 \neq j_2$ 处和自身内积。我们将 $\min\{\lambda_{j_1}, \lambda_{j_2}\}$ 表示为 $\lambda_{\min}$，将 $\max\{\lambda_{j_1}, \lambda_{j_2}\}$ 表示为 $\lambda_{\max}$。这里的计算类似于第 2.2.4 小节。等式左边是

$$\begin{aligned}
&\frac{1}{\epsilon} {}_y\langle v^{j_1}; \epsilon | v^{j_2}; \epsilon \rangle_y \\
=& \frac{1}{\epsilon} \int_0^\epsilon dz_1 \int_0^\epsilon dz_2 \, {}_y\langle v^{j_1}, 0 | e^{-i(z_2 - z_1) P'} | v^{j_2}, 0 \rangle_y \\
=& \frac{1}{\epsilon} \sum_{n_1 n_2} \int_0^\epsilon dz_1 \int_0^\epsilon dz_2 \hat{v}_{n_1}^{j_1 *}(\lambda_j z_1) \hat{v}_{n_2}^{j_2}(\lambda_j z_2) {}_y\langle n_1, \lambda_{j_1} z_1 | n_2, \lambda_{j_2} z_2 \rangle_y \\
=& \frac{1}{\epsilon} \sum_{n_1 n_2} \int_0^\epsilon dz_1 \int_0^\epsilon dz_2 \hat{v}_{n_1}^{j_1 *}(\lambda_j z_1) \hat{v}_{n_2}^{j_2}(\lambda_j z_2) \delta_{n_1 n_2} \delta(\lambda_{j_1} z_1 - \lambda_{j_2} z_2) \\
=& \frac{1}{\epsilon \lambda_{j_1} \lambda_{j_2}} \sum_{n_1 n_2} \int_0^\epsilon d(\lambda_{j_1} z_1) \int_0^\epsilon d(\lambda_{j_2} z_2) \hat{v}_{n_1}^{j_1 *}(\lambda_j z_1) \hat{v}_{n_2}^{j_2}(\lambda_j z_2) \delta_{n_1 n_2} \delta(\lambda_{j_1} z_1 - \lambda_{j_2} z_2) \\
=& \frac{1}{\epsilon \lambda_{j_1} \lambda_{j_2}} \sum_n \int_0^{\lambda_{j_1} \epsilon} d\tilde{z}_1 \int_0^{\lambda_{j_2} \epsilon} d\tilde{z}_2 \hat{v}_n^{j_1 *}(\tilde{z}_1) \hat{v}_n^{j_2}(\tilde{z}_2) \delta(\tilde{z}_1 - \tilde{z}_2) \\
=& \frac{1}{\epsilon \lambda_{\min} \lambda_{\max}} \sum_n \int_0^{\lambda_{\min} \epsilon} d\tilde{z} \hat{v}_n^{j_1 *}(\tilde{z}) \hat{v}_n^{j_2}(\tilde{z}).
\end{aligned} \quad (2\text{-}43)$$

再次地，和第 2.2.4 小节中一样，直到 $O(\epsilon)$ 阶的修正，我们可以使用近似 $\hat{v}(\tilde{z}) = \hat{v}$。然后我们得到

$$\frac{1}{\epsilon}{}_y\langle v^{j_1};\epsilon|v^{j_2};\epsilon\rangle_y = \frac{1}{\epsilon\lambda_{\min}\lambda_{\max}}\sum_n \hat{v}_n^{j_1*}\hat{v}_n^{j_2}\int_0^{\lambda_{\min}\epsilon} d\tilde{z} = \frac{\sum_n \hat{v}_n^{j_1*}\hat{v}_n^{j_2}}{\lambda_{\max}} = 0, \quad (2\text{-}44)$$

这里的最后一个等式使用了 $\hat{v}$ 的正交性。

这里我们假设所有的 $\lambda_j$ 都大于 0。因为我们对量子扭结感兴趣，$A$ 将是一个正标量加上一个耦合常数的幂指数压低的修正，所以这是小耦合的情况。

等式右边的计算类似

$$\begin{aligned}
0 &= \frac{1}{\epsilon\sqrt{\lambda_{j_1}\lambda_{j_2}}}\frac{|\hat{v}^{j_1}||\hat{v}^{j_2}|}{|v^{j_1}||v^{j_2}|}{}_x\langle v^{j_1};\epsilon|v^{j_2};\epsilon\rangle_x & (2\text{-}45)\\
&= \frac{1}{\epsilon\lambda_{j_1}\lambda_{j_2}}\int_0^\epsilon dx_1\int_0^\epsilon dx_2\, {}_x\langle v^{j_1},x_1|v^{j_2},x_2\rangle_x\\
&= \frac{1}{\epsilon\lambda_{j_1}\lambda_{j_2}}\sum_{n_1 n_2} v_{n_1}^{j_1*}v_{n_2}^{j_2}\int_0^\epsilon dx_1\int_0^\epsilon dx_2\, {}_x\langle n_1,x_1|n_2,x_2\rangle_x\\
&= \frac{1}{\epsilon\lambda_{j_1}\lambda_{j_2}}\sum_{n_1 n_2} v_{n_1}^{j_1*}v_{n_2}^{j_2}\int_0^\epsilon dx_1\int_0^\epsilon dx_2\, \delta_{n_1 n_2}\delta(x_1-x_2)\\
&= \frac{1}{\epsilon\lambda_{j_1}\lambda_{j_2}}\sum_n v_n^{j_1*}v_n^{j_2}\int_0^\epsilon dx = \frac{\sum_n v_n^{j_1*}v_n^{j_2}}{\lambda_{j_1}\lambda_{j_2}}.
\end{aligned}$$

从第二行到第三行，我们使用了等式 (2-35)。在从第一行到第二行的过程中，我们使用了稍后将予以证明的恒等式 (2-48)。所以 $v$ 也是正交的

$$\sum_n v_n^{j_1*}v_n^{j_2} = 0. \quad (2\text{-}46)$$

### 2.2.8 线性化的坐标：约化内积

最后，我们准备来计算 $|v^j\rangle$ 的约化模。约化模的平方定义为 $|v^j|^2$。以下操作在小 $y$ 下有效，原因是在小 $y$ 时 $y$-坐标微扰理论是才有效的。

$$\begin{aligned}
|v^j\rangle &= \sum_n\int dy\,\hat{v}_n^j(y)|ny\rangle_y = \frac{1}{\sqrt{\lambda_j}}|\hat{v}^j|\sum_n \frac{v_n^j}{|v^j|}\int dy|n,y/\lambda_j\rangle_x & (2\text{-}47)\\
&= \sqrt{\lambda_j}|\hat{v}^j|\sum_n \frac{v_n^j}{|v^j|}\int dx|n,x\rangle_x = \sqrt{\lambda_j}|\hat{v}^j|\sum_n \frac{v_n^j}{|v^j|}|e^n\rangle.
\end{aligned}$$

将结果和等式 (2-34) 对比，我们得到

$$\sqrt{\lambda_j}|\hat{v}^j| = |v^j|. \quad (2\text{-}48)$$

因此约化模是

$$\langle v^j | v^j \rangle_{\text{red}} = \lambda_j |\hat{v}^j|^2 \sum_{n_1 n_2} \frac{v_{n_1}^{j*} v_{n_2}^{j}}{|v^j|^2} \langle e^{n_1} | e^{n_2} \rangle_{\text{red}} \tag{2-49}$$

$$= \lambda_j |\hat{v}^j|^2 \sum_{n_1 n_2} \frac{v_{n_1}^{j*} v_{n_2}^{j}}{|v^j|^2} \delta_{n_1 n_2} = \lambda_j |\hat{v}^j|^2 = \sum_{mn} \hat{v}_m^{j*} A_{mn} \hat{v}_n^j.$$

类似地，如果 $j \neq k$ 那么约化内积是

$$\langle v^j | v^k \rangle_{\text{red}} = \sqrt{\lambda_j \lambda_k} |\hat{v}^j||\hat{v}^k| \sum_{n_1 n_2} \frac{v_{n_1}^{j*} v_{n_2}^{k}}{|v^j||v^k|} \delta_{n_1 n_2} \tag{2-50}$$

$$= \sqrt{\lambda_j \lambda_k} |\hat{v}^j||\hat{v}^k| \sum_{n} \frac{v_n^{j*} v_n^{k}}{|v^j||v^k|} = 0 = \lambda_k \sum_n \hat{v}_n^{j*} \hat{v}_n^k = \sum_{mn} \hat{v}_m^{j*} A_{mn} \hat{v}_n^k.$$

现在考虑任意两个平移不变态 $|\phi\rangle$ 和 $|\psi\rangle$。假设 $A$ 是厄米的，因而它的本征矢是 $|e^n\rangle$ 生成的向量空间的一组基。然后有

$$|\psi\rangle = \sum_n \int dy \hat{\psi}_n(y) |ny\rangle_y = \sum_n \int dy \left[ \hat{\psi}_n + O(y) \right] |ny\rangle_y$$

$$= \sum_n \hat{\psi}_n |n\rangle_y = \sum_{jn} \hat{\psi}_n \left( \hat{v}^{-1} \right)_n^j |v^j\rangle$$

$$|\phi\rangle = \sum_{jn} \hat{\phi}_n \left( \hat{v}^{-1} \right)_n^j |v^j\rangle, \tag{2-51}$$

其中我们定义了

$$|n\rangle_y = \int dy |ny\rangle_y. \tag{2-52}$$

这里我们舍去了 $O(y)$ 阶项，因为平移不变性意味着 $y$ 和 $x$ 右矢的匹配可以应用于任意的 $y$ 值，因而我们将它应用于 $y = 0$，在那里 $O(y)$ 阶修正为零。

则它们的约化内积是

$$\langle \phi | \psi \rangle_{\text{red}} = \sum_{n_1 n_2 j_1 j_2} \hat{\phi}_{n_1}^* \left( \hat{v}^{*-1} \right)_{n_1}^{j_1} \hat{\psi}_{n_2} \left( \hat{v}^{-1} \right)_{n_2}^{j_2} \langle v^{j_1} | v^{j_2} \rangle_{\text{red}} \tag{2-53}$$

$$= \hat{\phi}^* \left( \hat{v}^* \right)^{-1} \hat{v}^* A \hat{v} \hat{v}^{-1} \hat{\psi} = \hat{\phi}^* A \hat{\psi}.$$

### 2.2.9　对量子力学中约化内积的解释

让我们稍作暂停来解释一下我们的结果 (2-53)。由于对 $y$ 的积分，下面两个态

$$|\psi\rangle = \sum_n \int dy \hat{\psi}_n(y) |ny\rangle_y, \qquad |\phi\rangle = \sum_n \int dy \hat{\phi}_n(y) |ny\rangle_y \tag{2-54}$$

的内积是红外发散的。然而，这些内积仅以比值形式出现，因此只需考虑每个平移单位的内积再除以平移群的体积便足够了。

平移对称性可传递地作用于 $y$ 坐标，使态保持不变。因此我们可以在任意固定的 $y$ 的邻域内计算这个内积密度。考虑 $y = 0$，在这一点附近，我们可以使用如下近似

$$|\psi\rangle = \sum_n \hat{\psi}_n \int dy |ny\rangle_y, \qquad |\phi\rangle = \sum_n \hat{\phi}_n \int dy |ny\rangle_y. \tag{2-55}$$

它们的内积仍然是发散的，但在基点 $y = 0$ 附近有如下因子化

$$_y\langle \phi y_1 | A | \psi y_2 \rangle_y = \langle \phi | \psi \rangle_{\text{red}} \delta(y_1 - y_2). \tag{2-56}$$

我们看到，因为平移算符 $P'$ 和 $y$ 中的刚性平移之间存在差异，所以约化内积 $\langle \phi | \psi \rangle_{\text{red}}$ 由 $\hat{\psi}$ 和 $\hat{\phi}$ 的矢量内积，以及雅可比因子 $A$ 给出。因此，为了计算在集体坐标中直观地包含一个不变量 $\delta(x_1 - x_2)$ 的约化内积，我们首先需要如下分解 $\delta(x_1 - x_2) = A\delta(y_1 - y_2)$，然后截断 $\delta(y_1 - y_2)$ 因子（这个因子正是为了分解等式左边而添加）。

## 2.3 量子扭结的约化内积

### 2.3.1 量子场论中的记号约定

为了从量子力学过渡到允许量子扭结的量子场论，我们进行以下替换。首先，离散量子数 $n$ 被用一系列 $k$ 标记的连续模和形模的对称 $n$ 连续元组取代，这里 $k$ 对于连续模是实数，而对于形模是离散指标。现在 $n \geq 0$ 并且这些态表示扭结空间中的 $n$-介子福克空间。

我们令 $\phi_0$ 为本征值是 $y$ 的算符。它的对偶动量我们记为 $\pi_0$。我们引入简写记号

$$\Delta_{ij} = \int dx \mathfrak{g}_i(x) \mathfrak{g}'_j(x), \tag{2-57}$$

其中下标 $i$ 和 $j$ 可以是零模 $B$，连续模 $k$ 或者形模 $S$。

平移算符 $P'$ 现在由下式给出

$$P' = P + \sqrt{Q_0}\pi_0 \tag{2-58}$$

$$P = \sum \!\!\!\!\!\!\int \frac{dk}{2\pi} \Delta_{kB} \left[ i\phi_0 \left( -\omega_k B_k^\ddagger + \frac{B_{-k}}{2} \right) + \pi_0 \left( B_k^\ddagger + \frac{B_{-k}}{2\omega_k} \right) \right]$$

$$+ i \sum \!\!\!\!\!\!\int \frac{d^2k}{(2\pi)^2} \Delta_{k_1 k_2} \left[ \frac{\omega_{k_2} - \omega_{k_1}}{2} B_{k_1}^\ddagger B_{k_2}^\ddagger - \frac{1}{2}\left(1 + \frac{\omega_{k_1}}{\omega_{k_2}}\right) B_{k_1}^\ddagger B_{-k_2} + \frac{\omega_{k_1} - \omega_{k_2}}{8\omega_{k_1}\omega_{k_2}} B_{-k_1} B_{-k_2} \right].$$

直观上 $P$ 是介子的动量算符，而 $\sqrt{Q_0}\pi_0$ 是扭结的动量算符。只有 $P'$ 是守恒的。回顾我们在 (2-17) 中的旧的分解

$$P' = A\pi_0 + B + C\phi_0 \tag{2-59}$$

我们可以匹配 (2-58) 中的 $\pi_0$ 系数，从而得到

$$A = \sqrt{Q_0} + \sum\!\!\!\!\!\!\!\int \frac{dk}{2\pi} \Delta_{kB} \left( B_k^\ddagger + \frac{B_{-k}}{2\omega_k} \right). \tag{2-60}$$

我们对态进行如下分解

$$|\psi\rangle = \sum_{m,n=0}^{\infty} |\psi\rangle^{mn}, \qquad |k_1 \cdots k_n\rangle_0 = B_{k_1}^\ddagger \cdots B_{k_n}^\ddagger |0\rangle_0 \tag{2-61}$$

$$|\psi\rangle^{mn} = \phi_0^m \sum\!\!\!\!\!\!\!\int \frac{d^n k}{(2\pi)^n} \gamma_\psi^{mn}(k_1 \cdots k_n)|k_1 \cdots k_n\rangle_0, \qquad |0\rangle_0 = \int dy |y\rangle_y.$$

为了和第 2.2 节中的分解建立联系，注意到

$$|\psi\rangle^{mn} = \int dy |y, \psi\rangle_y^{mn}, \qquad |y, \psi\rangle_y^{mn} = y^m \sum\!\!\!\!\!\!\!\int \frac{d^n k}{(2\pi)^n} \gamma_\psi^{mn}(k_1 \cdots k_n)|y, k_1 \cdots k_n\rangle_y$$

$$|y, k_1 \cdots k_n\rangle_y = B_{k_1}^\ddagger \cdots B_{k_n}^\ddagger |y\rangle_y. \tag{2-62}$$

我们看到在这里，之前量子力学情形下的 $\hat{\psi}_n(y)$ 的角色现在在扭结量子场论中由下式取代

$$\hat{\psi}_n(y) \to \sum_m \gamma_\psi^{mn}(k_1 \cdots k_n) y^m. \tag{2-63}$$

离散的 $n$ 量子数被用一系列 $k$ 标记的形模和连续模的 $n$ 元组替换

$$|ny\rangle_y \to |y, k_1 \cdots k_n\rangle_y, \qquad \sum_n \to \sum_n \sum\!\!\!\!\!\!\!\int \frac{d^n k}{(2\pi)^n}. \tag{2-64}$$

回想一下，在量子力学中，原点处的系数是 $\hat{\psi}_n = \hat{\psi}_n(0)$。类似地，这里将 $y$ 取零我们得到 $\gamma^{0n}$

$$\hat{\psi}_n \to \gamma_\psi^{0n}(k_1 \cdots k_n). \tag{2-65}$$

### 2.3.2 约化内积

我们首先计算

$$A|0,\psi\rangle_y^{0n} = \left(\sqrt{Q_0} + \sum\!\!\!\!\!\!\!\int \frac{dk}{2\pi}\Delta_{kB}\left(B_k^{\ddagger} + \frac{B_{-k}}{2\omega_k}\right)\right)|0,\psi\rangle_y^{0n} \qquad (2\text{-}66)$$

$$= \sum\!\!\!\!\!\!\!\int \frac{d^n k}{(2\pi)^n}\gamma_\psi^{0n}(k_1\cdots k_n)\left(\sqrt{Q_0} + \sum\!\!\!\!\!\!\!\int \frac{dk'}{2\pi}\Delta_{k'B}\left(B_{k'}^{\ddagger} + \frac{B_{-k'}}{2\omega_{k'}}\right)\right)|0,k_1\cdots k_n\rangle_y$$

$$= \sqrt{Q_0}|0,\psi\rangle_y^{0n} + \sum\!\!\!\!\!\!\!\int \frac{d^{n+1}k}{(2\pi)^{n+1}}\gamma_\psi^{0n}(k_1\cdots k_n)\Delta_{k_{n+1},B}|0,k_1\cdots k_{n+1}\rangle_y$$

$$+ n\sum\!\!\!\!\!\!\!\int \frac{d^n k}{(2\pi)^n}\gamma_\psi^{0n}(k_1\cdots k_n)\frac{\Delta_{-k_n,B}}{2\omega_{k_n}}|0,k_1\cdots k_{n-1}\rangle_y,$$

其中，在最后一行中，我们假设了 $\gamma_\psi^{0n}$ 在其参数 $k_i$ 上对称。对所有 $n$ 求和我们得到

$$A\sum_n |0,\psi\rangle_y^{0n}$$

$$= \sum_n \sum\!\!\!\!\!\!\!\int \frac{d^n k}{(2\pi)^n}\left[\sqrt{Q_0}\gamma_\psi^{0n}(k_1\cdots k_n) + \gamma_\psi^{0,n-1}(k_1\cdots k_{n-1})\Delta_{k_n,B}\right.$$

$$\left. + (n+1)\sum\!\!\!\!\!\!\!\int \frac{dk_{n+1}}{2\pi}\gamma_\psi^{0,n+1}(k_1\cdots k_{n+1})\frac{\Delta_{-k_{n+1},B}}{2\omega_{k_{n+1}}}\right]|0,k_1\cdots k_n\rangle_y. \qquad (2\text{-}67)$$

利用 $B$ 和 $B^{\ddagger}$ 满足的振子代数以及 $_y\langle y_1|y_2\rangle_y = \delta(y_1 - y_2)$，我们得到

$$|a_i, y_i\rangle = \sum\!\!\!\!\!\!\!\int \frac{d^n k}{(2\pi)^n}a_i(k_1\cdots k_n)|y_i, k_1\cdots k_n\rangle_y \qquad (2\text{-}68)$$

的内积是

$$\langle a_1, y_1|a_2, y_2\rangle = n!\delta(y_1 - y_2)\sum\!\!\!\!\!\!\!\int \frac{d^n k}{(2\pi)^n}\frac{a_1^*(k_1\cdots k_n)a_2(k_1\cdots k_n)}{\prod_{i=1}^n (2\omega_{k_i})}, \qquad (2\text{-}69)$$

其中 $a_i$ 是对称的。

最后，我们希望将约化内积 (2-53) 推广到量子场论。为此，我们需要推广矢量 $\hat{v}$ 的矢量内积。我们的定义是，这个内积被解释为量子场论中的内积 (2-69)，其中不包括 $\delta(y_1 - y_2)$。这仅是等式 (2-56) 在量子场论中的推广。

然后我们得到我们的量子场论中约化内积的主要公式

$$\begin{aligned}\langle\phi|\psi\rangle_{\rm red} &= \sum_{n_1 n_2}{}^{0n_1}_y\langle y_1,\phi|A|y_2,\psi\rangle^{0n_2}_y|_{\delta(y_1-y_2)\text{ 在 }y_1=0\text{ 处的系数}} \\ &= \sum_n n! \sum\!\!\!\!\!\!\!\int \frac{d^n k}{(2\pi)^n} \frac{\gamma_\phi^{0n*}(k_1\cdots k_n)}{\prod_{i=1}^n(2\omega_{k_i})}\left[\sqrt{Q_0}\gamma_\psi^{0n}(k_1\cdots k_n) + \gamma_\psi^{0,n-1}(k_1\cdots k_{n-1})\Delta_{k_n,B}\right.\\ &\quad \left.+(n+1)\sum\!\!\!\!\!\!\!\int \frac{dk_{n+1}}{2\pi}\gamma_\psi^{0,n+1}(k_1\cdots k_{n+1})\frac{\Delta_{-k_{n+1},B}}{2\omega_{k_{n+1}}}\right].\end{aligned} \quad (2\text{-}70)$$

这是本章的主要结果。我们提醒读者，在应用此公式之前，必须对所有的 $\gamma$ 的参数进行对称化，否则 $n!$ 和 $(n+1)!$ 因子应替换为对 $S_n$ 和 $S_{n+1}$ 置换的求和。方括号中的第二项和第三项是 $A$ 的非对角部分产生的雅可比行列式项。两者都与 $\Delta_{kB}$ 成正比，其描述了扭结移动时零模和正规模之间的混合。

在下一节中，我们将看到这个公式满足一些基本的一致性检验。例如，我们将看到零介子态和单介子态的约化内积为零，而如果不包括 (2-70) 中的雅可比项，则会得到一个非零的结果。

## 2.4 约化内积计算示例

在本节中，我们将计算扭结基态和包含一个激发的扭结态（这个激发可以是一个连续的介子或者是一个被束缚的形模）的约化内积。我们将证明，直到被领头阶压低 $O(\lambda)$ 阶的修正，我们有

$$\langle 0|0\rangle_{\rm red} = \sqrt{Q_0} + O(\sqrt{\lambda}), \qquad \langle\mathfrak{K}_1|\mathfrak{K}_2\rangle_{\rm red} = \frac{2\pi\delta(\mathfrak{K}_1-\mathfrak{K}_2)}{2\omega_{\mathfrak{K}_1}}\sqrt{Q_0} + O(\sqrt{\lambda}). \quad (2\text{-}71)$$

若果真有 $O(\lambda^0)$ 阶修正，它相对于领头阶项仅会被 $\sqrt{\lambda}$ 的一次幂压低，这将使第 3 章和第 4 章中的主体计算无效，因此本节的检验对于全文是至关重要的。

我们将要把每个 $\gamma^{mn}$ 分解为

$$\gamma^{mn} = \sum_i Q_0^{-i/2}\gamma_i^{mn}. \quad (2\text{-}72)$$

### 2.4.1 基态的约化模

扭结基态在领头阶由以下这些系数表征[d]

$$\gamma_0^{00} = 1, \quad \gamma_1^{12}(k_1, k_2) = \frac{\left(\omega_{k_2} - \omega_{k_1}\right)\Delta_{k_1 k_2}}{2} \quad (2\text{-}73)$$

$$\gamma_1^{21}(k_1) = -\frac{\omega_{k_1}\Delta_{k_1 B}}{2}, \quad \gamma_1^{01}(k_1) = -\frac{\Delta_{k_1 B}}{2} - \frac{\sqrt{\lambda Q_0}}{2}\frac{V_{\mathcal{I}k_1}}{\omega_{k_1}}$$

$$\gamma_1^{03}(k_1, k_2, k_3) = -\frac{\sqrt{\lambda Q_0}}{6}\frac{V_{k_1 k_2 k_3}}{\omega_{k_1} + \omega_{k_2} + \omega_{k_3}},$$

这里我们定义了

$$V_{k_1 \cdots k_n} = \int dx V^{(n)}(gf(x))\mathfrak{g}_{k_1}(x) \cdots \mathfrak{g}_{k_n}(x) \quad (2\text{-}74)$$

$$V_{\mathcal{I}k_1 \cdots k_n} = \int dx V^{(n+2)}(gf(x))\mathcal{I}(x)\mathfrak{g}_{k_1}(x) \cdots \mathfrak{g}_{k_n}(x)$$

$$\mathcal{I}(x) = \int \frac{dk}{2\pi}\frac{|\mathfrak{g}_k(x)|^2 - 1}{2\omega_k} + \sum_S \frac{|\mathfrak{g}_S(x)|^2}{2\omega_k},$$

在前两行中，$k_i$ 的含义不仅包括连续动量，还包括形模。

约化模可以写成等式 (2-70) 中的 $n = 0, 1$ 和 3 三项之和

$$||0\rangle|_{n,\text{red}}^2 = n! \sum\!\!\!\!\!\!\!\int \frac{d^n k}{(2\pi)^n} \frac{\gamma^{0n*}(k_1 \cdots k_n)}{\prod_{i=1}^n (2\omega_{k_i})} \left[\sqrt{Q_0}\gamma^{0n}(k_1 \cdots k_n) + \gamma^{0,n-1}(k_1 \cdots k_{n-1})\Delta_{k_n, B}\right.$$

$$\left. + (n+1)\sum\!\!\!\!\!\!\!\int \frac{dk_{n+1}}{2\pi}\gamma^{0,n+1}(k_1 \cdots k_{n+1})\frac{\Delta_{-k_{n+1}, B}}{2\omega_{k_{n+1}}}\right]. \quad (2\text{-}75)$$

---

[d] 这些系数在文献 [18] 中被计算。由于本文中对 $\mathfrak{g}_B(x)$ 的正负号的约定 (1-10) 与文献 [18] 不同，此处介子和扭结动量对 $P'$ 的贡献具有异号的贡献。这会改变所有系数中所有 $\Delta$ 项的符号。此外，此处 $V^{(3)}[f(x), x]$ 的约定和文献 [18] 中的约定相差一个 $\sqrt{\lambda}$ 因子。

这些项分别是

$$\begin{aligned}
\||0\rangle|^2_{0,\text{red}} &= \sqrt{Q_0} + \sum\!\!\!\!\!\!\!\int \frac{dk_1}{2\pi} \gamma^{01}(k_1) \frac{\Delta_{-k_1 B}}{2\omega_{k_1}} \\
&= \sqrt{Q_0} - \frac{1}{4\sqrt{Q_0}} \sum\!\!\!\!\!\!\!\int \frac{dk_1}{2\pi} \left[ \Delta_{k_1 B} + \sqrt{\lambda Q_0} \frac{V_{\mathcal{I} k_1}}{\omega_{k_1}} \right] \frac{\Delta_{-k_1 B}}{\omega_{k_1}} \\
\||0\rangle|^2_{1,\text{red}} &= \sum\!\!\!\!\!\!\!\int \frac{dk_1}{2\pi} \frac{\gamma^{01*}(k_1)}{2\omega_{k_1}} \left[ \sqrt{Q_0}\gamma^{01}(k_1) + \Delta_{k_1 B} \right] \\
&= \frac{1}{8\sqrt{Q_0}} \sum\!\!\!\!\!\!\!\int \frac{dk_1}{2\pi} \left[ \frac{\lambda Q_0 |V_{\mathcal{I} k_1}|^2}{\omega_{k_1}^3} - \frac{|\Delta_{k_1 B}|^2}{\omega_{k_1}} \right] \\
\||0\rangle|^2_{3,\text{red}} &= \frac{3\sqrt{Q_0}}{4} \sum\!\!\!\!\!\!\!\int \frac{d^3k}{(2\pi)^3} \frac{|\gamma^{03}(k_1,k_2,k_3)|^2}{\prod_{i=1}^3 \omega_{k_i}} \\
&= \frac{\lambda\sqrt{Q_0}}{48} \sum\!\!\!\!\!\!\!\int \frac{d^3k}{(2\pi)^3} \frac{|V_{k_1 k_2 k_3}|^2}{\omega_{k_1}\omega_{k_2}\omega_{k_3}(\omega_{k_1}+\omega_{k_2}+\omega_{k_3})^2}.
\end{aligned}$$ (2-76)

加在一起，我们得到

$$\begin{aligned}
\||0\rangle|^2_{\text{red}} &= \sqrt{Q_0} + \frac{1}{8\sqrt{Q_0}} \sum\!\!\!\!\!\!\!\int \frac{dk_1}{2\pi} \frac{1}{\omega_{k_1}} \left( \sqrt{\lambda Q_0}\frac{V_{\mathcal{I} k_1}}{\omega_{k_1}} + \Delta_{k_1 B} \right) \left( \sqrt{\lambda Q_0}\frac{V_{\mathcal{I} -k_1}}{\omega_{k_1}} - 3\Delta_{-k_1 B} \right) \\
&\quad + \frac{\lambda\sqrt{Q_0}}{48} \sum\!\!\!\!\!\!\!\int \frac{d^3k}{(2\pi)^3} \frac{|V_{k_1 k_2 k_3}|^2}{\omega_{k_1}\omega_{k_2}\omega_{k_3}(\omega_{k_1}+\omega_{k_2}+\omega_{k_3})^2}.
\end{aligned}$$ (2-77)

注意，我们采取了约定 $\gamma_2^{00} = 0$。对 $\gamma_2^{00}$ 的其他约定会导致不同的模。这是自由选择整体归一化的最低阶表现，其在量子力学中已经存在。显然对每个态都有这样的选择自由。尽管所有的态的模都有这样一个约定问题，但对于一个给定约定，将模确定下来是关乎物理的，因为这一约定随后会固定所有的约化内积。

### 2.4.2 零介子态和单介子态的约化内积

#### 2.4.2.1 直到 $O(\sqrt{\lambda})$ 阶的单介子态

现在让我们考虑一个单介子哈密顿本征态 $|\mathfrak{K}\rangle$。在领头阶，它是 $|\mathfrak{K}\rangle_0$，其由下式表征

$$\gamma_{0\mathfrak{K}}^{01}(k_1) = 2\pi\delta(k_1 - \mathfrak{K}).$$ (2-78)

下一阶修正[e] 由相应的符号 $\gamma_{1\mathfrak{K}}^{mn}$ 总结如下

$$\begin{aligned}
\gamma_{1\mathfrak{K}}^{11}(k_1) &= \frac{1}{2}\Delta_{-\mathfrak{K}k_1}\left(1+\frac{\omega_{k_1}}{\omega_{\mathfrak{K}}}\right), \quad \gamma_{1\mathfrak{K}}^{13}(k_1,k_2,k_3)=\omega_{k_3}\Delta_{k_2 k_3}2\pi\delta(k_1-\mathfrak{K}) \\
\gamma_{1\mathfrak{K}}^{22}(k_1,k_2) &= -\frac{\omega_{k_2}}{2}\Delta_{k_2 B}2\pi\delta(k_1-\mathfrak{K}), \quad \gamma_{1\mathfrak{K}}^{00} = \frac{\sqrt{Q_0}\lambda V_{\mathcal{I},-\mathfrak{K}}}{4\omega_{\mathfrak{K}}^2} - \frac{\Delta_{-\mathfrak{K}B}}{4\omega_{\mathfrak{K}}} \\
\gamma_{1\mathfrak{K}}^{02}(k_1,k_2) &= -\frac{2\pi\delta(k_2-\mathfrak{K})}{4}\left(\Delta_{k_1 B}+\sqrt{Q_0}\lambda\frac{V_{\mathcal{I}k_1}}{\omega_{k_1}}\right) + \frac{\sqrt{Q_0}\lambda V_{-\mathfrak{K}k_1 k_2}}{4\omega_{\mathfrak{K}}\left(\omega_{\mathfrak{K}}-\omega_{k_1}-\omega_{k_2}\right)} \\
&\quad -\frac{2\pi\delta(k_1-\mathfrak{K})}{4}\left(\Delta_{k_2 B}+\sqrt{Q_0}\lambda\frac{V_{\mathcal{I}k_2}}{\omega_{k_2}}\right) \\
\gamma_{1\mathfrak{K}}^{04}(k_1\cdots k_4) &= -\frac{\sqrt{Q_0}\lambda V_{k_1 k_2 k_3}}{6\sum_{j=1}^3 \omega_{k_j}}2\pi\delta(k_4-\mathfrak{K}), \quad \gamma_{1\mathfrak{K}}^{20}=\frac{1}{4}\Delta_{-\mathfrak{K}B}.
\end{aligned} \quad (2\text{-}79)$$

### 2.4.2.2 约化内积

让我们计算直到 $O(\lambda^0)$ 阶的，扭结基态 $|0\rangle$ 和单扭结单介子态 $|\mathfrak{K}\rangle$ 的约化内积。在这一阶有两个贡献

$$\begin{aligned}
\langle 0|\mathfrak{K}\rangle_{n,\text{red}} &= n!\sum\!\!\!\!\!\!\int \frac{d^n k}{(2\pi)^n}\frac{\gamma^{0n*}(k_1\cdots k_n)}{\prod_{i=1}^n (2\omega_{k_i})}\left[\sqrt{Q_0}\gamma_{\mathfrak{K}}^{0n}(k_1\cdots k_n)+\gamma_{\mathfrak{K}}^{0,n-1}(k_1\cdots k_{n-1})\Delta_{k_n,B}\right.\\
&\quad \left.+(n+1)\sum\!\!\!\!\!\!\int \frac{dk'}{2\pi}\gamma_{\mathfrak{K}}^{0,n+1}(k_1\cdots k_n,k')\frac{\Delta_{-k'B}}{2\omega_{k'}}\right]. \quad (2\text{-}80)
\end{aligned}$$

具体来说它们分别是

$$\langle 0|\mathfrak{K}\rangle_{0,\text{red}} = \sqrt{Q_0}\gamma_{\mathfrak{K}}^{00}+\sum\!\!\!\!\!\!\int\frac{dk'}{2\pi}\gamma_{\mathfrak{K}}^{01}(k')\frac{\Delta_{-k'B}}{2\omega_{k'}} = \frac{\sqrt{Q_0}\lambda V_{\mathcal{I}-\mathfrak{K}}}{4\omega_{\mathfrak{K}}^2}+\frac{\Delta_{-\mathfrak{K}B}}{4\omega_{\mathfrak{K}}} \quad (2\text{-}81)$$

$$\langle 0|\mathfrak{K}\rangle_{1,\text{red}} = \sum\!\!\!\!\!\!\int\frac{dk_1}{2\pi}\frac{\gamma^{01*}(k_1)}{2\omega_{k_1}}\sqrt{Q_0}\gamma_{\mathfrak{K}}^{01}(k_1) = \frac{\gamma_1^{01*}(\mathfrak{K})}{2\omega_{\mathfrak{K}}} = -\frac{\Delta_{-\mathfrak{K}B}}{4\omega_{\mathfrak{K}}}-\frac{\sqrt{\lambda Q_0}}{2}\frac{V_{\mathcal{I}-\mathfrak{K}}}{2\omega_{\mathfrak{K}}^2}.$$

它们精确地抵消了，于是在 $O(\lambda^0)$ 阶有

$$\langle 0|\mathfrak{K}\rangle_{\text{red}} = 0. \quad (2\text{-}82)$$

这是意料之中的，因为 $|0\rangle$ 和 $|\mathfrak{K}\rangle$ 是 $H'$ 的具有不同本征值的本征态。注意，$A$ 中的非对角项对 $\langle 0|\mathfrak{K}\rangle_{0,\text{red}}$ 有贡献，并且这对于约化内积满足正交性是必要的。

---

[e] 它们在文献 [28] 中被计算，再次地，由于 (1-10) 的约定，所有 $\Delta$ 符号和文献 [28] 中相反，另外 $V$ 中也包含了 $\sqrt{\lambda}$。

### 2.4.3 两个单介子态的内积

接下来，我们将注意力转向两个单介子单扭结态 $|\mathfrak{K}_1\rangle$ 和 $|\mathfrak{K}_2\rangle$ 在 $O(\sqrt{\lambda})$ 阶的约化内积。

#### 2.4.3.1 $O(\lambda)$ 阶的系数

除了等式 (2-78) 中给出的 $O(\lambda^0)$ 阶系数和等式 (2-79) 中给出的 $O(\sqrt{\lambda})$ 阶系数外，我们还需要当 $k_1 \neq \mathfrak{K}$ 时的 $O(\lambda)$ 阶系数 $\gamma_{2\mathfrak{K}}^{01}(k_1)$。为了计算它，我们使用本征值方程

$$(H' - E)|\mathfrak{K}\rangle = 0. \tag{2-83}$$

它在 $O(\lambda)$ 阶包含五项

$$0 = H_4'|\mathfrak{K}\rangle_0 + H_3'|\mathfrak{K}\rangle_1 + H_2'|\mathfrak{K}\rangle_2 - E_2|\mathfrak{K}\rangle_0 - E_1|\mathfrak{K}\rangle_2. \tag{2-84}$$

这里 $E_n$ 是单扭结单介子态 $|\mathfrak{K}\rangle$ 能量中的第 $O(\lambda^{n-1})$ 阶能量项。特别地

$$E_1 = \omega_{\mathfrak{K}}, \qquad E_2 = \sigma_{\mathfrak{K}}, \qquad H_2' = \frac{\pi_0^2}{2} + \sum\!\!\!\!\!\!\int \frac{dk}{2\pi} \omega_k B_k^{\ddagger} B_k, \tag{2-85}$$

其中 $\sigma_{\mathfrak{K}}$ 在文献 [28] 中被计算。注意，由于 $H_0' = E_0 = Q_0$ 是一个标量，它们相互抵消，所以等式中并不包含 $H_0'$ 和 $E_0$ 这两项。

我们将等式 (2-84) 应用在这些项上，它们不依赖于 $\phi_0$ 且包含一个介子，换言之，它们是 $m = 0, n = 1$ 的项。这些只能由 $|\mathfrak{K}\rangle_2$ 中的下式部分产生

$$|\mathfrak{K}\rangle_2 \supset \frac{1}{Q_0} \sum\!\!\!\!\!\!\int \frac{dk_1}{2\pi} \left[\gamma_{2\mathfrak{K}}^{01}(k_1) + \phi_0^2 \gamma_{2\mathfrak{K}}^{21}(k_1)\right] |k_1\rangle_0. \tag{2-86}$$

等式 (2-84) 中的最后三项可以很容易地写成

$$H_2'|\mathfrak{K}\rangle_2 - E_2|\mathfrak{K}\rangle_0 - E_1|\mathfrak{K}\rangle_2 = \frac{1}{Q_0} \sum\!\!\!\!\!\!\int \frac{dk_1}{2\pi} \left[(\omega_{k_1} - \omega_{\mathfrak{K}})\gamma_{2\mathfrak{K}}^{01}(k_1) - \gamma_{2\mathfrak{K}}^{21}(k_1)\right] |k_1\rangle_0 - \sigma_{\mathfrak{K}}|\mathfrak{K}\rangle_0. \tag{2-87}$$

我们的策略是通过将 $|k_1\rangle_0$ 的系数和下式进行匹配来确定 $\gamma_{2\mathfrak{K}}^{01}(k_1)$

$$H_4'|\mathfrak{K}\rangle_0 + H_3'|\mathfrak{K}\rangle_1 = \frac{1}{Q_0} \sum\!\!\!\!\!\!\int \frac{dk_1}{2\pi} \rho_{\mathfrak{K}}(k_1)|k_1\rangle_0 + \hat{\sigma}_{\mathfrak{K}}|\mathfrak{K}\rangle_0, \tag{2-88}$$

其中 $\rho_{\mathfrak{K}}$ 将在下面计算。在这里，我们通过分解下式从 $\gamma_{2\mathfrak{K}}^{21}$ 中的 $\sigma_{\mathfrak{K}}$ 贡献中分离出 $\hat{\sigma}_{\mathfrak{K}}$

$$\gamma_{2\mathfrak{K}}^{21}(k_1) = \hat{\gamma}_{2\mathfrak{K}}^{21}(k_1) + 2\pi\delta(k_1 - \mathfrak{K})Q_0\left(\hat{\sigma}_{\mathfrak{K}} - \sigma_{\mathfrak{K}}\right), \tag{2-89}$$

其中 $\hat{\gamma}_{\mathfrak{K}}(k_1)$ 在 $k_1 = \mathfrak{K}$ 处连续。

通过匹配系数我们得到

$$\gamma_{2\mathfrak{K}}^{01}(k_1) = \frac{-\hat{\gamma}_{2\mathfrak{K}}^{21}(k_1) + \rho_\mathfrak{K}(k_1)}{\omega_\mathfrak{K} - \omega_{k_1}}. \qquad (2\text{-}90)$$

此式在位于 $k_1 = \pm\mathfrak{K}$ 的两个极点处未定义。$k_1 = \mathfrak{K}$ 处的歧义反映了对态 $|\mathfrak{K}\rangle$ 的归一化的选择。

$k_1 = -\mathfrak{K}$ 处的歧义是由于态 $|\mathfrak{K}\rangle$ 和 $|-\mathfrak{K}\rangle$ 具有相同的能量，并且都具有零动量（由 $P'$ 测量）。我们已经将 $|\mathfrak{K}\rangle$ 定义为 $H'$ 的本征态，它在领头阶是 $|\mathfrak{K}\rangle_0$，但是这个定义并没有解决它和 $|-\mathfrak{K}\rangle$ 在次领头阶的混合。我们将在本章的后续小节中看到，当我们讨论介子倍增时，对极点的定义的选择对应于扭结-介子散射中初始条件的选择。在以后的工作中，我们打算研究扭结-介子的弹性散射，其中间状态有下面几种组成方式：两个连续模，一个连续模和一个形模，或两个形模共振。我们期望对极点采用 $i\epsilon$ 平移，以确保初始介子始终向扭结移动。

### 2.4.3.2 计算 $\rho_\mathfrak{K}$

让我们从 $H_4'|\mathfrak{K}\rangle_0$ 开始。只有出现在维克定理 [55] 中的下面这一项会有贡献

$$\begin{aligned} H_4' &= \frac{\lambda}{24}\int dx V^{(4)}(gf(x)) :\phi^4(x):_a \supset \frac{\lambda}{4}\int dx V^{(4)}(gf(x))\left(\mathcal{I}(x):\phi^2(x):_b + \frac{\mathcal{I}^2(x)}{2}\right) \\ &\supset \frac{\lambda}{2}\sum\!\!\!\!\!\!\!\int \frac{d^2k}{(2\pi)^2} V_{\mathcal{I}k_1-k_2} B^{\ddagger}_{k_1}\frac{B_{k_2}}{2\omega_{k_2}} + \frac{\lambda V_{\mathcal{II}}}{8}, \qquad V_{\mathcal{II}} = \int dx V^{(4)}(gf(x))\mathcal{I}^2(x), \end{aligned} \qquad (2\text{-}91)$$

这里我们定义了正规序 $::_b$，它将 $B^{\ddagger}$ 放在 $B$ 之前。然后我们发现了下式的这部分贡献

$$H_4'|\mathfrak{K}\rangle_0 \supset \frac{\lambda V_{\mathcal{II}}}{8}|\mathfrak{K}\rangle_0 + \frac{\lambda}{4\omega_\mathfrak{K}}\sum\!\!\!\!\!\!\!\int \frac{dk_1}{2\pi}V_{\mathcal{I}k_1-\mathfrak{K}}|k_1\rangle_0. \qquad (2\text{-}92)$$

第一项对 $\hat{\sigma}_\mathfrak{K}$ 有贡献，第二项对 $\rho_\mathfrak{K}$ 有贡献。

$H_3'|\mathfrak{K}\rangle_1$ 产生了三个贡献。根据文献 [18] 第一个贡献是

$$H_3'|\mathfrak{K}\rangle_1^{00} = \frac{\lambda}{8}\left(\frac{V_{\mathcal{I}-\mathfrak{K}}}{\omega_\mathfrak{K}^2} - \frac{\Delta_{-\mathfrak{K}B}}{\omega_\mathfrak{K}\sqrt{\lambda Q_0}}\right)\sum\!\!\!\!\!\!\!\int \frac{dk_1}{2\pi}V_{\mathcal{I}k_1}|k_1\rangle_0.$$

第二个贡献是

$$\begin{aligned}H'_3|\mathfrak{K}\rangle_1^{02} &= \frac{\sqrt{\lambda}}{\sqrt{Q_0}}\sum\!\!\!\!\!\!\!\int\frac{dk_1}{2\pi}\Bigg[\sum\!\!\!\!\!\!\!\int\frac{d^2k'}{(2\pi)^2}\frac{\sqrt{\lambda Q_0}V_{-\mathfrak{K}k'_1k'_2}V_{-k'_1-k'_2k_1}}{16\omega_{\mathfrak{K}}\omega_{k'_1}\omega_{k'_2}\left(\omega_{\mathfrak{K}}-\omega_{k'_1}-\omega_{k'_2}\right)}\\ &\quad+\sum\!\!\!\!\!\!\!\int\frac{dk'}{2\pi}\left(\frac{\left(-\omega_{k'}\Delta_{k'B}-\sqrt{\lambda Q_0}V_{\mathcal{I}k'}\right)V_{-k'-\mathfrak{K}k_1}}{8\omega_{k'}^2\omega_{\mathfrak{K}}}+\frac{\sqrt{\lambda Q_0}V_{-\mathfrak{K}k'k_1}V_{\mathcal{I}-k'}}{8\omega_{\mathfrak{K}}\omega_{k'}\left(\omega_{\mathfrak{K}}-\omega_{k'}-\omega_{k_1}\right)}\right)\\ &\quad+\frac{\left(-\omega_{k_1}\Delta_{k_1B}-\sqrt{\lambda Q_0}V_{\mathcal{I}k_1}\right)V_{\mathcal{I}-\mathfrak{K}}}{8\omega_{\mathfrak{K}}\omega_{k_1}}\Bigg]|k_1\rangle_0\\ &\quad+\frac{\sqrt{\lambda}}{\sqrt{Q_0}}\Bigg[\sum\!\!\!\!\!\!\!\int\frac{dk'}{2\pi}\frac{\left(-\omega_{k'}\Delta_{k'B}-\sqrt{\lambda Q_0}V_{\mathcal{I}k'}\right)V_{\mathcal{I}-k'}}{8\omega_{k'}^2}\Bigg]|\mathfrak{K}\rangle_0.\end{aligned}$$

第三个贡献是

$$\begin{aligned}H'_3|\mathfrak{K}\rangle_1^{04} &= -\frac{\lambda}{16}\sum\!\!\!\!\!\!\!\int\frac{dk_1}{2\pi}\Bigg[\sum\!\!\!\!\!\!\!\int\frac{d^2k'}{(2\pi)^2}\frac{V_{k_1k'_1k'_2}V_{-\mathfrak{K}-k'_1-k'_2}}{\omega_{\mathfrak{K}}\omega_{k'_1}\omega_{k'_2}\left(\omega_{k_1}+\omega_{k'_1}+\omega_{k'_2}\right)}\Bigg]|k_1\rangle_0\\ &\quad-\frac{\lambda}{48}\Bigg[\sum\!\!\!\!\!\!\!\int\frac{d^3k'}{(2\pi)^3}\frac{V_{k'_1k'_2k'_3}V_{-k'_1-k'_2-k'_3}}{\omega_{k'_1}\omega_{k'_2}\omega_{k'_3}\left(\omega_{k'_1}+\omega_{k'_2}+\omega_{k'_3}\right)}\Bigg]|\mathfrak{K}\rangle_0.\end{aligned}$$

把这些加起来之后，我们可以读出

$$\begin{aligned}\hat{\sigma}_k &= \frac{\lambda V_{\mathcal{I}\mathcal{I}}}{8}+\frac{\sqrt{\lambda}}{\sqrt{Q_0}}\Bigg[\sum\!\!\!\!\!\!\!\int\frac{dk'}{2\pi}\frac{\left(-\omega_{k'}\Delta_{k'B}-\sqrt{\lambda Q_0}V_{\mathcal{I}k'}\right)V_{\mathcal{I}-k'}}{8\omega_{k'}^2}\Bigg]\\ &\quad-\frac{\lambda}{48}\Bigg[\sum\!\!\!\!\!\!\!\int\frac{d^3k'}{(2\pi)^3}\frac{V_{k'_1k'_2k'_3}V_{-k'_1-k'_2-k'_3}}{\omega_{k'_1}\omega_{k'_2}\omega_{k'_3}\left(\omega_{k'_1}+\omega_{k'_2}+\omega_{k'_3}\right)}\Bigg]\end{aligned}\qquad(2\text{-}93)$$

以及

$$\begin{aligned}
\rho_{\mathfrak{K}}(k_1) &= \frac{\lambda Q_0}{4\omega_{\mathfrak{K}}} V_{\mathcal{I}k_1-\mathfrak{K}} + \frac{\lambda Q_0}{8}\left(\frac{V_{\mathcal{I}-\mathfrak{K}}}{\omega_{\mathfrak{K}}^2} - \frac{\Delta_{-\mathfrak{K}B}}{\omega_{\mathfrak{K}}\sqrt{\lambda Q_0}}\right) V_{\mathcal{I}k_1} \\
&\quad + \sqrt{\lambda Q_0}\Bigg[\sum\!\!\!\!\!\!\!\int \frac{d^2 k'}{(2\pi)^2} \frac{\sqrt{\lambda Q_0} V_{-\mathfrak{K}k_1'k_2'} V_{-k_1'-k_2'k_1}}{16\omega_{\mathfrak{K}}\omega_{k_1'}\omega_{k_2'}\left(\omega_{\mathfrak{K}} - \omega_{k_1'} - \omega_{k_2'}\right)} \\
&\quad + \sum\!\!\!\!\!\!\!\int \frac{dk'}{2\pi}\left(\frac{\left(-\omega_{k_1'}\Delta_{k'B} - \sqrt{\lambda Q_0} V_{\mathcal{I}k'}\right) V_{-k'-\mathfrak{K}k_1}}{8\omega_{k'}^2\omega_{\mathfrak{K}}} + \frac{\sqrt{\lambda Q_0} V_{-\mathfrak{K}k'k_1} V_{\mathcal{I}-k'}}{8\omega_{\mathfrak{K}}\omega_{k'}\left(\omega_{\mathfrak{K}} - \omega_{k'} - \omega_{k_1}\right)}\right) \\
&\quad + \frac{\left(-\omega_{k_1}\Delta_{k_1 B} - \sqrt{\lambda Q_0} V_{\mathcal{I}k_1}\right) V_{\mathcal{I}-\mathfrak{K}}}{8\omega_{\mathfrak{K}}\omega_{k_1}}\Bigg] - \frac{\lambda Q_0}{16}\sum\!\!\!\!\!\!\!\int \frac{d^2 k'}{(2\pi)^2} \frac{V_{k_1 k_1' k_2'} V_{-\mathfrak{K}-k_1'-k_2'}}{\omega_{\mathfrak{K}}\omega_{k_1'}\omega_{k_2'}\left(\omega_{k_1} + \omega_{k_1'} + \omega_{k_2'}\right)}.
\end{aligned}$$

(2-94)

文献 [18] 中有

$$\begin{aligned}
\gamma_{2\mathfrak{K}}^{21}(k_1) &= 2\pi\delta(k_1 - \mathfrak{K})\Bigg[\sum\!\!\!\!\!\!\!\int \frac{dk'}{2\pi}\frac{\Delta_{-k'B}}{8}\left(\Delta_{k'B} - \frac{\sqrt{\lambda Q_0} V_{\mathcal{I}k'}}{\omega_{k'}}\right) \\
&\quad - \frac{1}{16}\sum\!\!\!\!\!\!\!\int \frac{d^2 k'}{(2\pi)^2} \frac{\left(\omega_{k_1'} - \omega_{k_2'}\right)^2}{\omega_{k_1'}\omega_{k_2'}}\Delta_{k_1'k_2'}\Delta_{-k_1',-k_2'}\Bigg] \\
&\quad + \frac{3}{8}\left(-1 + \frac{\omega_{k_1}}{\omega_{\mathfrak{K}}}\right)\Delta_{k_1 B}\Delta_{-\mathfrak{K}B} - \frac{1}{4}\sum\!\!\!\!\!\!\!\int \frac{dk'}{2\pi}\left(\frac{\omega_{k_1}}{\omega_{k'}} + \frac{\omega_{k'}}{\omega_{\mathfrak{K}}}\right)\Delta_{-\mathfrak{K},-k'}\Delta_{k_1 k'} \\
&\quad - \frac{\sqrt{\lambda Q_0}}{8\omega_{\mathfrak{K}}}\left(\omega_{k_1}\Delta_{k_1 B}\frac{V_{\mathcal{I}-\mathfrak{K}}}{\omega_{\mathfrak{K}}} + \omega_{\mathfrak{K}}\Delta_{-\mathfrak{K}B}\frac{V_{\mathcal{I}k_1}}{\omega_{k_1}}\right) + \frac{1}{8}\sum\!\!\!\!\!\!\!\int \frac{dk'}{2\pi}\frac{\sqrt{\lambda Q_0}\Delta_{-k'B}V_{-\mathfrak{K}k_1 k'}}{\omega_{\mathfrak{K}}\left(\omega_{\mathfrak{K}} - \omega_{k_1} - \omega_{k'}\right)}.
\end{aligned}$$

(2-95)

对它进行分解，即得

$$\begin{aligned}
\hat{\gamma}_{2\mathfrak{K}}^{21}(k_1) &= \frac{3}{8}\left(-1 + \frac{\omega_{k_1}}{\omega_{\mathfrak{K}}}\right)\Delta_{k_1 B}\Delta_{-\mathfrak{K}B} - \frac{1}{4}\sum\!\!\!\!\!\!\!\int \frac{dk'}{2\pi}\left(\frac{\omega_{k_1}}{\omega_{k'}} + \frac{\omega_{k'}}{\omega_{\mathfrak{K}}}\right)\Delta_{-\mathfrak{K},-k'}\Delta_{k_1 k'} \\
&\quad - \frac{\sqrt{\lambda Q_0}}{8\omega_{\mathfrak{K}}}\left(\omega_{k_1}\Delta_{k_1 B}\frac{V_{\mathcal{I}-\mathfrak{K}}}{\omega_{\mathfrak{K}}} + \omega_{\mathfrak{K}}\Delta_{-\mathfrak{K}B}\frac{V_{\mathcal{I}k_1}}{\omega_{k_1}}\right) + \frac{1}{8}\sum\!\!\!\!\!\!\!\int \frac{dk'}{2\pi}\frac{\sqrt{\lambda Q_0}\Delta_{-k'B}V_{-\mathfrak{K}k_1 k'}}{\omega_{\mathfrak{K}}\left(\omega_{\mathfrak{K}} - \omega_{k_1} - \omega_{k'}\right)}
\end{aligned}$$

(2-96)

$$\hat{\sigma}_{\mathfrak{K}} - \sigma_{\mathfrak{K}} = \frac{1}{Q_0} \left[ \sum\!\!\!\!\!\!\!\!\int \frac{dk'}{2\pi} \frac{\Delta_{-k'B}}{8} \left( \Delta_{k'B} - \frac{\sqrt{\lambda Q_0} V_{\mathcal{I}k'}}{\omega_{k'}} \right) \right.$$
$$\left. - \frac{1}{16} \sum\!\!\!\!\!\!\!\!\int \frac{d^2 k'}{(2\pi)^2} \frac{\left( \omega_{k_1'} - \omega_{k_2'} \right)^2}{\omega_{k_1'} \omega_{k_2'}} \Delta_{k_1' k_2'} \Delta_{-k_1', -k_2'} \right]. \qquad (2\text{-}97)$$

特别地，这意味着 $\sigma_{\mathfrak{K}} = Q_2$，其中 $Q_2$ 是对扭结基态质量的两圈修正，参见文献 [18]。

### 2.4.3.3 约化内积

$O(\lambda)$ 阶修正 $|\mathfrak{K}\rangle_2$ 与领头阶项 $|\mathfrak{K}\rangle_0$ 的内积产生

$$\langle \mathfrak{K}_1 | \mathfrak{K}_2 \rangle_{\text{red}} \supset \frac{1}{\sqrt{Q_0}} \frac{\gamma^{01}_{2\mathfrak{K}_2}(\mathfrak{K}_1)}{2\omega_{\mathfrak{K}_1}} + \frac{1}{\sqrt{Q_0}} \frac{\gamma^{01*}_{2\mathfrak{K}_1}(\mathfrak{K}_2)}{2\omega_{\mathfrak{K}_2}}$$
$$= \frac{-\omega_{\mathfrak{K}_2} \hat{\gamma}^{21}_{2\mathfrak{K}_2}(\mathfrak{K}_1) + \omega_{\mathfrak{K}_1} \hat{\gamma}^{21*}_{2\mathfrak{K}_1}(\mathfrak{K}_2)}{2\sqrt{Q_0} \omega_{\mathfrak{K}_1} \omega_{\mathfrak{K}_2} (\omega_{\mathfrak{K}_2} - \omega_{\mathfrak{K}_1})} + \frac{\omega_{\mathfrak{K}_2} \rho_{\mathfrak{K}_2}(\mathfrak{K}_1) - \omega_{\mathfrak{K}_1} \rho^*_{\mathfrak{K}_1}(\mathfrak{K}_2)}{2\sqrt{Q_0} \omega_{\mathfrak{K}_1} \omega_{\mathfrak{K}_2} (\omega_{\mathfrak{K}_2} - \omega_{\mathfrak{K}_1})}. \qquad (2\text{-}98)$$

由于反对称性，第二个分子中有很多项可以抵消

$$\omega_{\mathfrak{K}_2} \rho_{\mathfrak{K}_2}(\mathfrak{K}_1)$$
$$= \frac{\lambda Q_0}{4} V_{\mathcal{I} \mathfrak{K}_1 - \mathfrak{K}_2} + \frac{\lambda Q_0}{8} \left( \frac{V_{\mathcal{I} - \mathfrak{K}_2}}{\omega_{\mathfrak{K}_2}} - \frac{\Delta_{-\mathfrak{K}_2 B}}{\sqrt{\lambda Q_0}} \right) V_{\mathcal{I} \mathfrak{K}_1}$$
$$+ \frac{\lambda Q_0}{16} \sum\!\!\!\!\!\!\!\!\int \frac{d^2 k'}{(2\pi)^2} \frac{V_{-\mathfrak{K}_2 k_1' k_2'} V_{-k_1' -k_2' \mathfrak{K}_1}}{\omega_{k_1'} \omega_{k_2'} \left( \omega_{\mathfrak{K}_2} - \omega_{k_1'} - \omega_{k_2'} \right)}$$
$$+ \frac{\sqrt{\lambda Q_0}}{8} \sum\!\!\!\!\!\!\!\!\int \frac{dk'}{2\pi} \left( \frac{\left( -\omega_{k_1'} \Delta_{k'B} - \sqrt{\lambda Q_0} V_{\mathcal{I}k'} \right) V_{-k' - \mathfrak{K}_2 \mathfrak{K}_1}}{\omega_{k'}^2} + \frac{\sqrt{\lambda Q_0} V_{-\mathfrak{K}_2 k' \mathfrak{K}_1} V_{\mathcal{I} - k'}}{\omega_{k'} \left( \omega_{\mathfrak{K}_2} - \omega_{\mathfrak{K}_1} - \omega_{k'} \right)} \right)$$
$$+ \frac{\lambda Q_0}{8} \left( -\frac{\Delta_{k_1 B}}{\sqrt{\lambda Q_0}} - \frac{V_{\mathcal{I} \mathfrak{K}_1}}{\omega_{\mathfrak{K}_1}} \right) V_{\mathcal{I} - \mathfrak{K}_2} - \frac{\lambda Q_0}{16} \sum\!\!\!\!\!\!\!\!\int \frac{d^2 k'}{(2\pi)^2} \frac{V_{\mathfrak{K}_1 k_1' k_2'} V_{-\mathfrak{K}_2 - k_1' - k_2'}}{\omega_{k_1'} \omega_{k_2'} \left( \omega_{\mathfrak{K}_1} + \omega_{k_1'} + \omega_{k_2'} \right)}$$

(2-99)

因此

$$
\begin{aligned}
&\frac{\omega_{\mathfrak{K}_2}\rho_{\mathfrak{K}_2}(\mathfrak{K}_1) - \omega_{\mathfrak{K}_1}\rho^*_{\mathfrak{K}_1}(\mathfrak{K}_2)}{2\sqrt{Q_0}\omega_{\mathfrak{K}_1}\omega_{\mathfrak{K}_2}(\omega_{\mathfrak{K}_2}-\omega_{\mathfrak{K}_1})} \\
&= -\frac{\lambda\sqrt{Q_0}}{8}\frac{V_{\mathcal{I}-\mathfrak{K}_2}V_{\mathcal{I}\mathfrak{K}_1}}{\omega^2_{\mathfrak{K}_1}\omega^2_{\mathfrak{K}_2}} \\
&\quad -\frac{\lambda\sqrt{Q_0}}{32\omega_{\mathfrak{K}_1}\omega_{\mathfrak{K}_2}}\sum\!\!\!\!\!\!\int\frac{d^2k'}{(2\pi)^2}\frac{V_{-\mathfrak{K}_2 k'_1 k'_2}V_{-k'_1-k'_2\mathfrak{K}_1}}{\omega_{k'_1}\omega_{k'_2}\left(\omega_{\mathfrak{K}_2}-\omega_{k'_1}-\omega_{k'_2}\right)\left(\omega_{\mathfrak{K}_1}-\omega_{k'_1}-\omega_{k'_2}\right)} \\
&\quad +\frac{\lambda\sqrt{Q_0}}{8\omega_{\mathfrak{K}_1}\omega_{\mathfrak{K}_2}}\sum\!\!\!\!\!\!\int\frac{dk'}{2\pi}\frac{V_{-\mathfrak{K}_2 k'\mathfrak{K}_1}V_{\mathcal{I}-k'}}{\omega_{k'}\left[(\omega_{\mathfrak{K}_2}-\omega_{\mathfrak{K}_1})^2-\omega^2_{k'}\right]} \\
&\quad -\frac{\lambda\sqrt{Q_0}}{32\omega_{\mathfrak{K}_1}\omega_{\mathfrak{K}_2}}\sum\!\!\!\!\!\!\int\frac{d^2k'}{(2\pi)^2}\frac{V_{\mathfrak{K}_1 k'_1 k'_2}V_{-\mathfrak{K}_2-k'_1-k'_2}}{\omega_{k'_1}\omega_{k'_2}\left(\omega_{\mathfrak{K}_1}+\omega_{k'_1}+\omega_{k'_2}\right)\left(\omega_{\mathfrak{K}_2}+\omega_{k'_1}+\omega_{k'_2}\right)}.
\end{aligned}
$$
(2-100)

类似地，

$$
\begin{aligned}
&\omega_{\mathfrak{K}_2}\hat{\gamma}^{21}_{2\mathfrak{K}_2}(\mathfrak{K}_1) \\
&= \frac{3}{8}\left(\omega_{\mathfrak{K}_1}-\omega_{\mathfrak{K}_2}\right)\Delta_{\mathfrak{K}_1 B}\Delta_{-\mathfrak{K}_2 B} - \frac{1}{4}\sum\!\!\!\!\!\!\int\frac{dk'}{2\pi}\left(\frac{\omega_{\mathfrak{K}_1}\omega_{\mathfrak{K}_2}}{\omega_{k'}}+\omega_{k'}\right)\Delta_{-\mathfrak{K}_2,-k'}\Delta_{\mathfrak{K}_1 k'} \\
&\quad -\frac{\sqrt{\lambda Q_0}}{8}\left(\omega_{\mathfrak{K}_1}\Delta_{\mathfrak{K}_1 B}\frac{V_{\mathcal{I}-\mathfrak{K}_2}}{\omega_{\mathfrak{K}_2}}+\omega_{\mathfrak{K}_2}\Delta_{-\mathfrak{K}_2 B}\frac{V_{\mathcal{I}\mathfrak{K}_1}}{\omega_{\mathfrak{K}_1}}\right) \\
&\quad +\frac{1}{8}\sum\!\!\!\!\!\!\int\frac{dk'}{2\pi}\frac{\sqrt{\lambda Q_0}\Delta_{-k' B}V_{-\mathfrak{K}_2\mathfrak{K}_1 k'}}{\left(\omega_{\mathfrak{K}_2}-\omega_{\mathfrak{K}_1}-\omega_{k'}\right)}
\end{aligned}
$$
(2-101)

导出

$$
\begin{aligned}
&\frac{-\omega_{\mathfrak{K}_2}\hat{\gamma}^{21}_{2\mathfrak{K}_2}(\mathfrak{K}_1)+\omega_{\mathfrak{K}_1}\hat{\gamma}^{21*}_{2\mathfrak{K}_1}(\mathfrak{K}_2)}{2\sqrt{Q_0}\omega_{\mathfrak{K}_1}\omega_{\mathfrak{K}_2}\left(\omega_{\mathfrak{K}_2}-\omega_{\mathfrak{K}_1}\right)} \\
&= \frac{3\Delta_{\mathfrak{K}_1 B}\Delta_{-\mathfrak{K}_2 B}}{8\sqrt{Q_0}\omega_{\mathfrak{K}_1}\omega_{\mathfrak{K}_2}} - \frac{1}{8\sqrt{Q_0}\omega_{\mathfrak{K}_1}\omega_{\mathfrak{K}_2}}\sum\!\!\!\!\!\!\int\frac{dk'}{2\pi}\frac{\sqrt{\lambda Q_0}\Delta_{-k' B}V_{-\mathfrak{K}_2\mathfrak{K}_1 k'}}{\left[(\omega_{\mathfrak{K}_2}-\omega_{\mathfrak{K}_1})^2-\omega^2_{k'}\right]}.
\end{aligned}
$$
(2-102)

(2-100) 和 (2-102) 这两项贡献即是我们对约化内积 $\langle\mathfrak{K}_1|\mathfrak{K}_2\rangle_{\text{red}}$ 涉及到态的 $O(\lambda)$ 阶修正（对应 $\gamma_2$）的计算。

我们现在将计算仅涉及 $O(\lambda^0)$ 阶和 $O(\sqrt{\lambda})$ 阶的项的内积的贡献，它们分别对应着 $\gamma_0$ 和 $\gamma_1$。我们把 $\langle\mathfrak{K}_1|\mathfrak{K}_2\rangle_{n,\text{red}}$ 定义为把式 (2-70) 中的态 $\langle\phi|$ 和 $|\psi\rangle$ 换成

$\langle \mathfrak{K}_1 |$ 和 $|\mathfrak{K}_2\rangle$ 且对应于特定 $n$ 值的项。然后，使用式 (2-78) 和 (2-79) 中的系数，我们得到

$$
\begin{aligned}
\langle \mathfrak{K}_1 | \mathfrak{K}_2 \rangle_{1,\text{red}} &= \sum\!\!\!\!\!\!\!\int \frac{dk_1}{2\pi} \frac{\gamma^{01*}_{\mathfrak{K}_1}(k_1)}{(2\omega_{k_1})} \left[ \sqrt{Q_0}\gamma^{01}_{\mathfrak{K}_2}(k_1) + \Delta_{k_1 B}\gamma^{00}_{\mathfrak{K}_2} + 2\sum\!\!\!\!\!\!\!\int \frac{dk'}{2\pi}\frac{\Delta_{k'B}}{2\omega_{k'}}\gamma^{02}_{\mathfrak{K}_2}(-k', k_1) \right] \\
&= \frac{1}{2\omega_{\mathfrak{K}_1}} \left[ \sqrt{Q_0}\gamma^{01}_{\mathfrak{K}_2}(\mathfrak{K}_1) + \Delta_{\mathfrak{K}_1 B}\gamma^{00}_{\mathfrak{K}_2} + 2\sum\!\!\!\!\!\!\!\int \frac{dk'}{2\pi}\frac{\Delta_{k'B}}{2\omega_{k'}}\gamma^{02}_{\mathfrak{K}_2}(-k', \mathfrak{K}_1) \right] \\
&= \frac{\sqrt{Q_0}2\pi\delta(\mathfrak{K}_1 - \mathfrak{K}_2)}{2\omega_{\mathfrak{K}_1}} + \frac{1}{2\omega_{\mathfrak{K}_1}\sqrt{Q_0}} \left[ \Delta_{\mathfrak{K}_1 B}\left(\frac{\sqrt{Q_0}\lambda V_{\mathcal{I}-\mathfrak{K}_2}}{4\omega^2_{\mathfrak{K}_2}} - \frac{\Delta_{-\mathfrak{K}_2 B}}{4\omega_{\mathfrak{K}_2}}\right) \right. \\
&\quad + \sum\!\!\!\!\!\!\!\int \frac{dk'}{2\pi}\frac{\Delta_{k'B}}{\omega_{k'}}\left(-\frac{2\pi\delta(\mathfrak{K}_1-\mathfrak{K}_2)}{4}\left(\Delta_{-k'B} + \sqrt{Q_0}\lambda\frac{V_{\mathcal{I}-k'}}{\omega_{k'}}\right)\right. \\
&\quad \left. \left. + \frac{\sqrt{Q_0}\lambda V_{-\mathfrak{K}_2-k'\mathfrak{K}_1}}{4\omega_{\mathfrak{K}_2}\left(\omega_{\mathfrak{K}_2} - \omega_{k'} - \omega_{\mathfrak{K}_1}\right)} - \frac{2\pi\delta(k'+\mathfrak{K}_2)}{4}\left(\Delta_{\mathfrak{K}_1 B} + \sqrt{Q_0}\lambda\frac{V_{\mathcal{I}\mathfrak{K}_1}}{\omega_{\mathfrak{K}_1}}\right)\right) \right] \\
&= \frac{2\pi\delta(\mathfrak{K}_1 - \mathfrak{K}_2)}{2\omega_{\mathfrak{K}_1}} \left[ \sqrt{Q_0} - \frac{1}{\sqrt{Q_0}}\sum\!\!\!\!\!\!\!\int \frac{dk'}{2\pi}\frac{\Delta_{k'B}}{4\omega_{k'}}\left(\Delta_{-k'B} + \sqrt{Q_0}\lambda\frac{V_{\mathcal{I}-k'}}{\omega_{k'}}\right) \right] \\
&\quad + \frac{\Delta_{\mathfrak{K}_1 B}}{8\omega_{\mathfrak{K}_1}\omega_{\mathfrak{K}_2}\sqrt{Q_0}}\left(\frac{\sqrt{Q_0}\lambda V_{\mathcal{I}-\mathfrak{K}_2}}{\omega_{\mathfrak{K}_2}} - \Delta_{-\mathfrak{K}_2 B}\right) \\
&\quad - \frac{\Delta_{-\mathfrak{K}_2 B}}{8\omega_{\mathfrak{K}_1}\omega_{\mathfrak{K}_2}\sqrt{Q_0}}\left(\Delta_{\mathfrak{K}_1 B} + \sqrt{Q_0}\lambda\frac{V_{\mathcal{I}\mathfrak{K}_1}}{\omega_{\mathfrak{K}_1}}\right) \\
&\quad + \frac{\sqrt{\lambda}}{8\omega_{\mathfrak{K}_1}\omega_{\mathfrak{K}_2}}\sum\!\!\!\!\!\!\!\int \frac{dk'}{2\pi}\frac{\Delta_{k'B}V_{-\mathfrak{K}_2-k'\mathfrak{K}_1}}{\omega_{k'}(\omega_{\mathfrak{K}_2}-\omega_{k'}-\omega_{\mathfrak{K}_1})} \quad (2\text{-}103)
\end{aligned}
$$

以及

$$
\langle \mathfrak{K}_1 | \mathfrak{K}_2 \rangle_{0,\text{red}} = \frac{1}{16\sqrt{Q_0}\omega_{\mathfrak{K}_1}\omega_{\mathfrak{K}_2}}\left(\frac{\sqrt{Q_0}\lambda V_{\mathcal{I}\mathfrak{K}_1}}{\omega_{\mathfrak{K}_1}} - \Delta_{\mathfrak{K}_1 B}\right)\left(\frac{\sqrt{Q_0}\lambda V_{\mathcal{I}-\mathfrak{K}_2}}{\omega_{\mathfrak{K}_2}} + \Delta_{-\mathfrak{K}_2 B}\right) \tag{2-104}
$$

还有

$$\begin{aligned}
\langle \mathfrak{K}_1 | \mathfrak{K}_2 \rangle_{2,\text{red}} &= \sum \!\!\!\!\!\!\int \frac{d^2 k}{(2\pi)^2} \frac{\gamma^{02*}_{\mathfrak{K}_1}(k_1,k_2)}{4\omega_{k_1}\omega_{k_2}} \left[ \sqrt{Q_0} \gamma^{02}_{\mathfrak{K}_2}(k_1,k_2) + \Delta_{k_1 B} \gamma^{01}_{\mathfrak{K}_2}(k_2) + (k_1 \leftrightarrow k_2) \right] \\
&= \sum \!\!\!\!\!\!\int \frac{d^2 k}{(2\pi)^2} \frac{1}{16\omega_{k_1}\omega_{k_2}\sqrt{Q_0}} \left[ -2\pi \delta(k_2 - \mathfrak{K}_1) \left( \Delta_{-k_1 B} + \sqrt{Q_0} \lambda \frac{V_{\mathcal{I}-k_1}}{\omega_{k_1}} \right) \right. \\
&\quad + \frac{\sqrt{Q_0} \lambda V^*_{-\mathfrak{K}_1 k_1 k_2}}{\omega_{\mathfrak{K}_1}\left(\omega_{\mathfrak{K}_1} - \omega_{k_1} - \omega_{k_2}\right)} - 2\pi \delta(k_1 - \mathfrak{K}_1) \left( \Delta_{-k_2 B} + \sqrt{Q_0} \lambda \frac{V_{\mathcal{I}-k_2}}{\omega_{k_2}} \right) \bigg] \\
&\quad \times \left[ 2\pi \delta(k_2 - \mathfrak{K}_2) \left( \Delta_{k_1 B} - \sqrt{Q_0} \lambda \frac{V_{\mathcal{I} k_1}}{\omega_{k_1}} \right) + \frac{\sqrt{Q_0} \lambda V_{-\mathfrak{K}_2 k_1 k_2}}{2\omega_{\mathfrak{K}_2}\left(\omega_{\mathfrak{K}_2} - \omega_{k_1} - \omega_{k_2}\right)} \right] \\
&= \frac{2\pi \delta(\mathfrak{K}_1 - \mathfrak{K}_2)}{16\omega_{\mathfrak{K}_1} \sqrt{Q_0}} \int \frac{dk_1}{2\pi} \frac{1}{\omega_{k_1}} \left[ Q_0 \lambda \frac{|V_{\mathcal{I} k_1}|^2}{\omega_{k_1}^2} - |\Delta_{k_1 B}|^2 \right] \\
&\quad + \frac{1}{16\omega_{\mathfrak{K}_1}\omega_{\mathfrak{K}_2}\sqrt{Q_0}} \left( \sqrt{Q_0} \lambda \frac{V_{\mathcal{I}-\mathfrak{K}_2}}{\omega_{\mathfrak{K}_2}} + \Delta_{-\mathfrak{K}_2 B} \right) \left( \sqrt{Q_0} \lambda \frac{V_{\mathcal{I} \mathfrak{K}_1}}{\omega_{\mathfrak{K}_1}} - \Delta_{\mathfrak{K}_1 B} \right) \\
&\quad + \frac{\sqrt{\lambda}}{16\omega_{\mathfrak{K}_1}\omega_{\mathfrak{K}_2}} \sum \!\!\!\!\!\!\int \frac{dk'}{2\pi} \frac{V_{\mathfrak{K}_1 - \mathfrak{K}_2 - k'}}{\omega_{k'}\left(\omega_{\mathfrak{K}_1} - \omega_{\mathfrak{K}_2} - \omega_{k'}\right)} \left( \Delta_{k' B} - \sqrt{Q_0} \lambda \frac{V_{\mathcal{I} k'}}{\omega_{k'}} \right) \\
&\quad + \frac{\sqrt{\lambda}}{16\omega_{\mathfrak{K}_1}\omega_{\mathfrak{K}_2}} \sum \!\!\!\!\!\!\int \frac{dk'}{2\pi} \frac{V_{\mathfrak{K}_1 - \mathfrak{K}_2 - k'}}{\omega_{k'}\left(\omega_{\mathfrak{K}_1} - \omega_{\mathfrak{K}_2} + \omega_{k'}\right)} \left( \Delta_{k' B} + \sqrt{Q_0} \lambda \frac{V_{\mathcal{I} k'}}{\omega_{k'}} \right) \\
&\quad + \frac{\sqrt{Q_0}\lambda}{32\omega_{\mathfrak{K}_1}\omega_{\mathfrak{K}_2}} \sum \!\!\!\!\!\!\int \frac{d^2 k'}{(2\pi)^2} \frac{V_{\mathfrak{K}_1 - k'_1 - k'_2} V_{-\mathfrak{K}_2 k'_1 k'_2}}{\omega_{k'_1}\omega_{k'_2}\left(\omega_{\mathfrak{K}_1} - \omega_{k'_1} - \omega_{k'_2}\right)\left(\omega_{\mathfrak{K}_2} - \omega_{k'_1} - \omega_{k'_2}\right)}.
\end{aligned}$$
(2-105)

第二行的 $V_{\mathcal{I}-\mathfrak{K}_2} V_{\mathcal{I}\mathfrak{K}_1}$ 项加上等式 (2-104) 中的 $V_{\mathcal{I}-\mathfrak{K}_2} V_{\mathcal{I}\mathfrak{K}_1}$ 项，抵消了等式 (2-100) 中的第一行。第二行的 $\Delta_{-\mathfrak{K}_2 B}\Delta_{\mathfrak{K}_1 B}$ 项，加上等式 (2-104) 中的 $\Delta_{-\mathfrak{K}_2 B}\Delta_{\mathfrak{K}_1 B}$ 项，以及等式 (2-103) 中的两个 $\Delta_{-\mathfrak{K}_2 B}\Delta_{\mathfrak{K}_1 B}$ 项，恰好抵消了等式 (2-102) 的第一项。将第三行和第四行的 $V_{\mathfrak{K}_1-\mathfrak{K}_2-k'}\Delta_{k'B}$ 项和等式 (2-103) 的最后一行相加所得之和，精确地和等式 (2-102) 中的另一项抵消。

将第三行和第四行相加，$V_{\mathfrak{K}_1-\mathfrak{K}_2 k'} V_{\mathcal{I} k'}$ 项抵消了等式 (2-100) 的第三行。最后一行和等式 (2-100) 的第二行抵消。最后，第二行的 $\Delta_{\mathfrak{K}_1} V_{\mathcal{I} \mathfrak{K}_2}$ 项，和等式 (2-104) 中的 $\Delta_{\mathfrak{K}_1} V_{\mathcal{I}\mathfrak{K}_2}$ 项，以及和等式 (2-103) 的最后一个表达式的第二行中的 $\Delta_{\mathfrak{K}_1} V_{\mathcal{I}\mathfrak{K}_2}$ 项相加之后抵消，同样地，与它结构相同但 $\mathfrak{K}_1 \leftrightarrow \mathfrak{K}_2$ 的项 $\Delta_{\mathfrak{K}_2} V_{\mathcal{I}\mathfrak{K}_1}$ 也抵消了。

最后，$n = 4$ 的贡献是

$$
\begin{aligned}
\langle \mathfrak{K}_1 | \mathfrak{K}_2 \rangle_{4,\text{red}} &= 2\pi \delta(\mathfrak{K}_1 - \mathfrak{K}_2) \frac{\lambda \sqrt{Q_0}}{96 \omega_{\mathfrak{K}_1}} \sum\!\!\!\!\!\!\int \frac{d^3 k}{(2\pi)^3} \frac{|V_{k_1 k_2 k_3}|^2}{\omega_{k_1} \omega_{k_2} \omega_{k_3} (\omega_{k_1} + \omega_{k_2} + \omega_{k_3})^2} \\
&\quad + \frac{\lambda \sqrt{Q_0}}{32 \omega_{\mathfrak{K}_1} \omega_{\mathfrak{K}_2}} \sum\!\!\!\!\!\!\int \frac{d^2 k}{(2\pi)^2} \frac{V^*_{\mathfrak{K}_2 k_1 k_2} V_{\mathfrak{K}_1 k_1 k_2}}{\omega_{k_1} \omega_{k_2} (\omega_{\mathfrak{K}_1} + \omega_{k_1} + \omega_{k_2})(\omega_{\mathfrak{K}_2} + \omega_{k_1} + \omega_{k_2})}.
\end{aligned}
$$
(2-106)

第二行，也就是当 $\mathfrak{K}_1 \neq \mathfrak{K}_2$ 时唯一留下的项，和等式 (2-100) 的最后一行抵消，于是等式 (2-100) 中的项的抵消全部完成。

#### 2.4.3.4 总结

因此我们得出结论，在 $\mathfrak{K}_1 \neq \mathfrak{K}_2$ 处约化内积为零。这是必然的，因为它们代表了 $H'$ 的不同本征态。因此，这是对我们的主要结果 (2-70) 的一致性检验。

我们的推导不适用于 $\mathfrak{K}_1 = -\mathfrak{K}_2$ 处的 $O(\sqrt{\lambda})$ 修正，因为在分子和分母中都存在带有 $(\omega_{\mathfrak{K}_1} - \omega_{\mathfrak{K}_2})$ 的项并且我们将它们抵消了。事实上，态 $|\mathfrak{K}_1\rangle$ 和态 $|-\mathfrak{K}_1\rangle$ 具有相同的能量，因而有可能会出现混合。

问题不是因为我们不够细致，而是确实存在简并的本征空间，其上可以自由定义 $|\mathfrak{K}_1\rangle$ 以与 $|-\mathfrak{K}_1\rangle$ 有任意重叠。然而，对于给定的物理问题，在极点 $\omega_{\mathfrak{K}_1} = \omega_{\mathfrak{K}_2}$ 处可能存在更有用的处理方法。当我们在本章后续部分考虑介子倍增时，我们将看到这样的物理原理如何确定一个相关的极点。在未来的工作中，我们打算使用扭结-介子弹性散射来确定用于在 $\mathfrak{K}_1 = -\mathfrak{K}_2$ 处定义极点的处理方法。

同样，我们的推导在 $\mathfrak{K}_1 = \mathfrak{K}_2$ 处也不可靠，因为同样的操作是缺乏明确定义的。这仅是我们自由选择 $|\mathfrak{K}_1\rangle$ 的归一化的反映。例如，我们可以对所有大于 0 的 $i$ 固定 $\gamma^{01}_{i\mathfrak{K}}(\mathfrak{K}) = 0$，这类似于我们在计算单扭结零介子态的约化模时施加的条件 $\gamma^{00}_2 = 0$。

### 2.5 对初态和末态的修正

#### 2.5.1 初、末态修正的研究动机

在实验中，任何初始条件都是允许的。初始条件的选择由实验者自行决定，原因是它取决于实验的设置方式。同样，每个检测道中末态的选择也是由实验者决定，原因是它取决于探测器的设计。在第 3 章中，我们考虑将初态波包构造为 $H'_2$ 本征态 $|k_1\rangle_0$ 的叠加，其中每一个 $|k_1\rangle_0$ 对应于所需态的领头阶半经典近

似。换言之，初始单介子态仅在自由扭结哈密顿量 $H'_2$ 的单介子福克空间构造。类似地，概率的计算需要投影到自由哈密顿量的双介子福克空间上，该空间由态 $|k_2 k_3\rangle_0$ 生成。这个过程是明确定义的，并且可以和一些实验的结果对应。

但是其实我们已经做了一个选择。我们也可以另外地使用完整哈密顿量 $H'$ 的本征态 $|k_1\rangle$ 来构建波包。完整哈密顿量 $H'$ 的单介子福克空间中的每个元素都包含自由哈密顿量 $H'_2$ 的各种 $n$-介子本征态 $|k_1 \cdots k_n\rangle_0$ 的叠加。这种选择有些随意，因为波包本身不会是上述两个哈密顿量中任何一个的本征态。然而，人们可能会问，最终的概率是否取决于这一选择。这一点在实验上很重要，因为如果概率取决于选择，则需要确定给定的制备方法和探测器对应于何种选择。从理论上讲这也是很重要的，因为如果结果不同，其中的一个选择可能与 LSZ 约化定理兼容，而另一个可能不兼容。

需要说明的是，本节的部分结果和第 3 章的第 3.2 节重复（除此之外，本节还包括了重要的对极点贡献的研究）。然而，在第 3 章第 3.2 节的推导更为简单，并且那里的处理方式更接近人们熟悉的相互作用表象。

有两种计算反射/透射系数的方法：

（1）从入射波开始，令它随时间演化，然后看出射的结果，这即是第 3 章的第 3.2 节中的方法，其中入射波是真空哈密顿量的本征态。

（2）找到完整哈密顿量的本征态，利用右侧没有入射波这个边界条件（我们令所有入射介子都从左侧的远处入射），直接从两侧的出射波中读出反射和透射系数。因为全哈密顿量的本征态是时间无关的，所以这里没有时间演化，这是本节中使用的方法。

简言之，这两节的区别在于，在这一节我们使用的是完整哈密顿量的本征态（这类似于普通量子力学中的散射），而在第 3 章的第 3.2 节我们使用的是真空哈密顿量的本征态。

### 2.5.2 初态和末态条件

在第 3 章中，我们将要计算的是单扭结单介子的初态演化到单扭结两介子的末态的振幅，我们称这个过程为介子倍增。虽然初态和末态不涉及 $\lambda$ 的幂，但相互作用包含 $\sqrt{\lambda}$，因此振幅为 $O(\sqrt{\lambda})$ 阶。然而，如果初态包含 $O(\sqrt{\lambda})$ 的量子修正，则它可以通过不含 $\lambda$ 的 $H'_2$ 演化到末态，并产生同阶的贡献。类似地，如果允许的末态中包含一个 $O(\sqrt{\lambda})$ 阶修正，其与随 $H'_2$ 演化的初态有一个 $O(1)$ 阶内积，它也将在相同阶给出贡献。如果我们的初态或投影算符被构造为完整哈密顿量的本征态的叠加，就会出现这样的修正。

因此，我们来考虑一个无反射性扭结，以便远离扭结的正规模可以很好地用平面波近似，我们将在式 (2-118) 中快速回顾其形式。让初始介子波包具有与第 3 章中相同的叠加系数

$$\alpha_{k_1} = 2\sigma\sqrt{\pi}\mathcal{B}_{k_1}e^{-\sigma^2(k_1-k_0)^2}e^{i(k_0-k_1)x_0} \tag{2-107}$$

但这一次是作为完整扭结哈密顿量 $H'$ 的本征态的单介子态 $|k_1\rangle$ 的叠加。我们的初态是

$$|\Phi\rangle = \int \frac{dk_1}{2\pi}\alpha_{k_1}|k_1\rangle \tag{2-108}$$

这里与在第 3 章中的不同之处在于，在那里，$H'$ 的本征态 $|k_1\rangle$ 被替换为了 $H'_2$ 的本征态 $|k_1\rangle_0$

$$|\Phi\rangle_0 = \int \frac{dk_1}{2\pi}\alpha_{k_1}|k_1\rangle_0. \tag{2-109}$$

注意，在这两种情况下，都是对连续模 $k_1$ 进行积分，而不对束缚模（包括零模和形模）求和，因为这些模在远离扭结的地方呈指数消失（我们假设了 $|x_0| \gg 1/m$）。

与第 3 章中将要计算的矩阵元 $_0\langle k_2k_3|e^{-it(H'_2+H'_3)}|\Phi\rangle_0$ 不同，我们现在对如下矩阵元感兴趣

$$_{\text{vac}}\langle k_2k_3|e^{-itH'}|\Phi\rangle. \tag{2-110}$$

这里 $|k_2k_3\rangle_{\text{vac}}$ 不是扭结哈密顿量的本征态 $|k_2k_3\rangle$。假设它是，那么演化算符中的 $H'$ 只会将它乘以一个相位，然后矩阵元仅通过简单的相位旋转演化，而我们感兴趣的是从零开始演化为非零的概率值。相反，它是通过将 $f(x)$ 替换为 $f(-\infty)$ 或 $f(+\infty)$ 的远离扭结左侧或右侧的 $H'$ 的平移不变本征态，参考定义式 (1-1) 和 (1-4)。

在非反射性扭结的情况下，在领头阶，$|k_2k_3\rangle_{\text{vac}}$ 中唯一相关的量子修正是

$$|k_2k_3\rangle_{\text{vac}} = |k_2k_3\rangle_0 + \frac{\sqrt{\lambda}V^{(3)}(\sqrt{\lambda}f(-\infty))\mathcal{B}_{-k_2}\mathcal{B}_{-k_3}\mathcal{B}_{k_2+k_3}}{4\omega_{k_2}\omega_{k_3}(\omega_{k_2}+\omega_{k_3}-\omega_{k_2+k_3})}|k_2+k_3\rangle_0 \tag{2-111}$$

和

$$|k_2k_3\rangle_{\text{vac}} = |k_2k_3\rangle_0 + \frac{\sqrt{\lambda}V^{(3)}(\sqrt{\lambda}f(+\infty))\mathcal{D}_{-k_2}\mathcal{D}_{-k_3}\mathcal{D}_{k_2+k_3}}{4\omega_{k_2}\omega_{k_3}(\omega_{k_2}+\omega_{k_3}-\omega_{k_2+k_3})}|k_2+k_3\rangle_0. \tag{2-112}$$

其中前者用于波包位于 $x \ll 0$ 时的内积计算，后者用于波包位于 $x \gg 0$ 时的内积计算。投影算符由对位于 $x \ll 0$ 和 $x \gg 0$ 的 $|k_2k_3\rangle_{\text{vac}}$ 的波包的积分组装而成，其分别由 (2-111) 和 (2-112) 的叠加组成。注意，只有 (2-111) 和 (2-112) 中的第一项与初态修正有关，而第二项与末态修正有关。

### 2.5.3 初末态修正计算

时刻 $t$ 的态是

$$|t\rangle = \int \frac{dk_1}{2\pi} \alpha_{k_1} e^{-itH'} |k_1\rangle = \int \frac{dk_1}{2\pi} \alpha_{k_1} e^{-it\tilde{\omega}_{k_1}} |k_1\rangle, \qquad (2\text{-}113)$$

其中 $\tilde{\omega}_{k_1}$ 是 $|k_1\rangle$ 的量子修正能量。它等于 $\omega_{k_1}$ 加上 $O(\lambda)$ 阶的修正 [18]。由于这里只考虑 $O(\sqrt{\lambda})$ 阶的修正，我们可以忽略这些修正而只使用 $\omega_{k_1}$。接下来，由于 $\alpha_{k_1}$ 位于 $k_1 = k_0$ 附近，我们可以作如下展开

$$\tilde{\omega}_{k_1} = \omega_{k_1} = \omega_{k_0} + \frac{k_0}{\omega_{k_0}}(k_1 - k_0). \qquad (2\text{-}114)$$

然后我们得到

$$
\begin{aligned}
|t\rangle &= \int \frac{dk_1}{2\pi} 2\sigma\sqrt{\pi} \mathcal{B}_{k_1} e^{-\sigma^2(k_1-k_0)^2} e^{i(k_0-k_1)x_0} e^{-it\left(\omega_{k_0} + \frac{k_0}{\omega_{k_0}}(k_1-k_0)\right)} |k_1\rangle \\
&= 2\sigma\sqrt{\pi} \mathcal{B}_{k_0} e^{-i\omega_{k_0}t} \int \frac{dk_1}{2\pi} e^{-\sigma^2(k_1-k_0)^2} e^{-i(k_1-k_0)\left(x_0 + \frac{k_0}{\omega_{k_0}}t\right)} |k_1\rangle. \quad (2\text{-}115)
\end{aligned}
$$

现在，让我们考虑在 $O(\sqrt{\lambda})$ 阶的对 $|k_1\rangle$ 的一个特定贡献

$$
\begin{aligned}
|k_1\rangle &\supset \frac{1}{\sqrt{Q_0}} \int \frac{dk_2}{2\pi} \int \frac{dk_3}{2\pi} \gamma^{02}_{1k_1}(k_2, k_3) |k_2 k_3\rangle_0 \\
&\supset \frac{\sqrt{\lambda}}{4\omega_{k_1}} \int \frac{dk_2}{2\pi} \int \frac{dk_3}{2\pi} \frac{V_{-k_1 k_2 k_3}}{\left(\omega_{k_1} - \omega_{k_2} - \omega_{k_3}\right)} |k_2 k_3\rangle_0. \quad (2\text{-}116)
\end{aligned}
$$

$k_2$ 或 $k_3$ 是束缚模的情况很有趣，那种情况将成为第 4 章中的（反）斯托克斯散射的研究主题，这里我们将仅考虑连续模 $k_2$ 和 $k_3$，然后直接将结果推广到（反）斯托克斯散射情形。在 $\omega_{k_1} = \omega_{k_2} + \omega_{k_3}$ 处有一个极点。这个极点当然很重要，因为介子倍增就发生在极点处。但是让我们首先考虑远离这个极点的 $k_2$ 和 $k_3$（与 $1/\sigma$ 的尺度相比），在第 2.5.4 节我们再回到对极点的研究。于是在这里我们可以在分母中将 $k_1$ 取为 $k_0$ 并且式 (2-116) 贡献了

$$
\begin{aligned}
|t\rangle \supset{} & \frac{\sqrt{\lambda}\mathcal{B}_{k_0} e^{-i\omega_{k_0}t}}{4\omega_{k_0}} \int \frac{dk_2}{2\pi} \int \frac{dk_3}{2\pi} \frac{1}{\omega_{k_0} - \omega_{k_2} - \omega_{k_3}} \int dx V^{(3)}(gf(x)) \mathfrak{g}_{k_2}(x) \mathfrak{g}_{k_3}(x) \\
& \times \left[ 2\sigma\sqrt{\pi} \int \frac{dk_1}{2\pi} e^{-\sigma^2(k_1-k_0)^2} e^{-i(k_1-k_0)\left(x_0 + \frac{k_0}{\omega_{k_0}}t\right)} \mathfrak{g}_{-k_1}(x) \right] |k_2 k_3\rangle_0. \quad (2\text{-}117)
\end{aligned}
$$

让我们尝试计算 $x \gg 0$ 和 $x \ll 0$ 时方括号中的积分，其中

$$\mathfrak{g}_k(x) = \begin{cases} \mathcal{B}_k e^{-ikx} & \text{if} \quad x \ll -1/m \\ \mathcal{D}_k e^{-ikx} & \text{if} \quad x \gg 1/m \end{cases} \quad (2\text{-}118)$$

$$|\mathcal{B}_k|^2 = |\mathcal{D}_k|^2 = 1, \quad \mathcal{B}_k^* = \mathcal{B}_{-k}, \quad \mathcal{D}_k^* = \mathcal{D}_{-k}.$$

于是方括号中的积分是

$$2\sigma\sqrt{\pi} \int \frac{dk_1}{2\pi} e^{-\sigma^2(k_1-k_0)^2} e^{-i(k_1-k_0)\left(x_0 + \frac{k_0}{\omega_{k_0}}t\right)} \mathfrak{g}_{-k_1}(x)$$

$$= e^{ik_0 x}\mathrm{Exp}\left[-\frac{\left(-x + x_0 + \frac{k_0}{\omega_{k_0}}t\right)^2}{4\sigma^2}\right] \begin{cases} \mathcal{B}_{-k_0} & \text{if} \quad x \ll -1/m \\ \mathcal{D}_{-k_0} & \text{if} \quad x \gg 1/m. \end{cases}$$

(2-119)

我们看到 $x$ 在 $x_t$ 附近达到峰值，其中

$$x_t = x_0 + \frac{k_0}{\omega_{k_0}}t. \quad (2\text{-}120)$$

当 $|x_t| \gg 0$ 时，高斯项位于 $|x| \gg 0$ 处。这里 $f(x)$ 趋于一个常数，因此 $V^{(3)}(gf(x))$ 也趋于一个常数，其对应于理论的两个真空之一中势能的三阶导数。常数的值取决于 $x_t$ 的符号。现在让我们转向 $x$ 的被积函数。具体而言，让我们考虑 $t$ 远小于介子波包撞到扭结的时间，因此 $x \ll 0$，则

$$\int dx V^{(3)}(gf(x))\mathfrak{g}_{k_2}(x)\mathfrak{g}_{k_3}(x)e^{ik_0 x}\mathrm{Exp}\left[-\frac{\left(-x + x_0 + k_0 t/\omega_{k_0}\right)^2}{4\sigma^2}\right]\mathcal{B}_{-k_0}$$

$$= 2\sigma\sqrt{\pi}V^{(3)}(\sqrt{\lambda}f(-\infty))\mathcal{B}_{-k_0}\mathcal{B}_{k_2}\mathcal{B}_{k_3}e^{-\sigma^2(k_0-k_2-k_3)^2}e^{ix_t(k_0-k_2-k_3)}.$$

(2-121)

当 $t$ 很大时，$x_t \gg 0$，只需将相位 $\mathcal{B}$ 更改为 $\mathcal{D}$ 并且 $V^{(3)}$ 在扭结右边的真空处计算。

总结一下，介子和扭结碰撞之后

$$|t\rangle \supset \frac{2\sigma\sqrt{\pi}V^{(3)}(\sqrt{\lambda}f(+\infty))\sqrt{\lambda}\mathcal{B}_{k_0}\mathcal{D}_{-k_0}e^{-i\omega_{k_0}t}}{4\omega_{k_0}}$$

$$\times \int \frac{dk_2}{2\pi} \int \frac{dk_3}{2\pi} \frac{\mathcal{D}_{k_2}\mathcal{D}_{k_3}}{\omega_{k_0}-\omega_{k_2}-\omega_{k_3}} e^{-\sigma^2(k_0-k_2-k_3)^2} e^{ix_t(k_0-k_2-k_3)} |k_2 k_3\rangle_0$$

$$= \frac{2\sigma\sqrt{\pi}V^{(3)}(\sqrt{\lambda}f(+\infty))\sqrt{\lambda}\mathcal{B}_{k_0}\mathcal{D}_{-k_0}e^{-i\omega_{k_0}t+ik_0 x_t}}{4\omega_{k_0}}$$

$$\times \int \frac{dk_2}{2\pi} \int \frac{dk_3}{2\pi} \frac{\mathfrak{g}_{k_2}(x_t)\mathfrak{g}_{k_3}(x_t)}{\omega_{k_0}-\omega_{k_2}-\omega_{k_3}} e^{-\sigma^2(k_0-k_2-k_3)^2} |k_2 k_3\rangle_0$$

$$= \frac{V^{(3)}(\sqrt{\lambda}f(+\infty))\sqrt{\lambda}\mathcal{B}_{k_0}\mathcal{D}_{-k_0}e^{-i\omega_{k_0}t}}{4\omega_{k_0}}$$

$$\times \int \frac{dk_2}{2\pi} \int \frac{dk_3}{2\pi} \frac{\mathcal{D}_{k_2}\mathcal{D}_{k_3}}{\omega_{k_0}-\omega_{k_2}-\omega_{k_3}} 2\pi\delta(k_0-k_2-k_3) |k_2 k_3\rangle_0. \qquad (2\text{-}122)$$

在最后一个等式中，我们考虑了极限 $\sigma \to \infty$。碰撞之前则是

$$|t\rangle \supset \frac{2\sigma\sqrt{\pi}V^{(3)}(\sqrt{\lambda}f(-\infty))\sqrt{\lambda}\mathcal{B}_{k_0}\mathcal{B}_{-k_0}e^{-i\omega_{k_0}t}}{4\omega_{k_0}}$$

$$\times \int \frac{dk_2}{2\pi} \int \frac{dk_3}{2\pi} \frac{\mathcal{B}_{k_2}\mathcal{B}_{k_3}}{\omega_{k_0}-\omega_{k_2}-\omega_{k_3}} e^{-\sigma^2(k_0-k_2-k_3)^2} e^{ix_t(k_0-k_2-k_3)} |k_2 k_3\rangle_0.$$

$$= \frac{2\sigma\sqrt{\pi}V^{(3)}(\sqrt{\lambda}f(-\infty))\sqrt{\lambda}e^{-i\omega_{k_0}t+ik_0 x_t}}{4\omega_{k_0}}$$

$$\times \int \frac{dk_2}{2\pi} \int \frac{dk_3}{2\pi} \frac{\mathfrak{g}_{k_2}(x_t)\mathfrak{g}_{k_3}(x_t)}{\omega_{k_0}-\omega_{k_2}-\omega_{k_3}} e^{-\sigma^2(k_0-k_2-k_3)^2} |k_2 k_3\rangle_0$$

$$= \frac{V^{(3)}(\sqrt{\lambda}f(-\infty))\sqrt{\lambda}e^{-i\omega_{k_0}t}}{4\omega_{k_0}}$$

$$\times \int \frac{dk_2}{2\pi} \int \frac{dk_3}{2\pi} \frac{\mathcal{B}_{k_2}\mathcal{B}_{k_3}}{\omega_{k_0}-\omega_{k_2}-\omega_{k_3}} 2\pi\delta(k_0-k_2-k_3) |k_2 k_3\rangle_0. \qquad (2\text{-}123)$$

注意，在任何一种情况下，$k_0$ 和 $k_2 + k_3$ 的差值在 $1/\sigma$ 阶，并且我们的假设是它远小于 $m$。因此 $\omega_{k_0}$ 与 $\omega_{k_2} + \omega_{k_3}$ 相差甚远，并且从初始波包产生的任何能量为 $\omega_{k_2}$ 和 $\omega_{k_3}$ 介子都将极度离壳。因此，我们预计这些项不会影响介子倍增概率，接着我们来看是否果真如此。为此，我们只需计算 $|t\rangle$ 和式 (2-111) 中 $|k_2 k_3\rangle_{\text{vac}}$ 的约化内积。

碰撞后，直到 $O(\sqrt{\lambda})$ 阶，$|k_2 k_3\rangle_{\text{vac}}$ 由式 (2-112) 给出。我们可以很容易地从态中读出它们的系数 $\gamma$。我们将始终考虑极限 $\sigma \to \infty$，首先，让我们考虑碰撞后

的内积。

$$\gamma_t^{01}(k) = \mathcal{B}_k e^{-i\omega_k t} 2\pi\delta(k-k_0) \tag{2-124}$$

$$\gamma_t^{02}(k_2' k_3') = \frac{\sqrt{\lambda} V^{(3)}(\sqrt{\lambda}f(+\infty))\mathcal{B}_{k_0}\mathcal{D}_{-k_0}\mathcal{D}_{k_2'}\mathcal{D}_{k_3'}e^{-i\omega_{k_0}t}2\pi\delta(k_0-k_2'-k_3')}{4\omega_{k_0}\left(\omega_{k_0}-\omega_{k_2'}-\omega_{k_3'}\right)}$$

$$\gamma_{k_2 k_3,\text{vac}}^{02}(k_2' k_3') = 2\pi\delta(k_2'-k_2)2\pi\delta(k_3'-k_3)$$

$$\gamma_{k_2 k_3,\text{vac}}^{01}(k) = \frac{\sqrt{\lambda} V^{(3)}(\sqrt{\lambda}f(+\infty))\mathcal{D}_{-k_2}\mathcal{D}_{-k_3}\mathcal{D}_k 2\pi\delta(k-k_2-k_3)}{4\omega_{k_2}\omega_{k_3}(\omega_{k_2}+\omega_{k_3}-\omega_k)}.$$

现在我们再次使用我们的主要公式 (2-70) 来计算直到 $O(\sqrt{\lambda})$ 阶的约化内积

$$_{\text{vac}}\langle k_2 k_3 | t \rangle_{\text{red}}$$

$$= \sum\!\!\!\!\!\!\!\int \frac{dk}{2\pi} \frac{\gamma_{k_2 k_3,\text{vac}}^{01*}(k)}{2\omega_k}\sqrt{Q_0}\gamma_t^{01}(k) + 2\sum\!\!\!\!\!\!\!\int \frac{dk_2' dk_3'}{(2\pi)^2} \frac{\gamma_{k_2 k_3,\text{vac}}^{02*}(k_2' k_3')}{4\omega_{k_2'}\omega_{k_3'}}\sqrt{Q_0}\gamma_t^{02}(k_2' k_3')$$

$$= \sum\!\!\!\!\!\!\!\int \frac{dk}{2\pi} \frac{\sqrt{Q_0}}{2\omega_k}\frac{\sqrt{\lambda}V^{(3)}(\sqrt{\lambda}f(+\infty))\mathcal{D}_{k_2}\mathcal{D}_{k_3}\mathcal{D}_{-k}2\pi\delta(k-k_2-k_3)}{4\omega_{k_2}\omega_{k_3}(\omega_{k_2}+\omega_{k_3}-\omega_k)}\mathcal{B}_k e^{-i\omega_k t}2\pi\delta(k-k_0)$$

$$+2\sum\!\!\!\!\!\!\!\int \frac{dk_2' dk_3'}{(2\pi)^2}\frac{\sqrt{Q_0}}{4\omega_{k_2'}\omega_{k_3'}}2\pi\delta(k_2'-k_2)2\pi\delta(k_3'-k_3)$$

$$\times\frac{\sqrt{\lambda}V^{(3)}(\sqrt{\lambda}f(+\infty))\mathcal{B}_{k_0}\mathcal{D}_{-k_0}\mathcal{D}_{k_2'}\mathcal{D}_{k_3'}e^{-i\omega_{k_0}t}2\pi\delta(k_0-k_2'-k_3')}{4\omega_{k_0}\left(\omega_{k_0}-\omega_{k_2'}-\omega_{k_3'}\right)}$$

$$= \frac{\sqrt{\lambda Q_0}V^{(3)}(\sqrt{\lambda}f(+\infty))\mathcal{B}_{k_2+k_3}\mathcal{D}_{k_2}\mathcal{D}_{k_3}\mathcal{D}_{-k_2-k_3}e^{-i\omega_{k_2+k_3}t}2\pi\delta(k_0-k_2-k_3)}{8\omega_{k_2}\omega_{k_3}\omega_{k_2+k_3}(\omega_{k_2}+\omega_{k_3}-\omega_{k_2+k_3})}$$

$$+\frac{\sqrt{\lambda Q_0}V^{(3)}(\sqrt{\lambda}f(+\infty))\mathcal{B}_{k_2+k_3}\mathcal{D}_{k_2}\mathcal{D}_{k_3}\mathcal{D}_{-k_2-k_3}e^{-i\omega_{k_2+k_3}t}2\pi\delta(k_0-k_2-k_3)}{8\omega_{k_2}\omega_{k_3}\omega_{k_2+k_3}(\omega_{k_2+k_3}-\omega_{k_2}-\omega_{k_3})}$$

$$= 0. \tag{2-125}$$

碰撞前的内积的计算与之相似。这里出现在投影算符中以及因此出现在矩阵元中的 $|k_2 k_3\rangle_{\text{vac}}$ 由式 (2-111) 给出。因此式 (2-124) 中的其中两个 $\gamma$ 需要更换为

$$\gamma_t^{02}(k_2' k_3') = \frac{\sqrt{\lambda}V^{(3)}(\sqrt{\lambda}f(-\infty))\mathcal{B}_{k_2'}\mathcal{B}_{k_3'}e^{-i\omega_{k_0}t}2\pi\delta(k_0-k_2'-k_3')}{4\omega_{k_0}\left(\omega_{k_0}-\omega_{k_2'}-\omega_{k_3'}\right)}$$

$$\gamma_{k_2 k_3,\text{vac}}^{01}(k) = \frac{\sqrt{\lambda}V^{(3)}(\sqrt{\lambda}f(-\infty))\mathcal{B}_{-k_2}\mathcal{B}_{-k_3}\mathcal{B}_k 2\pi\delta(k-k_2-k_3)}{4\omega_{k_2}\omega_{k_3}(\omega_{k_2}+\omega_{k_3}-\omega_k)}.$$

再次地，直到 $O(\sqrt{\lambda})$ 阶

$$\begin{aligned}
&{}_{\text{vac}}\langle k_2 k_3|t\rangle_{\text{red}} \\
&= \sum\!\!\!\!\!\!\!\int \frac{dk}{2\pi} \frac{\gamma^{01*}_{k_2k_3,\text{vac}}(k)}{2\omega_k}\sqrt{Q_0}\gamma_t^{01}(k) + 2\sum\!\!\!\!\!\!\!\int \frac{dk_2'dk_3'}{(2\pi)^2} \frac{\gamma^{02*}_{k_2k_3,\text{vac}}(k_2'k_3')}{4\omega_{k_2}\omega_{k_3}}\sqrt{Q_0}\gamma_t^{02}(k_2'k_3') \\
&= \sum\!\!\!\!\!\!\!\int \frac{dk}{2\pi}\frac{\sqrt{Q_0}}{2\omega_k} \frac{\sqrt{\lambda}V^{(3)}(\sqrt{\lambda}f(-\infty))\mathcal{B}_{k_2}\mathcal{B}_{k_3}\mathcal{B}_{-k}2\pi\delta(k-k_2-k_3)}{4\omega_{k_2}\omega_{k_3}(\omega_{k_2}+\omega_{k_3}-\omega_k)}\mathcal{B}_k e^{-i\omega_k t}2\pi\delta(k-k_0) \\
&\quad + 2\sum\!\!\!\!\!\!\!\int \frac{dk_2'dk_3'}{(2\pi)^2}\frac{\sqrt{Q_0}}{4\omega_{k_2}\omega_{k_3}}2\pi\delta(k_2'-k_2)2\pi\delta(k_3'-k_3) \\
&\quad \times \frac{\sqrt{\lambda}V^{(3)}(\sqrt{\lambda}f(-\infty))\mathcal{B}_{k_2'}\mathcal{B}_{k_3'}e^{-i\omega_{k_0}t}2\pi\delta(k_0-k_2'-k_3')}{4\omega_{k_0}\left(\omega_{k_0}-\omega_{k_2'}-\omega_{k_3'}\right)} \\
&= \frac{\sqrt{\lambda Q_0}V^{(3)}(\sqrt{\lambda}f(+\infty))\mathcal{B}_{k_2}\mathcal{B}_{k_3}e^{-i\omega_{k_2+k_3}t}2\pi\delta(k_0-k_2-k_3)}{8\omega_{k_2}\omega_{k_3}\omega_{k_2+k_3}(\omega_{k_2}+\omega_{k_3}-\omega_{k_2+k_3})} \\
&\quad + \frac{\sqrt{\lambda Q_0}V^{(3)}(\sqrt{\lambda}f(-\infty))\mathcal{B}_{k_2}\mathcal{B}_{k_3}e^{-i\omega_{k_2+k_3}t}2\pi\delta(k_0-k_2-k_3)}{8\omega_{k_2}\omega_{k_3}\omega_{k_2+k_3}(\omega_{k_2+k_3}-\omega_{k_2}-\omega_{k_3})} = 0.
\end{aligned} \quad (2\text{-}126)$$

我们看到内积同样地消失了。因此，在远离极点处，初态和末态的修正不会在这一阶产生贡献。当然这是意料之中的，因为这个过程只有在极点处发生才会保证能量守恒。

### 2.5.4 极点处的贡献

在式 (2-117) 中，我们计算了式 (2-116) 对介子倍增振幅的贡献并发现它为零。然而，我们忽略了 $\omega_{k_1} = \omega_{k_2} + \omega_{k_3}$ 处的极点贡献。更准确地说，我们在分母中令 $k_1$ 为 $k_0$，尽管通常它们的差异在 $O(1/\sigma)$ 阶。这种近似是合理的，而在极点的大小为 $1/\sigma$ 的邻域内则例外。在极限 $\sigma \to \infty$ 下这样的近似是有效的，而在极点的无穷小邻域内则例外。因此，我们期待引入的误差仅取决于无穷小邻域中的被积函数，特别是仅取决于极点的留数。

在这一极点能量守恒，因此它对介子倍增的贡献是在壳的。让我们重写式 (2-117) 来看它对振幅的贡献，现在我们保留来自极点的贡献

$$|t\rangle \supset \frac{\sqrt{\lambda}\mathcal{B}_{k_0}e^{-i\omega_{k_0}t}}{4}\int\frac{dk_2}{2\pi}\int\frac{dk_3}{2\pi}\int dx V^{(3)}(gf(x))\mathfrak{g}_{k_2}(x)\mathfrak{g}_{k_3}(x)$$
$$\times 2\sigma\sqrt{\pi}\left[\int\frac{dk_1}{2\pi}\frac{e^{-\sigma^2(k_1-k_0)^2}e^{-i(k_1-k_0)x_t}}{\omega_{k_1}-\omega_{k_2}-\omega_{k_3}}\frac{\mathfrak{g}_{-k_1}(x)}{\omega_{k_1}}\right]|k_2 k_3\rangle_0. \quad (2\text{-}127)$$

正如我们所写下的，积分在极点处没有定义。现在，让我们使用积分主值来定义它。

让我们通过 $k_1 = k_I$ 定义极点的位置，使得

$$\omega_{k_I} = \omega_{k_2} + \omega_{k_3}, \qquad k_I > 0. \tag{2-128}$$

在 $k_1 = -k_I$ 处还有另一个极点。我们可以使用 Sokhotski–Plemelj 定理计算极点处方括号中项的贡献。我们已经论证了远离极点处对振幅没有贡献，所以我们只需要考虑 $\pm i\pi$ 乘以每个极点的留数。在极点 $k_1 = -k_I$，留数包含因子 $e^{-\sigma^2(k_I+k_0)^2}$，它在极限 $\sigma \to \infty$ 下消失，因此我们不再需要考虑那个极点。

如果 $x_t > x$ 那么围道应该向下关闭从而得到 $-\pi i$ 乘以留数，否则应该向上关闭从而得到 $\pi i$ 乘以留数。注意，朴素地看，高斯项在这样的围道上发散。然而，在 $\sigma \to \infty$ 极限下，它只对极点的 $1/\sigma$ 邻域的积分贡献一个常数因子，因此可以在进行积分之前简单地将高斯中的 $k_1$ 取为其在极点处的值。这会影响实轴上的积分值，但正如我们已经论证过的，只有极点附近的积分才能对振幅产生影响。式 (2-127) 中方括号中的项因此变为

$$-\operatorname{sign}(x_t - x) \frac{i}{2k_I} e^{-\sigma^2(k_I - k_0)^2} e^{-i(k_I - k_0)x_t} \mathfrak{g}_{-k_I}(x). \tag{2-129}$$

不使用围道积分的另一种推导方法如下。在大 $|x|$ 处，方括号中的 $\mathfrak{g}_{-k_1}(x)$ 仅仅是 $e^{ik_1 x}$（相差一个常数相位相位 $\mathcal{B}_{-k_1}$ 或 $\mathcal{D}_{-k_1}$）。将其与 $e^{-i(k_1-k_0)x_t}$ 合并得到

$$e^{-i(k_1-k_0)x_t} e^{ik_1 x} = e^{-i(k_1-k_I)(x_t-x)} e^{-i(k_I-k_0)x_t} e^{ik_I x}. \tag{2-130}$$

右边的第三项连同相位 $\mathcal{B}_{-k_1}$ 或 $\mathcal{D}_{-k_1}$ 一起变成了式 (2-129) 中的 $\mathfrak{g}_{-k_I}(x)$。第二项同样出现在式 (2-129) 中。在大 $\sigma$ 的情况下，我们可以将方括号中项的分母 $(\omega_{k_1} - \omega_{k_I})\omega_{k_1}$ 以 $(k_1 - k_I)$ 线性展开，得到 $(k_1 - k_I)k_1$。这个分母对于 $(k_1 - k_I)$ 是奇的，因此只有式 (2-130) 右边第一项中的奇项有贡献。将其除以 $(k_1 - k_I)k_1$ 可以得到原生 delta 函数

$$\lim |x_t - x| \to \infty - \frac{\sin\left[(k_1 - k_I)(x_t - x)\right]}{(k_1 - k_I)k_1} = -\pi \operatorname{sign}(x_t - x) \frac{\delta(k_1 - k_I)}{k_1} \tag{2-131}$$

它可用于在式 (2-127) 中的方括号中计算 $k_1$ 积分，再次得到 (2-129)。

这个结果却不是我们期望看到的。它关于时间对称（相差一个符号），因此在遥远的过去和遥远的未来观察到两个介子的概率是相同的。

让我们通过改变我们对 $\gamma_{1\mathfrak{K}}^{02}$ 中的极点解释来破缺时间反演对称性。现在我们不用积分主值的定义方法，我们来尝试

$$\gamma_{1\mathfrak{K}}^{02}(k_1, k_2) = \frac{2\pi\delta(k_2 - \mathfrak{K})}{2}\left(-\Delta_{k_1 B} - \sqrt{Q_0}\lambda\frac{V_{Ik_1}}{\omega_{k_1}}\right) + \frac{\sqrt{Q_0}\lambda V_{-\mathfrak{K}k_1k_2}}{4\omega_{\mathfrak{K}}\left(\omega_{\mathfrak{K}} - \omega_{k_1} - \omega_{k_2} + i\epsilon\right)}. \tag{2-132}$$

在下一小节中，我们将解释为什么这样的平移会导致另一个具有相同能量的哈密顿本征态，因而是被允许的。

现在，复 $k_1$ 平面上的极点处于无穷小的负虚数值。因此，如果 $x_t < x$ 则极点不包含在围道中。现在方括号中的项变为

$$-\Theta(x_t - x)\frac{i}{k_I}e^{-\sigma^2(k_I - k_0)^2}e^{-i(k_I - k_0)x_t}\mathfrak{g}_{-k_I}(x), \tag{2-133}$$

其中 $\Theta$ 是 Heaviside 阶跃函数。它对 $|t\rangle$ 的相应贡献是

$$|t\rangle \supset -\frac{\sqrt{\lambda}\mathcal{B}_{k_0}e^{-i\omega_{k_0}t}}{4}\int\frac{dk_2}{2\pi}\int\frac{dk_3}{2\pi}\int_{-\infty}^{x_t}dx V^{(3)}(gf(x))\mathfrak{g}_{k_2}(x)\mathfrak{g}_{k_3}(x)$$
$$\times 2\sigma\sqrt{\pi}\left[\frac{i}{k_I}e^{-\sigma^2(k_I - k_0)^2}e^{-i(k_I - k_0)x_t}\mathfrak{g}_{-k_I}(x)\right]|k_2 k_3\rangle_0. \tag{2-134}$$

现在如果 $x_t \ll 0$，使得介子波包还没有到达扭结，那么 $x$-积分将只覆盖到渐近区域，在这个渐近区域内 $\mathfrak{g}_{k_2}(x)\mathfrak{g}_{k_3}(x)\mathfrak{g}_{-k_I}(x) \sim e^{ix(k_I - k_2 - k_3)}$ 快速振荡，指数地压低振幅。另一方面，在碰撞之后 $x_t \gg 0$，积分等于 $V_{-k_I k_2 k_3}$（相差一个指数压低的修正）。态则是

$$|t\rangle \supset -\Theta(x_t)\frac{i\sigma\sqrt{\pi\lambda}\mathcal{B}_{k_0}e^{-i\omega_{k_0}t}}{2}\int\frac{dk_2}{2\pi}\int\frac{dk_3}{2\pi}e^{-\sigma^2(k_I - k_0)^2}e^{-i(k_I - k_0)x_t}\frac{V_{-k_I k_2 k_3}}{k_I}|k_2 k_3\rangle_0. \tag{2-135}$$

使用式 (2-71) 计算分母，这导致约化矩阵元

$$\frac{{}_{\text{vac}}\langle k_2 k_3|t\rangle_{\text{red}}}{\langle 0|0\rangle_{\text{red}}} \supset -\Theta(x_t)\frac{i\sigma\sqrt{\pi\lambda}\mathcal{B}_{k_0}e^{-i\omega_{k_0}t}}{4\omega_{k_2}\omega_{k_3}k_I}e^{-\sigma^2(k_I - k_0)^2}e^{-i(k_I - k_0)x_t}V_{-k_I k_2 k_3} \tag{2-136}$$

加上 $O(\lambda^{3/2})$ 阶修正，与第 3 章中的振幅 (3-56) 一致。然而，这里我们考虑了平移不变的初态和末态。因此，我们得出结论，可以保证平移不变性的对初态和末态的高阶修正，在 $O(\sqrt{\lambda})$ 阶并不会影响介子倍增振幅。由于计算过程十分类似，这个结果也可以直接推广到斯托克斯散射和反斯托克斯散射情形，即式 (2-136) 和式 (3-56) 在领头阶的等同也表明了我们在第 4 章中的计算所得到的结果是正确的。

## 2.5.5 简并的本征态

我们将 $|k_1\rangle$ 定义为被 $P'$ 湮灭的 $H'$ 本征态，其领头阶项为 $|k_1\rangle_0$。这并不能完全表征这个态，因为还有其他具有相同能量的平移不变态。考虑任意使得 $\tilde{\omega}_{k_2} + \tilde{\omega}_{k_3} = \tilde{\omega}_{k_1}$ 的 $k_2$ 和 $k_3$。回想一下，直到 $O(\lambda)$ 阶修正（我们并不考虑到这一阶），这个条件是 $\omega_{k_2} + \omega_{k_3} = \omega_{k_1}$。然后态 $|k_2 k_3\rangle$ 与 $|k_1\rangle$ 具有相同的能量，并且根据构造，它也是平移不变的。

让我们平移对 $|k_1\rangle$ 的定义

$$|k_1\rangle \longrightarrow |k_1\rangle + c_{k_1 k_2 k_3} \sqrt{\lambda} |k_2 k_3\rangle, \qquad (2\text{-}137)$$

其中 $c$ 的阶数为 $O(\lambda^0)$ 且仅当 $\tilde{\omega}_{k_2} + \tilde{\omega}_{k_3} = \tilde{\omega}_{k_1}$ 时才是非零的。现在能量本征值在这个新项中匹配了，所以上面使用的论证不再有效（上面使用的论证表明了 $\gamma_{1k_1}^{m2}$ 中对 $|k_1\rangle$ 的贡献对振幅没有贡献）。

$|k_1\rangle$ 的新选择同样满足我们的定义。然而，它与旧选择的不同之处在于 $\gamma_1^{20}(k_2, k_3)$ 的变化。事实上，$\gamma_1^{20}(k_2, k_3)$ 的任意值都对应于 $c_{k_1 k_2 k_3}$ 的某个选择，只需它当 $\omega_{k_1} \neq \omega_{k_2} + \omega_{k_3}$ 时与式 (2-79) 中的旧值一致。直觉上，人们可能只会添加与 $\delta(\omega_{k_1} - \omega_{k_2} - \omega_{k_3})$ 成正比的项，式 (2-132) 中极点的无穷小平移正是如此。

因此，我们认为第 3 章中的正确初始条件对应于式 (2-79) 且把其中的 $\gamma_1^{21}$ 替换为式 (2-132)。而在介子波包从另一侧撞向扭结并散射的情形，有 $x_0 > 0$ 和 $k_0 < 0$。在这种情况下，我们希望积分在 $x_t < x$ 处为零，以便 $k_1$ 围道在复平面底部闭合。这需要将极点平移 $+i\epsilon$。然而，由于 $k_0 < 0$，这仍然对应于 $\omega_{k_1}$ 的负虚部，因此仍然对应于同样的改动 (2-132)。我们提醒读者，该态与第 3 章中定义的态具有相同的能量、动量和 $O(\lambda^0)$ 阶项，但这里的态不会导致在初始波包 $|\Phi\rangle$ 中出现双介子分量。

## 2.6 本章小结

给定一个静态的扭结解，我们可以计算它的正规模甚至它们之间的相互作用 [56]。每年都会有新的模型被如此研究 [57–59]，包括最近的引力扭结方面的工作 [60, 61]。有了这些正规模，就可以构建与扭结相对应的量子态。最近人们甚至在非拓扑孤子的量子处理方面取得了进展 [62, 63]。然而，由于平移群的无限体积，每一种处理方法中都需要对待平移不变的扭结态是不可归一化的这一事实。

针对这个问题有许多解决方案被提出，每一种都是在某些情况下适用。而许多方案都有一个缺点即破坏了平移不变性，它们使得局域量不守恒，或者它们会

导致有限的位移。在本章中，我们提出了另一种处理此问题的方法，即用被平移群除的约化内积来替换内积。在我们看来，这是一种合理的方法，因为平移群的体积同时出现在可观测量的分子和分母中，因而相互抵消。我们已经发现这样能够使许多计算得到极大简化，因为我们能固定平移对称性，以使所有具有零模 $\phi_0$ 的项都为零。

我们选择的坐标 $y$ 使问题在一开始变得复杂，因为该坐标 $y$ 定义为 $\phi_0$ 的本征值。直观上，这可以理解如下。令 $f(x)$ 是一个经典扭结解。集体坐标的移动将 $f(x)$ 变换为 $f(x-x_0)$，因此它对位置是线性作用。另一方面，$y$ 上的平移将 $f(x)$ 变换为 $f(x) - y_0 f'(x_0)/\sqrt{Q_0}$。这并不对应于一个平移的扭结解，除非同时平移正规模来进行补偿。因此，由平移对称性对应的商带来的雅可比因子的非对角部分与 $\Delta_{Bk}$ 成正比，它是当平移 $x$ 时零模 $\mathfrak{g}_B(x)$ 和其他正规模之间的混合。然而，在 $y_0$ 较小时，这两个变换通过一个简单的比例因子 $\sqrt{Q_0}$ 相关联，这使我们能够轻松地定义匹配条件并计算必要的雅可比项。

文献 [18] 中研究了对单介子态 $|\mathfrak{K}\rangle$ 的领头阶修正，这在本章 (2-79) 中进行了总结。然而，如果 $\omega_\mathfrak{K} \geq 2m$ 则这个态与某些两介子态具有相同的能量和动量。该态始终含有离壳两介子态的云。在 (2-79) 中，简并态是在壳的。研究两介子态产生的物理上正确的初始条件，是从不包含具有两个在壳介子的态开始。在式 (2-132) 我们展示了一个这样的态，它和文献 [18] 中是一样的，含有由于简并两介子态产生的次领头阶贡献。这个次领头阶贡献是通过包含了极点的无穷小虚位移来加入的。更普遍地，我们猜想态中的这些极点代表了简并态的在壳贡献，这些贡献可以而且通常应该通过这样的虚位移来消除。换句话说，我们猜想用于计算此类极点的处理方法通常对应于简并的本征空间中的哈密顿本征态的物理选择。

# 第 3 章　介子倍增

## 3.1　引言

使用文献 [64] 中引入的经典辐射运动方程的微扰方法，文献 [11, 12] 研究了扭结上的入射辐射。结果发现，如果扭结是无反射的，并且辐射是频率为 $\omega$ 的单色辐射，则一些透射辐射的频率将为 $2\omega$，这种倍频会对扭结施加负压。在量子化的模型中这很容易理解，它表示过程扭结 +2 个介子 → 扭结 +1 个介子。人们可以证明介子之间的能量守恒（其在领头阶是精确的）意味着末态介子比两个融合的介子具有更大的动量，其差异导致扭结的负反冲。这（包括更高阶的介子的介子融合）是经典无反射扭结情况下唯一允许的过程。在反射性扭结的情况下，文献 [13] 发现也存在着介子反射，其对压力产生正贡献。

与经典体系相比，介子-扭结相互作用在量子体系中具有非常不同的特征，前者导致负压，而后者导致正压。在某种程度上，这并不奇怪，因为由 $N$ 个介子组成的初态将产生与 $N$ 成比例的介子倍增事例数，而介子融合的概率将是 $O(N^2)$ 阶。因此，对于足够强的介子源，我们期待介子融合现象成为主导。

本章中，我们首先在第 3.2 节中处理了对初态和末态的量子修正，这是要求它们在远离扭结时既正常地运动又不发生演化所必需的，我们发现，在我们所计算到的这一阶，这些修正对介子倍增的概率没有贡献。接着在第 3.3 节计算了一般性的 1+1 维标量场论中介子倍增的概率。在第 3.4 节中，我们将所得结果应用于两个无反射扭结：Sine-Gordon 孤子和 $\phi^4$ 扭结。当然，由于可积性，介子倍增现象不会发生在 Sine-Gordon 的情况下（这可以作为一致性检验）。在第 3.5 节中，我们给出了 $\phi^4$ 模型中与介子倍增相关的各种概率的数值结果，例如概率密度和反冲概率。

## 3.2　初态和末态

在这一节我们尝试和第 2 章第 2.5 节不同的另一套方法，即从真空哈密顿量的角度来尝试理解对初态和末态的选择。

### 3.2.1　振幅修正

我们在之后计算的振幅 (3-56) 是 $O(\sqrt{\lambda})$ 阶的结果。它由初态波函数、$H_3'$ 中的一项和末态波函数的乘积产生，它们分别是 $O(\lambda^0)$ 阶、$O(\sqrt{\lambda})$ 阶和 $O(\lambda^0)$ 阶。然而，我们需要考虑来自初态和末态的 $O(\sqrt{\lambda})$ 阶修正，它们带来的修正和结果

是同阶的（为了产生同阶的修正我们需要使用 $O(\lambda^0)$ 阶的自由哈密顿量 $H'_2$ 来演化）。

自由哈密顿量使介子量子数守恒，因此这些贡献来自对包含两介子的初态的量子修正和对包含单介子的末态的量子修正。本章之后的主要计算部分将此类修正简单地设为了零（即没有考虑这些修正的贡献）。虽然这样的初始条件是允许的，并且可以强制定义这样的两介子态，但这样做都太不自然，因为它们并不是与该问题中的任何机制相关的相互作用的哈密顿量的本征态。事实上，如果 $O(\sqrt{\lambda})$ 阶修正不包含在初态中，那么它们将动态地生成，并随着波包传播而振荡。然而，这种次领头阶的振荡并不影响波包在远离扭结处不加速的结论，在我们计算到的阶也不会影响介子倍增的概率。

我们已经在第 2 章证明了这一点。在本节中，我们将做出对初态和末态的量子修正更有物理动机的另一个选择（从真空哈密顿量的角度），并证明它直到我们关心的阶数，不会影响本章对介子倍增振幅计算的有效性，以及可以直接推广到第 4 章对斯托克斯散射和反斯托克斯散射的振幅计算的有效性，从而为我们形式上简化了很多的计算提供了坚实的有效性保障。

### 3.2.2 构建初态

初态是一个单介子波包，它从左侧朝向扭结运动。在扭结的左侧，经典扭结解 $f(x)$ 接近势 $V(\sqrt{\lambda}f(x))$ 的最小值 $f_L = f(-\infty)$。

#### 3.2.2.1 对构建的概述

在 $t < 0$ 时，介子一直处于左侧真空，并且一直都不会靠近扭结。因此，我们希望在 $t = 0$ 时刻构造一个初态，使得介子波包是左真空哈密顿量 $H_L$（而不是全扭结哈密顿量 $H'$）的本征态的近单色叠加

$$H_L = \mathcal{D}_L^\dagger H \mathcal{D}_L, \qquad \mathcal{D}_L = \mathrm{Exp}\left[-i \int dx f_L \pi(x)\right]. \tag{3-1}$$

左真空演化算符 $e^{-iH_L t}$ 作用于我们从 $H_L$ 本征态构造的介子波包的方式是严格地平移这个介子波包，并不会有加速或者由于寻常的色散导致的形变。由于两个哈密顿量 $H_L$ 和 $H'$ 对远离扭结的左侧的介子作用相同，因此在介子靠近扭结之前，真正的演化算符 $e^{-iH't}$ 同样是通过严格的平移作用于介子波包，并没有加速或变形。因此，这个构造将为我们的散射问题定义一个合适的单介子单扭结渐近态。由于这个波包不是扭结哈密顿量的本征态（其会随着时间演化），一旦介子波包到达扭结处，它就会非平庸地演化。

构建这样的态很简单。在文献 [28] 中，任意经典解 $f(x)$ 描述的空间中的单介子哈密顿量本征态被构造出来。虽然在大多数场景，$f(x)$ 取在扭结空间中，但推导实际上适用于任何空间中的任何静态经典解 $f(x)$。特别地，它同样适用于左真空解 $f(x) = f_L$ 或右真空解 $f(x) = f_R$。人们可以重复文献 [28] 的所有论证，只需将 $f(x)$ 替换为 $f_L$ 即可在左真空标架中获得左真空中的单介子态。主动变换 $\mathcal{D}_f \mathcal{D}_L^\dagger$ 在原点加入了一个扭结，这样就产生了单介子单扭结态（同时它们是在左真空标架中的态）。然后执行被动变换 $\mathcal{D}_L \mathcal{D}_f^\dagger$，被动变换使得态不变，但将希尔伯特空间的标架从左真空标架更改为扭结标架。将这两个变换放在一起我们发现，在没有变换的情况下，我们可以直接将左真空标架中如此构造的单介子态解释为扭结标架中的单介子单扭结态。当然，前者是左真空标架中演化算符 $H_L$ 的本征态，而后者并不是扭结标架演化算符 $H'$ 的本征态，这正是我们可以得到任意的动力学的原因。

我们构造的 $H_L$ 本征态与 $H'$ 本征态完全不同。然而，它们可以在 $x \ll 0$ 处组装成局域波包，并且只要它们在 $x \ll 0$ 处保持局域化，它们就会表现为自由粒子，这是因为在那里，左真空哈密顿量和扭结哈密顿量作用于它们之上是效果是一样的。实际上，在 $x \ll 0$ 处，扭结哈密顿量 $H'$ 和左真空哈密顿量之间的差异以 $m|x|$ 被指数压低。

要使其适合作为初始条件，这种波包需要具有三个特性。首先，它们是使用无扭结的哈密顿量 $H_L$ 定义的，这正如预期的那样，因为介子波包尚未与扭结发生相互作用。其次，正如我们将在下面看到的那样，在使用扭结哈密顿量 $H'$ 演化的过程中，它们在到达扭结之前进行严格的平移传播（速度恒定且无形变）。最后，在领头阶，它们是来自下面的 3.3 节中的波包 (3-19)。在该处，波包 (3-19) 仅在领头阶通过 $e^{-iH't}$ 进行严格平移，而在 $O(\sqrt{\lambda})$ 阶它在演化时发生形变。

#### 3.2.2.2  明确地构造

让我们明确地看一下领头阶修正。任意态 $|\psi\rangle$ 都可以用一些系数函数 $\gamma_\psi$ 展开为

$$|\psi\rangle = \sum_{mn} \phi_0^m \sum \int \frac{d^n k}{(2\pi)^n} \gamma_\psi^{mn}(k_1 \cdots k_n) |k_1 \cdots k_n\rangle_0. \tag{3-2}$$

然后，本章通篇使用的单扭结单介子态 $|\mathfrak{K}\rangle$ 中的领头阶项是

$$\gamma_\mathfrak{K}^{01}(k_1) = 2\pi \delta(k_1 - \mathfrak{K}). \tag{3-3}$$

在一般的空间，有许多阶修正。然而在真空空间，正规模为平面波。为简单起见，让我们考虑一个无反射性扭结，因此这些平面波可以和很左边的连续正规模

$\mathfrak{g}_k(x) = \mathcal{B}_k e^{-ikx}$ 等同,其中 $\mathcal{B}_k$ 是相因子。然后文献 [28] 中的所有修正除了下面两个以外都为零

$$\gamma_{\mathfrak{K}}^{02}(k_1, k_2) = \frac{\sqrt{\lambda}V^{(3)}(\sqrt{\lambda}f_L)\mathcal{B}_{k_1}\mathcal{B}_{k_2}\mathcal{B}_{-k_1-k_2}2\pi\delta(k_1 + k_2 - \mathfrak{K})}{4\omega_{\mathfrak{K}}\left(\omega_{\mathfrak{K}} - \omega_{k_1} - \omega_{k_2}\right)} \quad (3\text{-}4)$$

$$\gamma_{\mathfrak{K}}^{04}(k_1 \cdots k_4) = -\frac{\sqrt{\lambda}V^{(3)}(\sqrt{\lambda}f_L)\mathcal{B}_{k_1}\mathcal{B}_{k_2}\mathcal{B}_{k_3}2\pi\delta(k_1 + k_2 + k_3)}{6\sum_{j=1}^{3}\omega_{k_j}}2\pi\delta(k_4 - \mathfrak{K}).$$

这些对单介子态的贡献都来自三介子顶点。第一个出现在顶点将一个介子转换为两个的情况,第二个出现在它产生三个介子而只留下已经存在的介子的情况。

总而言之,我们建议在下面的初态 (3-19) 的构造中,将裸 $|k_1\rangle_0$ 替换为

$$|k_1\rangle_L = |k_1\rangle_0 + \frac{\sqrt{\lambda}V^{(3)}(\sqrt{\lambda}f_L)}{4\omega_{k_1}}\int\frac{dk_2}{2\pi}\frac{\mathcal{B}_{k_1-k_2}\mathcal{B}_{k_2}\mathcal{B}_{-k_1}|k_2, k_1 - k_2\rangle_0}{\omega_{k_1} - \omega_{k_2} - \omega_{k_1-k_2}} \quad (3\text{-}5)$$

$$-\frac{\sqrt{\lambda}V^{(3)}(\sqrt{\lambda}f_L)}{6}\int\frac{dk_2}{2\pi}\int\frac{dk_3}{2\pi}\frac{\mathcal{B}_{k_2}\mathcal{B}_{k_3}\mathcal{B}_{-k_2-k_3}|k_1, k_2, k_3, -k_2 - k_3\rangle_0}{\omega_{k_2} + \omega_{k_3} + \omega_{k_2+k_3}}.$$

这些是 $O(\sqrt{\lambda})$ 阶修正,它们是唯一与本章中处理的 $O(\sqrt{\lambda})$ 阶振幅相关的修正。在更高阶上,修正可以同样地像文献 [28] 中那样导出,其中 $f(x)$ 替换为 $f_L$。

### 3.2.3 初态的早期演化

我们已经提出了初态

$$|\Phi\rangle_L = \int\frac{dk_1}{2\pi}\alpha_{k_1}|k_1\rangle_L. \quad (3\text{-}6)$$

我们认为,在离碰撞时刻相当长的时间之前,这个形式的初态,与后面的 (3-19) 不同,是在全扭结哈密顿量下以一个恒定速度简单平移的演化,因此它适用对本章及第 4 章的过程作一个传统的散射解释。现在让我们证明情况确实如此。

为简洁起见,我们将忽略态的四介子部分,因为它可以通过在 $H_I$ 中包含 $H_3$ 中的项和三个 $B^{\ddagger}$ 算符来与双介子部分同样地处理。现在我们想要找到 $e^{-iH't}|\Phi\rangle_L$ 中的 $O(\sqrt{\lambda})$ 阶贡献。其中大部分贡献将在式 (3-27) 中得到。和那里

的结果相比,唯一的额外项来自对态的双介子修正的自由演化

$$
\begin{aligned}
&e^{-iH_{\text{free}}t}\left(|\Phi\rangle_L - |\Phi\rangle_0\right) \\
&= e^{-iH_{\text{free}}t}\int\frac{d^2k}{(2\pi)^2}\alpha_{k_1}\frac{\sqrt{\lambda}V^{(3)}(\sqrt{\lambda}f_L)}{4\omega_{k_1}}\frac{\mathcal{B}_{k_1-k_2}\mathcal{B}_{k_2}\mathcal{B}_{-k_1}|k_2,k_1-k_2\rangle_0}{\omega_{k_1}-\omega_{k_2}-\omega_{k_1-k_2}} \\
&= \frac{\sqrt{\lambda}V^{(3)}(\sqrt{\lambda}f_L)}{4}\int\frac{d^2k}{(2\pi)^2}\frac{\alpha_{k_1}}{\omega_{k_1}}\frac{e^{-it(\omega_{k_2}+\omega_{k_1-k_2})}\mathcal{B}_{k_1-k_2}\mathcal{B}_{k_2}\mathcal{B}_{-k_1}|k_2,k_1-k_2\rangle_0}{\omega_{k_1}-\omega_{k_2}-\omega_{k_1-k_2}}.
\end{aligned}
$$

(3-7)

这一项可以被写为

$$
\begin{aligned}
|\Phi(t)\rangle_L &= e^{-iH't}|\Phi\rangle_L \supset e^{-iH_{\text{free}}t}\left(|\Phi\rangle_L - |\Phi\rangle_0\right) = \int\frac{dk_1}{2\pi}e^{-i\omega_{k_1}t}\alpha_{k_1}\left(|k_1\rangle_L - |k_1\rangle_0\right) \\
&+ \frac{\sqrt{\lambda}V^{(3)}(\sqrt{\lambda}f_L)}{4}\int\frac{d^2k}{(2\pi)^2}\frac{\alpha_{k_1}}{\omega_{k_1}}\frac{\left(e^{-it(\omega_{k_2}+\omega_{k_1-k_2})}-e^{-i\omega_{k_1}t}\right)\mathcal{B}_{k_1-k_2}\mathcal{B}_{k_2}\mathcal{B}_{-k_1}|k_2,k_1-k_2\rangle_0}{\omega_{k_1}-\omega_{k_2}-\omega_{k_1-k_2}}.
\end{aligned}
$$

(3-8)

右边的第一项对应于无形变的刚性运动,来看第二项。要获得全演化 $|\Phi(t)\rangle_L$,还需要加上后文中式 (3-27) 中的贡献。将 (3-8) 的第二行修正加到 (3-27) 上,发现 (3-27) 以如下替换被修改

$$V_{k_1k_2k_3} \to V_{k_1k_2k_3} - V^{(3)}(\sqrt{\lambda}f_L)\mathcal{B}_{k_1}\mathcal{B}_{k_2}\mathcal{B}_{k_3}2\pi\delta(k_1+k_2+k_3). \quad (3\text{-}9)$$

三介子相互作用 $V_{k_1k_2k_3}$ 中的这种替换精确消除了左真空中在此阶存在的唯一相互作用对演化的贡献——动量守恒的三介子顶点。

我们将在第 3.3.2 节中论述,除了 $V_{-k_1k_2k_3}$ 中的 $\delta(-k_1+k_2+k_3)$ 项,在介子到达扭结之前振幅不会演化。我们将会发现 $k_1 = k_2 + k_3$ 处 (3-27) 的明显演化被更高阶修正 (3-5) 的演化 (3-7) 抵消到初始条件 (3-6)。只有 (3-8) 中的第一行没有被 (3-27) 抵消。因此,当具体到在 $x \ll 0$ 处有支撑的波函数时,修正后的态 $|k_1\rangle$ 在全扭结哈密顿演化算符 $e^{-iH't}$ 下演化为 $e^{-i\omega_{k_1}t}|k_1\rangle$,这和我们之前认为的一样。

这与我们上面提出的物理图像相一致。事实上,在 $x_0 \ll 0$ 处,单介子到双介子的过程只能发生在 $k_1 + k_2 + k_3 = 0$,因为扭结太远而无法与介子交换动量。因此介子系统本身在早期有着守恒的动量。

我们看到我们的初始波包 $|\Phi\rangle_L$ 在遥远的过去有着明确的定义和恒定的动量,并且在介子波包到达扭结之前(包括其领头阶量子修正)保持不变。每个 $k_1$

扭结-介子非弹性散射处的简单的相位旋转 (3-8) 与第 3.3.2 节中的动量中心为 $k_0$ 的波包的刚性运动相对应（相差一个寻常的传播效应）。特别是，尽管扭结的存在即使在无穷大的距离内也会影响介子的自相互作用，但这些相互作用是平移不变的。事实上，它们是在真空空间的相互作用。因此不会有远距离加速度，这意味着使用通常的散射矩阵是不被明确定义的 [65, 66]。在这种情况下，扭结可能会影响远处的介子，导致记忆效应 [67]，特别是态中的远距离信息 [68, 69]。

### 3.2.4 末态修正

我们在上面已经论证了，我们将用来在式 (3-19) 中构建初始波包的单介子态 $|k_1\rangle_0$ 并不是理想的选择，因为在介子到达扭结之前会产生 $O(\sqrt{\lambda})$ 阶量子修正。我们也找到了对初态进行量子修正的方法，以使得它在到达扭结之前不受干扰。我们将量子修正后的初态称为 $|k_1\rangle_L$。

概率由初始条件、哈密顿量和触发探测器的末态（末态由投影算符投影过去）决定。我们已经考虑了对前两个的量子修正。在本小节中，我们将考虑对投影算符进行量子修正。我们将表明在第 3.3 节中简单地考虑了未修正的投影算符 (3-60) 和其他的简化近似结合起来可以给出同阶的正确结果。更一般地，如果 $|\alpha\rangle$ 是希尔伯特空间子空间的正交基，则当投影算符

$$\mathcal{P} = \int d\alpha |\alpha\rangle\langle\alpha| \qquad (3-10)$$

夹在态和自身之间时，产生该态在由态 $|\alpha\rangle$ 张成的子空间中的概率。

这里 $\alpha$ 是可以触发探测器的末态的基 $|\alpha\rangle$ 的抽象指标。这些态需要满足什么性质？原则上，任何选择都对应于某个探测器，因此会产生明确定义的概率。然而，我们将通过对这些末态 $|\alpha\rangle$ 施加三个条件来定义介子倍增。首先，在领头阶它们应该由两介子态 $|k_2 k_3\rangle_0$ 组成。其次，在遥远的过去和遥远的未来，投影算符的行为应当独立于时间。在遥远的过去和遥远的未来，态由位于扭结左侧或右侧的波包描述。因此，投影算符应当由扭结两侧的与时间无关的态构成。换句话说，这些态应当是扭结每一侧真空空间的哈密顿量的两介子态。这些哈密顿量是用 (3-1) 定义的 $H_L$，以及用同样方式定义但在定义中把 $f_L$ 替换为 $f_R$ 的 $H_R$。

但是，态 $|\alpha\rangle$ 如何由两个不同的、非对易的哈密顿量 $H_L$ 和 $H_R$ 的本征态构成？人们可以用局域波包态的基构建投影算符，这些局域波包态在扭结的左侧和右侧分别是左右真空哈密顿量的本征态 $|k_1 k_2\rangle_L$ 和 $|k_1 k_2\rangle_R$ 的叠加。

最后，我们要求在实验开始时观察到双介子末态的概率必须等于 0。因此，我们需要选择量子修正使得投影算符湮灭我们的初态。

5858扭结-介子非弹性散射处的简单的相位旋转 (3-8) 与第 3.3.2 节中的动量中心为 $k_0$ 的波包的刚性运动相对应（相差一个寻常的传播效应）。特别是，尽管扭结的存在即使在无穷大的距离内也会影响介子的自相互作用，但这些相互作用是平移不变的。事实上，它们是在真空空间的相互作用。因此不会有远距离加速度，这意味着使用通常的散射矩阵是不被明确定义的 [65, 66]。在这种情况下，扭结可能会影响远处的介子，导致记忆效应 [67]，特别是态中的远距离信息 [68, 69]。

### 3.2.4 末态修正

我们在上面已经论证了，我们将用来在式 (3-19) 中构建初始波包的单介子态 $|k_1\rangle_0$ 并不是理想的选择，因为在介子到达扭结之前会产生 $O(\sqrt{\lambda})$ 阶量子修正。我们也找到了对初态进行量子修正的方法，以使得它在到达扭结之前不受干扰。我们将量子修正后的初态称为 $|k_1\rangle_L$。

概率由初始条件、哈密顿量和触发探测器的末态（末态由投影算符投影过去）决定。我们已经考虑了对前两个的量子修正。在本小节中，我们将考虑对投影算符进行量子修正。我们将表明在第 3.3 节中简单地考虑了未修正的投影算符 (3-60) 和其他的简化近似结合起来可以给出同阶的正确结果。更一般地，如果 $|\alpha\rangle$ 是希尔伯特空间子空间的正交基，则当投影算符

$$\mathcal{P} = \int d\alpha |\alpha\rangle\langle\alpha| \qquad (3-10)$$

夹在态和自身之间时，产生该态在由态 $|\alpha\rangle$ 张成的子空间中的概率。

这里 $\alpha$ 是可以触发探测器的末态的基 $|\alpha\rangle$ 的抽象指标。这些态需要满足什么性质？原则上，任何选择都对应于某个探测器，因此会产生明确定义的概率。然而，我们将通过对这些末态 $|\alpha\rangle$ 施加三个条件来定义介子倍增。首先，在领头阶它们应该由两介子态 $|k_2 k_3\rangle_0$ 组成。其次，在遥远的过去和遥远的未来，投影算符的行为应当独立于时间。在遥远的过去和遥远的未来，态由位于扭结左侧或右侧的波包描述。因此，投影算符应当由扭结两侧的与时间无关的态构成。换句话说，这些态应当是扭结每一侧真空空间的哈密顿量的两介子态。这些哈密顿量是用 (3-1) 定义的 $H_L$，以及用同样方式定义但在定义中把 $f_L$ 替换为 $f_R$ 的 $H_R$。

但是，态 $|\alpha\rangle$ 如何由两个不同的、非对易的哈密顿量 $H_L$ 和 $H_R$ 的本征态构成？人们可以用局域波包态的基构建投影算符，这些局域波包态在扭结的左侧和右侧分别是左右真空哈密顿量的本征态 $|k_1 k_2\rangle_L$ 和 $|k_1 k_2\rangle_R$ 的叠加。

最后，我们要求在实验开始时观察到双介子末态的概率必须等于 0。因此，我们需要选择量子修正使得投影算符湮灭我们的初态。

58处的简单的相位旋转 (3-8) 与第 3.3.2 节中的动量中心为 $k_0$ 的波包的刚性运动相对应（相差一个寻常的传播效应）。特别是，尽管扭结的存在即使在无穷大的距离内也会影响介子的自相互作用，但这些相互作用是平移不变的。事实上，它们是在真空空间的相互作用。因此不会有远距离加速度，这意味着使用通常的散射矩阵是不被明确定义的 [65, 66]。在这种情况下，扭结可能会影响远处的介子，导致记忆效应 [67]，特别是态中的远距离信息 [68, 69]。

### 3.2.4 末态修正

我们在上面已经论证了，我们将用来在式 (3-19) 中构建初始波包的单介子态 $|k_1\rangle_0$ 并不是理想的选择，因为在介子到达扭结之前会产生 $O(\sqrt{\lambda})$ 阶量子修正。我们也找到了对初态进行量子修正的方法，以使得它在到达扭结之前不受干扰。我们将量子修正后的初态称为 $|k_1\rangle_L$。

概率由初始条件、哈密顿量和触发探测器的末态（末态由投影算符投影过去）决定。我们已经考虑了对前两个的量子修正。在本小节中，我们将考虑对投影算符进行量子修正。我们将表明在第 3.3 节中简单地考虑了未修正的投影算符 (3-60) 和其他的简化近似结合起来可以给出同阶的正确结果。更一般地，如果 $|\alpha\rangle$ 是希尔伯特空间子空间的正交基，则当投影算符

$$\mathcal{P} = \int d\alpha |\alpha\rangle\langle\alpha| \qquad (3\text{-}10)$$

夹在态和自身之间时，产生该态在由态 $|\alpha\rangle$ 张成的子空间中的概率。

这里 $\alpha$ 是可以触发探测器的末态的基 $|\alpha\rangle$ 的抽象指标。这些态需要满足什么性质？原则上，任何选择都对应于某个探测器，因此会产生明确定义的概率。然而，我们将通过对这些末态 $|\alpha\rangle$ 施加三个条件来定义介子倍增。首先，在领头阶它们应该由两介子态 $|k_2 k_3\rangle_0$ 组成。其次，在遥远的过去和遥远的未来，投影算符的行为应当独立于时间。在遥远的过去和遥远的未来，态由位于扭结左侧或右侧的波包描述。因此，投影算符应当由扭结两侧的与时间无关的态构成。换句话说，这些态应当是扭结每一侧真空空间的哈密顿量的两介子态。这些哈密顿量是用 (3-1) 定义的 $H_L$，以及用同样方式定义但在定义中把 $f_L$ 替换为 $f_R$ 的 $H_R$。

但是，态 $|\alpha\rangle$ 如何由两个不同的、非对易的哈密顿量 $H_L$ 和 $H_R$ 的本征态构成？人们可以用局域波包态的基构建投影算符，这些局域波包态在扭结的左侧和右侧分别是左右真空哈密顿量的本征态 $|k_1 k_2\rangle_L$ 和 $|k_1 k_2\rangle_R$ 的叠加。

最后，我们要求在实验开始时观察到双介子末态的概率必须等于 0。因此，我们需要选择量子修正使得投影算符湮灭我们的初态。

这并没有完全固定投影算符，也没有固定 $|\alpha\rangle$ 态。然而，由于我们只想找到 $|\alpha\rangle$ 中的 $O(\sqrt{\lambda})$ 阶部分，我们对矩阵元 $\langle\alpha|e^{-iH't}|k_1\rangle_L$ 的 $O(\sqrt{\lambda})$ 阶部分感兴趣，我们只需要考虑与 $|k_1\rangle_L$ 的 $O(\lambda^0)$ 阶的内积部分，即 $|k_1\rangle_0$。一般来说，需要注意此类论证中零模的贡献，但在第 2 章中，我们找到了此类内积的精确公式，并且在那里我们表明虽然对这种朴素计算的修正是非零的，但修正被 $O(\sqrt{\lambda})$ 阶的幂压低（尽管它们混合了介子数相差 1 的空间）。因此，这些修正将不会影响 $O(\sqrt{\lambda})$ 阶的振幅。

对于位于 $x \ll 0$ 的波包，在式 (3-5) 中，我们要求对 $|k_1\rangle_L$ 的领头阶修正具有一个特定形式。让我们定义另一组态 $|k_1\rangle_R$，它们具有类似的修正，但这次是对应于右侧的真空，其中 $f_R = f(\infty)$

$$|k_1\rangle_R = |k_1\rangle_0 + \frac{\sqrt{\lambda}V^{(3)}(\sqrt{\lambda}f_R)}{4\omega_{k_1}}\int\frac{dk_2}{2\pi}\frac{\mathcal{D}_{k_1-k_2}\mathcal{D}_{k_2}\mathcal{D}_{-k_1}|k_2,k_1-k_2\rangle_0}{\omega_{k_1}-\omega_{k_2}-\omega_{k_1-k_2}} \qquad (3\text{-}11)$$

$$-\frac{\sqrt{\lambda}V^{(3)}(\sqrt{\lambda}f_R)}{6}\int\frac{dk_2}{2\pi}\int\frac{dk_3}{2\pi}\frac{\mathcal{D}_{k_2}\mathcal{D}_{k_3}\mathcal{D}_{-k_2-k_3}|k_1,k_2,k_3,-k_2-k_3\rangle_0}{\omega_{k_2}+\omega_{k_3}+\omega_{k_2+k_3}}.$$

这里 $\mathcal{D}_k$ 是一个相位，它使得在 $x \gg 0$，$\mathfrak{g}_k(x) = \mathcal{D}_k e^{-ikx}$。

分别与扭结的左侧和右侧相关的修正 $|k_1\rangle_L - |k_1\rangle_0$ 和修正 $|k_1\rangle_R - |k_1\rangle_0$，与 $|k_2 k_3\rangle_0$ 的内积分别是

$$\frac{{}_0\langle k_2 k_3|\left(|k_1\rangle_L - |k_1\rangle_0\right)}{{}_0\langle 0|0\rangle_0} = \frac{\sqrt{\lambda}V^{(3)}(\sqrt{\lambda}f_L)\mathcal{B}_{k_2}\mathcal{B}_{k_3}\mathcal{B}_{-k_2-k_3}2\pi\delta(k_2+k_3-k_1)}{8\omega_{k_2}\omega_{k_3}\omega_{k_1}\left(\omega_{k_1}-\omega_{k_2}-\omega_{k_3}\right)}$$

$$\frac{{}_0\langle k_2 k_3|\left(|k_1\rangle_R - |k_1\rangle_0\right)}{{}_0\langle 0|0\rangle_0} = \frac{\sqrt{\lambda}V^{(3)}(\sqrt{\lambda}f_R)\mathcal{D}_{k_2}\mathcal{D}_{k_3}\mathcal{D}_{-k_2-k_3}2\pi\delta(k_2+k_3-k_1)}{8\omega_{k_2}\omega_{k_3}\omega_{k_1}\left(\omega_{k_1}-\omega_{k_2}-\omega_{k_3}\right)}.$$

$$(3\text{-}12)$$

要抵消它们，需要在 $|k_2 k_3\rangle_0$ 的修正中包括

$$|k_2 k_3\rangle_L = |k_2 k_3\rangle_0 + \frac{\sqrt{\lambda}V^{(3)}(\sqrt{\lambda}f_L)\mathcal{B}_{-k_2}\mathcal{B}_{-k_3}\mathcal{B}_{k_2+k_3}}{4\omega_{k_2}\omega_{k_3}(\omega_{k_2}+\omega_{k_3}-\omega_{k_2+k_3})}|k_2+k_3\rangle_0 \qquad (3\text{-}13)$$

$$|k_2 k_3\rangle_R = |k_2 k_3\rangle_0 + \frac{\sqrt{\lambda}V^{(3)}(\sqrt{\lambda}f_R)\mathcal{D}_{-k_2}\mathcal{D}_{-k_3}\mathcal{D}_{k_2+k_3}}{4\omega_{k_2}\omega_{k_3}(\omega_{k_2}+\omega_{k_3}-\omega_{k_2+k_3})}|k_2+k_3\rangle_0$$

其中我们使用了性质 $\mathcal{B}_k^* = \mathcal{B}_{-k}$ 和 $\mathcal{D}_k^* = \mathcal{D}_{-k}$。在 $n$-介子福克空间中对其它项的修正也是允许的，但以上是在这一阶仅有的非零的与 $|k_1\rangle_0$ 的内积修正，因此是仅有的对末态修正有贡献的项。

注意，投影算符 $\mathcal{P}$ 不是通过对所有 $|k_2 k_3\rangle_{LL}\langle k_2 k_3|$ 和 $|k_2 k_3\rangle_{RR}\langle k_2 k_3|$ 求和而构造，而是以一组局域波包基构造的，当局域在 $x \ll 0$ 时，由 $|k_2 k_3\rangle_L$ 构造，当局域在 $x \gg 0$ 时，由 $|k_2 k_3\rangle_R$ 构造。我们没有必要包括位于扭结附近的介子波包的态，因为这些态永远不会出现在渐近的过去或未来。实际操作中，这些具有局域态的两介子态的内积可以通过简单地插入 $|k_2 k_3\rangle_L$ 或 $|k_2 k_3\rangle_R$ 的公式 (3-13) 来获得（取决于态的局域化的位置），其不精确程度被指数压低。

### 3.2.5 对振幅的修正

在第 3.3 节中我们将要计算振幅

$$\frac{{}_0\langle k_2 k_3 | e^{-iH't} | k_1\rangle_0}{{}_0\langle 0|0\rangle_0}. \tag{3-14}$$

我们现在对出现在

$$\frac{{}_L\langle k_2 k_3 | e^{-iH_{\text{free}}t} | k_1\rangle_L}{{}_0\langle 0|0\rangle_0} \quad \text{and} \quad \frac{{}_R\langle k_2 k_3 | e^{-iH_{\text{free}}t} | k_1\rangle_R}{{}_0\langle 0|0\rangle_0} \tag{3-15}$$

中的修正感兴趣。在时刻 $t$ 的概率的初态和末态修正是根据位于位置 $x_0 + k_0 t/\omega_{k_0}$ 附近的波包的矩阵元计算的。因此，(3-15) 中的第一项和较早时刻 $t \ll -x_0 \omega_{k_0}/k_0$ 相关，而第二项和较晚时刻 $t \gg -x_0 \omega_{k_0}/k_0$ 相关。

将以上结果组合起来，对 (3-15) 中第一个表达式的相应初态和末态的修正分别为

$$\frac{\sqrt{\lambda} V^{(3)}(\sqrt{\lambda} f_L) \mathcal{B}_{k_2} \mathcal{B}_{k_3} \mathcal{B}_{-k_2-k_3}}{4\omega_{k_2}\omega_{k_3}(\omega_{k_2}+\omega_{k_3}-\omega_{k_2+k_3})} {}_0\langle k_2 + k_3 | e^{-iH_{\text{free}}t}|k_1\rangle_0$$

$$= \frac{\sqrt{\lambda} V^{(3)}(\sqrt{\lambda} f_L) e^{(-i\omega_{k_2+k_3}t)} \mathcal{B}_{k_2} \mathcal{B}_{k_3} \mathcal{B}_{-k_1} 2\pi\delta(k_2+k_3-k_1)}{8\omega_{k_2}\omega_{k_3}\omega_{k_2+k_3}(\omega_{k_2}+\omega_{k_3}-\omega_{k_2+k_3})}$$

$${}_0\langle k_2 k_3|e^{(-iH_{\text{free}}t)}\frac{\sqrt{\lambda}V^{(3)}(\sqrt{\lambda}f_L)\mathcal{B}_{k_1-k'}\mathcal{B}_{k'}\mathcal{B}_{-k_1}}{4\omega_{k_1}}\int\frac{dk'}{2\pi}\frac{|k',k_1-k'\rangle_0}{\omega_{k_1}-\omega_{k'}-\omega_{k_1-k'}}$$

$$= -\frac{\sqrt{\lambda}V^{(3)}(\sqrt{\lambda}f_L)e^{(-i\omega_{k_2+k_3}t)}\mathcal{B}_{k_2}\mathcal{B}_{k_3}\mathcal{B}_{-k_1}2\pi\delta(k_2+k_3-k_1)}{8\omega_{k_2}\omega_{k_3}\omega_{k_2+k_3}(\omega_{k_2}+\omega_{k_3}-\omega_{k_2+k_3})}.$$

$$\tag{3-16}$$

人们可能会观察到这两个修正恰好抵消，因此碰撞前的介子倍增概率不受初态和末态修正的影响。对于 (3-15) 中的第二项，使用在扭结右侧有效的矩阵元进行类似的计算，因此碰撞后的介子倍增概率也不受初态和末态修正的影响。粗略地

说，我们已经证明 (3-15) 和 (3-14) 在 $O(\sqrt{\lambda})$ 阶上是相等的。我们得出结论，下文使用的绝热近似 (3-19) 在领头阶可以导出正确的介子倍增振幅。

这个结果是显然的。远离扭结处，介子之间保持动量和能量守恒，因此介子倍增在运动学上是被禁止的。另一方面，初态和末态修正分别在介子与扭结相互作用之前或之后，分别由介子倍增和介子融合产生。因此我们将在本章的后续部分以及第 4 章直接使用初末态的领头阶以及朴素的内积（出现分子和分母上的内积的无穷大相互抵消），本节以及第 2 章的论述表明了在我们所关心的领头阶，后续计算得到的结果是可靠的（而考虑更高阶修正时，则不可以使用后续的简化计算方法）。

## 3.3 介子倍增

### 3.3.1 高斯波包

我们的初始条件是以 $x_0$ 为中心的介子波包

$$\Phi(x) = \mathrm{Exp}\left[-\frac{(x-x_0)^2}{4\sigma^2} + ixk_0\right], \quad x_0 \ll -\frac{1}{m}, \quad \frac{1}{k_0}, \frac{1}{m} \ll \sigma \ll |x_0|. \quad (3\text{-}17)$$

对 $x_0$ 和 $|x_0|$ 范围的限定确保了起始于 $x = x_0$ 的初始波包不会与以 $x = 0$ 为中心的扭结重叠。$\sigma$ 的下限确保介子的动量在峰值足够大，以使所有分量都向扭结移动，并且我们可以近似认为波包是单色的。

经过傅里叶变换后，波包的演化会更简单

$$\Phi(x) = \int \frac{dk}{2\pi} \alpha_k \mathfrak{g}_k^*(x), \quad \alpha_k = \int dx \Phi(x) \mathfrak{g}_k(x). \quad (3\text{-}18)$$

这种变换与平面波无关（平面波是真空空间中自由运动方程的解），而是与正规模有关，它们是单扭结空间中的解。形模和零模不需要包含在该变换中，因为它们在 $|x|$ 处是 $O(1/m)$ 阶，在那里 $\Phi(x)$ 小到可以忽略不计。

初始的单扭结单介子态 $|\Phi\rangle_0$ 可以在扭结标架中根据自由扭结基态 $|0\rangle_0$ 构造为

$$|\Phi\rangle_0 = \int dx \Phi(x) |x\rangle_0 = \int \frac{dk}{2\pi} \alpha_k |k\rangle_0, \quad |k\rangle_0 = B_k^\ddagger |0\rangle_0, \quad |x\rangle_0 = \int \frac{dk}{2\pi} \mathfrak{g}_k(x) |k\rangle_0. \quad (3\text{-}19)$$

(3-19) 是我们对初态的一种选择。在本节中，我们的策略将是简单地假设这个初始条件并计算末态在希尔伯特空间的某个类似任意定义的子空间中的概率，这是有良好定义的。然而，初态和末态的选择也可能与介子倍增的概率有关。特

别是对这些初态和末态的量子修正和我们将要计算的振幅同阶的情况下。这就是我们为何首先在第 3.2 节中讨论了初态和末态修正，并论证了在这里的我们对初态和末态的选择在我们所计算到的阶并不会影响结果。

### 3.3.2 时间演化

扭结标架中的相互作用由式 (1-7) 中的哈密顿项给出。不同的哈密顿项对应着不同的 $\sqrt{\lambda}$ 的幂级数。在 $O(\lambda^0)$ 阶，只有 $H_{\text{free}}$ 对演化有贡献

$$|\Phi(t)\rangle_0|_{O(\lambda^0)} = e^{-iH_{\text{free}}t}|\Phi\rangle_0 = \int \frac{dk}{2\pi}\alpha_k e^{-i\omega_k t}|k\rangle_0 = \int dx \int \frac{dk}{2\pi}\alpha_k e^{-i\omega_k t}\mathfrak{g}_{-k}(x)|x\rangle_0. \tag{3-20}$$

其中系数

$$\Phi(x,t) = \int \frac{dk}{2\pi}\alpha_k e^{-i\omega_k t}\mathfrak{g}_{-k}(x) = \int dy\,\Phi(y)\int \frac{dk}{2\pi}\mathfrak{g}_k(y)e^{-i\omega_k t}\mathfrak{g}_{-k}(x) \tag{3-21}$$

在这一阶是介子波包的构型，我们可以用传播子 $G$ 来表示它

$$\Phi(x,t) = \int dy\,\Phi(y)G(x,y,t), \qquad G(x,y,t) = \int \frac{dk}{2\pi}\mathfrak{g}_k(y)e^{-i\omega_k t}\mathfrak{g}_{-k}(x). \tag{3-22}$$

为明确起见，我们来考虑一个无反射性的扭结。然后我们将在下面看到在 $x \ll -1/m$ 处

$$\alpha_k \mathfrak{g}_{-k}(x) = 2\sigma\sqrt{\pi}e^{-\sigma^2(k-k_0)^2}e^{ik_0 x + i(k-k_0)(x-x_0)}. \tag{3-23}$$

使用 $\omega_k$ 在 $k \sim k_0$ 处的线性展开（见式 (3-38)），我们得到

$$\begin{aligned}\Phi(x,t) &= 2\sigma\sqrt{\pi}e^{ik_0 x - i\omega_{k_0}t}\int \frac{dk}{2\pi}e^{-i(k-k_0)\frac{k_0 t}{\omega_{k_0}}}e^{-\sigma^2(k-k_0)^2}e^{i(k-k_0)(x-x_0)}\\ &= e^{ik_0 x - i\omega_{k_0}t}\text{Exp}\left[-\frac{1}{4\sigma^2}\left(x-x_0-\frac{k_0 t}{\omega_{k_0}}\right)^2\right].\end{aligned} \tag{3-24}$$

由此我们看出 $x_0 + k_0 t/\omega_{k_0}$ 是 $t$ 时刻局域波包的领头阶部分的位置。特别地，在接近扭结之前，介子波包以 $k_0/\omega_{k_0}$ 的恒定速度移动，并不会加速。

在下一阶 $O(\sqrt{\lambda})$ 中，唯一对介子倍增有贡献的项是 [a]

$$\begin{aligned}H_I &= \frac{\sqrt{\lambda}}{4}\int \frac{dk_1}{2\pi}\frac{dk_2}{2\pi}\frac{dk_3}{2\pi}V_{-k_1 k_2 k_3}\frac{1}{\omega_{k_1}}B^{\ddagger}_{k_2}B^{\ddagger}_{k_3}B_{k_1} \tag{3-25}\\ V_{-k_1 k_2 k_3} &= \int dx V^{(3)}(\sqrt{\lambda}f(x))\mathfrak{g}_{-k_1}(x)\mathfrak{g}_{k_2}(x)\mathfrak{g}_{k_3}(x).\end{aligned}$$

---

[a] 在这里，我们根据 (1-7) 和 (1-11) 中的定义交换了 $k$ 和 $x$ 积分的顺序。这些积分实际上并不是可交换的，因此 $V_{-k_1 k_2 k_3}$ 看似是不可积函数的积分。我们需要记住，要使这个积分有意义，首先需要对 $k$ 积分。而事实证明，这等效于首先使用将在式 (3-75) 中定义的积分主值来对 $x$ 积分。

$H_I$ 将单介子态转换为双介子态

$$H_I|k_1\rangle_0 = \frac{\sqrt{\lambda}}{4\omega_{k_1}} \int \frac{dk_2}{2\pi}\frac{dk_3}{2\pi} V_{-k_1 k_2 k_3} |k_2 k_3\rangle_0. \qquad (3\text{-}26)$$

在 $t$ 时刻，在 $O(\sqrt{\lambda})$ 阶，波包演化为

$|\Phi(t)\rangle_0$

$= e^{-i(H_{\text{free}}+H_I)t}|_{O(\sqrt{\lambda})}|\Phi\rangle_0$

$= \sum_{n=1}^{\infty} \frac{(-it)^n}{n!} (H_{\text{free}}+H_I)^n|_{O(\sqrt{\lambda})}|\Phi\rangle_0 = \sum_{n=1}^{\infty} \frac{(-it)^n}{n!} \sum_{m=0}^{n-1} H_{\text{free}}^m H_I H_{\text{free}}^{n-m-1} |\Phi\rangle_0$

$= \int \frac{dk_1}{2\pi}\frac{dk_2}{2\pi}\frac{dk_3}{2\pi} \frac{\sqrt{\lambda}}{4} \alpha_{k_1} V_{-k_1 k_2 k_3} \sum_{n=1}^{\infty} \frac{(-it)^n}{n!} \sum_{m=0}^{n-1} (\omega_{k_2}+\omega_{k_3})^m \omega_{k_1}^{n-m-2} |k_2 k_3\rangle_0$

$= -\frac{i\sqrt{\lambda}}{4} \int \frac{dk_1}{2\pi}\frac{dk_2}{2\pi}\frac{dk_3}{2\pi} \frac{\alpha_{k_1}}{\omega_{k_1}} V_{-k_1 k_2 k_3} \text{Exp}\left[-i\frac{\omega_{k_1}+\omega_{k_2}+\omega_{k_3}}{2}t\right] \frac{\sin\left(\frac{\omega_{k_2}+\omega_{k_3}-\omega_{k_1}}{2}t\right)}{(\omega_{k_2}+\omega_{k_3}-\omega_{k_1})/2} |k_2 k_3\rangle_0.$

$(3\text{-}27)$

这里我们丢掉了 $O(\lambda^0)$ 项，因为它对之后的矩阵元没有贡献。

我们可以通过下式定义对应于单扭结双介子态 (1-17) 的狄拉克左矢

$$_0\langle k_2 k_3| = \left(B_{k_2}^{\ddagger} B_{k_3}^{\ddagger}|0\rangle_0\right)^{\dagger} = {}_0\langle 0|\frac{B_{k_2}}{2\omega_{k_2}}\frac{B_{k_3}}{2\omega_{k_3}}. \qquad (3\text{-}28)$$

这导致了如下归一化[b]

$$_0\langle k_2 k_3|k_2' k_3'\rangle_0 = \frac{{}_0\langle 0|0\rangle_0}{4\omega_{k_2}\omega_{k_3}} \left(2\pi\delta(k_2'-k_2) 2\pi\delta(k_3'-k_3) + 2\pi\delta(k_2'-k_3) 2\pi\delta(k_3'-k_2)\right).$$

$(3\text{-}29)$

然后我们得到计算非归一化介子倍增振幅的主要公式

$$\frac{{}_0\langle k_2 k_3|\Phi(t)\rangle_0}{{}_0\langle 0|0\rangle_0} = -\frac{i\sqrt{\lambda}}{8\omega_{k_2}\omega_{k_3}} \int \frac{dk_1}{2\pi} \frac{\alpha_{k_1}}{\omega_{k_1}} V_{-k_1 k_2 k_3} \text{Exp}\left[-i\frac{\omega_{k_1}+\omega_{k_2}+\omega_{k_3}}{2}t\right] \frac{\sin\left(\frac{\omega_{k_2}+\omega_{k_3}-\omega_{k_1}}{2}t\right)}{(\omega_{k_2}+\omega_{k_3}-\omega_{k_1})/2}.$$
$(3\text{-}30)$

---

[b]矩阵元 ${}_0\langle k_2 k_3|k_2' k_3'\rangle_0$ 和 ${}_0\langle 0|0\rangle_0$ 都是无限的，但只有它们的比值会出现在介子倍增的概率上。我们已经表明了，在领头阶，该比值与式 (3-29) 中的朴素计算一致。

### 3.3.3 有限时间演化的振幅

把振幅写成

$$_0\langle k_2 k_3|\Phi(t)\rangle_0 = \frac{\sqrt{\lambda}}{8\omega_{k_2}\omega_{k_3}}\int \frac{dk_1}{2\pi}\frac{\alpha_{k_1}}{\omega_{k_1}}V_{-k_1 k_2 k_3}\frac{e^{-i(\omega_{k_2}+\omega_{k_3})t}-e^{-i\omega_{k_1}t}}{(\omega_{k_2}+\omega_{k_3}-\omega_{k_1})}\,_0\langle 0|0\rangle_0. \quad (3\text{-}31)$$

我们可以不去管总的相因子和常数

$$A_{k_2 k_3}(t) = \frac{e^{i(\omega_{k_2}+\omega_{k_3})t}}{_0\langle 0|0\rangle_0}\,_0\langle k_2 k_3|\Phi(t)\rangle_0 = \frac{\sqrt{\lambda}}{8\omega_{k_2}\omega_{k_3}}\int \frac{dk_1}{2\pi}\frac{\alpha_{k_1}}{\omega_{k_1}}V_{-k_1 k_2 k_3}\frac{1-e^{i(\omega_{k_2}+\omega_{k_3}-\omega_{k_1})t}}{(\omega_{k_2}+\omega_{k_3}-\omega_{k_1})}.$$
$$(3\text{-}32)$$

在 $t=0$ 时刻，式 (3-30) 分子中的正弦项为零，矩阵元也随之为零。对时间取导数我们得到

$$\dot{A}_{k_2 k_3}(t) = -i\frac{\sqrt{\lambda}}{8\omega_{k_2}\omega_{k_3}}\int \frac{dk_1}{2\pi}\frac{\alpha_{k_1}}{\omega_{k_1}}V_{-k_1 k_2 k_3}e^{i(\omega_{k_2}+\omega_{k_3}-\omega_{k_1})t}. \quad (3\text{-}33)$$

这可以通过一些良好的近似来简化。

#### 3.3.3.1 无反射性扭结

首先，由于 $|x_0|\gg\sigma$ 和 $|x_0|\gg 1/m$，$\alpha_{k_1}$ 中的高斯因子位于大 $|x|$ 区域。我们首先考虑无反射性扭结的情况，在这种情况下

$$\mathfrak{g}_k(x) = \begin{cases} \mathcal{B}_k e^{-ikx} & \text{if } x \ll -1/m \\ \mathcal{D}_k e^{-ikx} & \text{if } x \gg 1/m \end{cases} \quad (3\text{-}34)$$

$$|\mathcal{B}_k|^2 = |\mathcal{D}_k|^2 = 1, \quad \mathcal{B}_k^* = \mathcal{B}_{-k}, \quad \mathcal{D}_k^* = \mathcal{D}_{-k}$$

其中相位 $\mathcal{B}_k$ 和 $\mathcal{D}_k$ 在 $k$ 空间中以 $O(m)$ 阶尺度变化

$$\frac{\partial_k \mathcal{B}_k}{\mathcal{B}_k} \sim \frac{\partial_k \mathcal{D}_k}{\mathcal{D}_k} \sim O\left(\frac{1}{m}\right). \quad (3\text{-}35)$$

由于 $x_0 \ll -1/m$，这个近似导致

$$\alpha_{k_1} = 2\sigma\sqrt{\pi}\mathcal{B}_{k_1}e^{-\sigma^2(k_1-k_0)^2}e^{i(k_0-k_1)x_0}. \quad (3\text{-}36)$$

接下来，让我们考虑 $t \gg 1/m$。我们并不会假设介子到达扭结的时间足够长。因此，通过这种近似，该过程将大致在壳，因此 $\omega_{k_1}$ 可以替换为 $\omega_{k_2}+\omega_{k_3}$。我们需要小心地处理这一点，因为在 $\omega_{k_2}+\omega_{k_3}-\omega_{k_1}$ 这一阶的项已经在不同的地方用到了。每个表达式都应当被视为 $\omega_{k_2}+\omega_{k_3}-\omega_{k_1}$ 的幂展开。然而，这个替换

能够可靠地在式 (3-33) 的分母中的 $\omega_{k_1}$ 上完成，因为这一项对于 $\omega_{k_2} + \omega_{k_3} - \omega_{k_1}$ 是零阶的。

使用这两个近似，我们得到

$$\dot{A}_{k_2k_3}(t) = -i2\sigma\sqrt{\pi}\frac{\sqrt{\lambda}}{8\omega_{k_2}\omega_{k_3}(\omega_{k_2}+\omega_{k_3})}\int\frac{dk_1}{2\pi}\mathcal{B}_{k_1}e^{-\sigma^2(k_1-k_0)^2}e^{i(k_0-k_1)x_0}$$
$$\times\left[\int dy V^{(3)}(\sqrt{\lambda}f(y))\mathfrak{g}_{-k_1}(y)\mathfrak{g}_{k_2}(y)\mathfrak{g}_{k_3}(y)\right]e^{i(\omega_{k_2}+\omega_{k_3}-\omega_{k_1})t}. \quad (3\text{-}37)$$

因为 $\sigma \gg 1/m$，$k_1$ 总是接近于 $k_0$，因此我们可以作如下展开

$$\omega_{k_1} = \omega_{k_0} + (k_1-k_0)\frac{k_0}{\omega_{k_0}}, \qquad \mathcal{B}_{k_1} = \mathcal{B}_{k_0}, \qquad \mathfrak{g}_{-k_1} = \mathfrak{g}_{-k_0}. \quad (3\text{-}38)$$

将 (3-38) 代入 (3-37)，

$$\dot{A}_{k_2k_3}(t) = -i2\sigma\sqrt{\pi}\mathcal{B}_{k_0}\frac{\sqrt{\lambda}e^{i(\omega_{k_2}+\omega_{k_3}-\omega_{k_0})t}}{8\omega_{k_2}\omega_{k_3}(\omega_{k_2}+\omega_{k_3})}\left[\int dy V^{(3)}(\sqrt{\lambda}f(y))\mathfrak{g}_{-k_0}(y)\mathfrak{g}_{k_2}(y)\mathfrak{g}_{k_3}(y)\right]$$
$$\times\int\frac{dk_1}{2\pi}e^{-\sigma^2(k_1-k_0)^2}e^{i(k_0-k_1)(x_0+\frac{k_0}{\omega_{k_0}}t)}$$
$$= -i\mathcal{B}_{k_0}\frac{\sqrt{\lambda}e^{i(\omega_{k_2}+\omega_{k_3}-\omega_{k_0})t}}{8\omega_{k_2}\omega_{k_3}(\omega_{k_2}+\omega_{k_3})}\mathrm{Exp}\left[-\frac{(x_0+\frac{k_0}{\omega_{k_0}}t)^2}{4\sigma^2}\right]V_{-k_0k_2k_3}. \quad (3\text{-}39)$$

注意，在将 $V_{-k_1k_2k_3}$ 替换为 $V_{-k_0k_2k_3}$ 时，我们假设了 $V$ 对 $k_1$ 的依赖性在尺度上比 $1/\sigma$ 宽很多。如果 $V^{(3)}(gf(x))$ 没有紧支撑，这个假设将在 $k_1 + k_2 + k_3 = 0$ 附近失效（因为 $V$ 可能有一个 $\delta$ 函数项和一个极点）。这些发生在远离质壳的地方，因此不会反映任何有趣的动力学过程，它们其实是我们的初始条件 (3-19) 没有包括使得介子在远离扭结处严格平移的量子修正导致的，我们已经在第 3.2 节中解释了这一点。

#### 3.3.3.2 反射性扭结

到目前为止，我们只考虑了无反射性扭结（无反射性扭结的例子有 Sine-Gordon 孤子和 $\phi^4$ 扭结）。然而，更一般的扭结是具有反射性的，因此正规模具有以下的渐近形式

$$\mathfrak{g}_k(x) = \begin{cases} \mathcal{B}_k e^{-ikx} + C_k e^{ikx} & \text{if} \quad x \ll -1/m \\ \mathcal{D}_k e^{-ikx} + \mathcal{E}_k e^{ikx} & \text{if} \quad x \gg 1/m \end{cases} \quad (3\text{-}40)$$

$$|\mathcal{B}_k|^2 + |C_k|^2 = |\mathcal{D}_k|^2 + |\mathcal{E}_k|^2 = 1$$

$$\mathcal{B}_k^* = \mathcal{B}_{-k}, \qquad C_k^* = C_{-k}, \qquad \mathcal{D}_k^* = \mathcal{D}_{-k}, \qquad \mathcal{E}_k^* = \mathcal{E}_{-k}.$$

同样，我们的初始波包位于 $x_0 \ll -1/m$ 处，因此这种近似允许我们简化系数 $\alpha_{k_1}$

$$\alpha_{k_1} = 2\sigma\sqrt{\pi}\left[\mathcal{B}_{k_1}e^{-\sigma^2(k_1-k_0)^2}e^{i(k_0-k_1)x_0} + \mathcal{C}_{k_1}e^{-\sigma^2(k_1+k_0)^2}e^{i(k_0+k_1)x_0}\right]. \tag{3-41}$$

将其代入 (3-33) 我们得到

$$\dot{A}_{k_2k_3}(t) = -i2\sigma\sqrt{\pi}\frac{\sqrt{\lambda}}{8\omega_{k_2}\omega_{k_3}(\omega_{k_2}+\omega_{k_3})}\int\frac{dk_1}{2\pi}V_{-k_1k_2k_3}e^{i(\omega_{k_2}+\omega_{k_3}-\omega_{k_1})t}$$

$$\times \left[\mathcal{B}_{k_1}e^{-\sigma^2(k_1-k_0)^2}e^{i(k_0-k_1)x_0} + \mathcal{C}_{k_1}e^{-\sigma^2(k_1+k_0)^2}e^{i(k_0+k_1)x_0}\right]. \tag{3-42}$$

回想一下，我们已经固定了 $k_0 > 0$ 所以波包向右朝向扭结移动。在无反射情况下，这意味着 $k_1 > 0$。现在我们看到有两个高斯因子，第一个位于 $k_1 \sim k_0$，而第二个位于 $k_1 \sim -k_0$。因此，虽然介子的初始运动总是向右，在有反射情况下，这对应于单介子福克空间中的两个不同区域。

因此，我们需要考虑 $k_1$ 在 $k_0$ 和 $-k_0$ 处展开这两种情况，这会导致相应的频率展开

$$\omega_{k_1} = \omega_{k_0} + (\pm k_1 - k_0)\frac{k_0}{\omega_{k_0}}. \tag{3-43}$$

将这两个展开代入式 (3-42)，我们得到

$$\dot{A}_{k_2k_3}(t) = -i2\sigma\sqrt{\pi}\frac{\sqrt{\lambda}e^{i(\omega_{k_2}+\omega_{k_3}-\omega_{k_0})t}}{8\omega_{k_2}\omega_{k_3}(\omega_{k_2}+\omega_{k_3})}\int\frac{dk_1}{2\pi}V_{-k_1k_2k_3}$$

$$\times \left[\mathcal{B}_{k_1}e^{-\sigma^2(k_1-k_0)^2}e^{i(k_0-k_1)(x_0+\frac{k_0}{\omega_{k_0}}t)} + \mathcal{C}_{k_1}e^{-\sigma^2(k_1+k_0)^2}e^{i(k_1+k_0)(x_0+\frac{k_0}{\omega_{k_0}}t)}\right]$$

$$= -i\frac{\sqrt{\lambda}e^{i(\omega_{k_2}+\omega_{k_3}-\omega_{k_0})t}}{8\omega_{k_2}\omega_{k_3}(\omega_{k_2}+\omega_{k_3})}\text{Exp}\left[-\frac{(x_0+\frac{k_0}{\omega_{k_0}}t)^2}{4\sigma^2}\right]\tilde{V}_{-k_0k_2k_3}, \tag{3-44}$$

其中我们定义了简写

$$\tilde{V}_{-k_0k_2k_3} = \mathcal{B}_{k_0}V_{-k_0k_2k_3} + \mathcal{C}^*_{k_0}V_{k_0k_2k_3}. \tag{3-45}$$

#### 3.3.3.3 关于有限时间演化的小结

由于高斯因子的存在，振幅的时间导数只有在指数因子

$$x_t = x_0 + \frac{k_0}{\omega_{k_0}}t \tag{3-46}$$

很小时才是显著的，它发生在时刻

$$t \sim t_1 = -\frac{\omega_{k_0}}{k_0}x_0, \tag{3-47}$$

其正是介子撞上扭结的时刻。

特别地，由于 $t \geq 0$，我们看到这需要 $k_0$ 和 $x_0$ 具有相反的符号，这当然是介子朝向扭结移动所必需的。由于 $A(0) = 0$，我们也能得出在碰撞前，$t \ll t_1$ 时，振幅 $A(t)$ 为零。

### 3.3.4 渐近未来的振幅

#### 3.3.4.1 大时间极限

我们对介子已经与扭结发生散射的大时间极限感兴趣。在大 $t$ 下，我们可以对式 (3-44) 进行积分，得到

$$\begin{aligned}
\lim_{t\to\infty} A_{k_2 k_3}(t) &= -i\frac{\sqrt{\lambda}\tilde{V}_{-k_0 k_2 k_3}}{8\omega_{k_2}\omega_{k_3}(\omega_{k_2}+\omega_{k_3})}\int_{-\infty}^{\infty} dt \operatorname{Exp}\left[-\frac{(x_0+\frac{k_0}{\omega_{k_0}}t)^2}{4\sigma^2}\right]e^{i(\omega_{k_2}+\omega_{k_3}-\omega_{k_0})t} \\
&= -i\frac{\sqrt{\lambda}\tilde{V}_{-k_0 k_2 k_3}}{4\omega_{k_2}\omega_{k_3}(\omega_{k_2}+\omega_{k_3})}\sigma\sqrt{\pi}\frac{\omega_{k_0}}{k_0} \\
&\quad\times \operatorname{Exp}\left[-\sigma^2\frac{\omega_{k_0}^2}{k_0^2}\left(\omega_{k_2}+\omega_{k_3}-\omega_{k_0}\right)^2 - i\left(\omega_{k_2}+\omega_{k_3}-\omega_{k_0}\right)\frac{\omega_{k_0}}{k_0}x_0\right].
\end{aligned}$$
(3-48)

因此

$$\lim_{t\to\infty} \frac{|{}_0\langle k_2 k_3|\Phi(t)\rangle_0|^2}{|{}_0\langle 0|0\rangle_0|^2} = \frac{\pi\lambda\sigma^2\left|\tilde{V}_{-k_0 k_2 k_3}\right|^2}{16\omega_{k_2}^2\omega_{k_3}^2(\omega_{k_2}+\omega_{k_3})^2}\left(\frac{\omega_{k_0}}{k_0}\right)^2 \operatorname{Exp}\left[-2\sigma^2\frac{\omega_{k_0}^2}{k_0^2}\left(\omega_{k_2}+\omega_{k_3}-\omega_{k_0}\right)^2\right].$$
(3-49)

让我们定义在壳初始动量 $k_I$

$$k_I \equiv \sqrt{\left(\omega_{k_2}+\omega_{k_3}\right)^2 - m^2}$$
(3-50)

所以有 $\omega_{k_I} = \omega_{k_2} + \omega_{k_3}$。式 (3-49) 中的高斯因子在 $\omega_{k_0} \sim \omega_{k_I}$ 有贡献。因此，由于 $k_0$ 和 $k_I$ 都被定义为正数，在 $k_2 - k_3$ 相空间中对概率贡献最大的区域在 $k_0 \sim k_I$。因此我们使用展开

$$k_0 = k_I + (k_0 - k_I)$$
(3-51)

并且只保留每个表达式中的领头阶非零项。这导致

$$\lim_{t\to\infty} \frac{|{}_0\langle k_2 k_3|\Phi(t)\rangle_0|^2}{|{}_0\langle 0|0\rangle_0|^2} = \frac{\pi\lambda\sigma^2\left|\tilde{V}_{-k_I k_2 k_3}\right|^2}{16\omega_{k_2}^2\omega_{k_3}^2 k_I^2}\operatorname{Exp}\left[-2\sigma^2\frac{\omega_{k_I}^2}{k_I^2}\left(\omega_{k_I}-\omega_{k_0}\right)^2\right].$$
(3-52)

使用与式 (3-43) 中同样的展开，这进一步简化为

$$\lim_{t\to\infty} \frac{|_0\langle k_2 k_3|\Phi(t)\rangle_0|^2}{|_0\langle 0|0\rangle_0|^2} = \frac{\pi\lambda\sigma^2 \left|\tilde{V}_{-k_I k_2 k_3}\right|^2}{16\omega_{k_2}^2 \omega_{k_3}^2 k_I^2} e^{-2\sigma^2(k_I-k_0)^2}. \tag{3-53}$$

### 3.3.4.2 更快的推导

一种更快的方法是直接取式 (3-30) 的 $t \to \infty$ 极限，只不过这样我们无法了解中间时刻的演化。使用等式

$$\lim_{t\to\infty} \frac{\sin\left(\frac{\omega_{k_2}+\omega_{k_3}-\omega_{k_1}}{2}t\right)}{\left(\omega_{k_2}+\omega_{k_3}-\omega_{k_1}\right)/2} = 2\pi\delta\left(\omega_{k_2}+\omega_{k_3}-\omega_{k_1}\right) = \frac{\omega_{k_I}}{k_I}\left(2\pi\delta\left(k_1-k_I\right) + 2\pi\delta\left(k_1+k_I\right)\right) \tag{3-54}$$

振幅可以被简化为

$$\lim_{t\to\infty} \frac{_0\langle k_2 k_3|\Phi(t)\rangle_0}{_0\langle 0|0\rangle_0} = -\frac{i\sqrt{\lambda}}{8\omega_{k_2}\omega_{k_3}k_I} e^{-i\omega_{k_I}t}\left(\alpha_{k_I} V_{-k_I k_2 k_3} + \alpha_{-k_I} V_{k_I k_2 k_3}\right). \tag{3-55}$$

由于 $k_I$ 和 $k_0$ 都很大且为正，式 (3-41) 中和 $(k_I + k_0)$ 有关的的高斯项被指数压低，使得 $\alpha_{k_I}$ 中只留下 $\mathcal{B}_{k_I}$ 项，$\alpha_{-k_I}$ 中只留下 $C_{k_I}^*$ 项。把这些都加起来我们得到

$$\lim_{t\to\infty} \frac{_0\langle k_2 k_3|\Phi(t)\rangle_0}{_0\langle 0|0\rangle_0} = -\frac{i\sigma\sqrt{\pi\lambda}}{4\omega_{k_2}\omega_{k_3}k_I} e^{i(k_0-k_I)x_0} e^{-i\omega_{k_I}t} e^{-\sigma^2(k_0-k_I)^2} \tilde{V}_{-k_I k_2 k_3}. \tag{3-56}$$

所得结果和上面较长的推导一致。并且我们再次提醒读者此结果和 (2-136) 的领头阶一致。因此我们证明了本节的简单计算在领头阶是有效的（并可以直接推广到第 4 章的斯托克斯散射和反斯托克斯散射情形）。

### 3.3.5 介子倍增的概率和概率密度

$|\Phi(t)\rangle_0$（时刻 $t$ 的态）在希尔伯特空间的给定子空间中的概率 $P$ 由下式给出

$$P = \frac{_0\langle\Phi(t)|\mathcal{P}|\Phi(t)\rangle_0}{_0\langle\Phi(t)|\Phi(t)\rangle_0} \tag{3-57}$$

其中 $\mathcal{P}$ 是投影到该子空间的投影算符。

我们对末态有两个介子的概率 $P_{\text{tot}}$ 感兴趣，它对应于投影算符

$$\mathcal{P}_{\text{tot}}|k_2 k_3\rangle_0 = |k_2 k_3\rangle_0, \qquad k_2, k_3 \in \mathbb{R}. \tag{3-58}$$

我们还对末态介子具有动量 $k_2$ 和 $k_3$ 的相应概率密度 $P_{\text{diff}}(k_2, k_3)$ 感兴趣。它和总概率由下式联系

$$P_{\text{tot}} = \frac{1}{2}\int dk_2 dk_3 P_{\text{diff}}(k_2, k_3) \tag{3-59}$$

其中 1/2 因子是由于 $|k_2 k_3\rangle$ 和 $|k_3 k_2\rangle$ 代表相同的态。$\mathcal{P}_{\text{diff}}$ 由类似于 (3-57) 的公式定义，其中算符 $\mathcal{P}_{\text{diff}}$ 湮灭所有的 $k$ 不等于 $k_2$ 和 $k_3$ 的态。因为它具有无穷大的本征值所以它并不是投影算符。这两个方程很容易求解，于是我们得到算符

$$\mathcal{P}_{\text{diff}}(k_2, k_3) = \frac{\omega_{k_2}\omega_{k_3}}{\pi^2{}_0\langle 0|0\rangle_0}|k_2 k_3\rangle_{00}\langle k_2 k_3|$$

$$\mathcal{P}_{\text{tot}} = \frac{1}{{}_0\langle 0|0\rangle_0}\int dk_2 dk_3 \frac{\omega_{k_2}\omega_{k_3}}{2\pi^2}|k_2 k_3\rangle_{00}\langle k_2 k_3|. \tag{3-60}$$

考虑具有式 (3-41) 中 $\alpha_{k_1}$ 形式的一般反射性扭结。

$${}_0\langle \Phi(t)|\Phi(t)\rangle_0 = {}_0\langle \Phi|\Phi\rangle_0 = \int \frac{dk_1}{2\pi}\alpha_{k_1}\alpha_{k_1}^* \frac{{}_0\langle 0|0\rangle_0}{2\omega_{k_1}} = \sqrt{2\pi}\sigma \frac{{}_0\langle 0|0\rangle_0}{2\omega_{k_0}}, \tag{3-61}$$

其中我们使用了 $\omega_{k_1} \sim \omega_{k_0}$。

在大 $t$ 时刻的概率密度是

$$P_{\text{diff}}(k_2, k_3) = \lim_{t\to\infty} \frac{{}_0\langle \Phi(t)|\mathcal{P}_{\text{diff}}(k_2, k_3)|\Phi(t)\rangle_0}{{}_0\langle \Phi(t)|\Phi(t)\rangle_0} = \lim_{t\to\infty} \frac{\sqrt{2}\omega_{k_0}\omega_{k_2}\omega_{k_3}}{\pi^{5/2}\sigma}\frac{|{}_0\langle k_2 k_3|\Phi(t)\rangle_0|^2}{|{}_0\langle 0|0\rangle_0|^2}$$

$$= \frac{\lambda\sigma\omega_{k_0}\left|\tilde{V}_{-k_I k_2 k_3}\right|^2}{8\sqrt{2}\pi^{3/2}\omega_{k_2}\omega_{k_3}k_I^2}e^{-2\sigma^2(k_I-k_0)^2}. \tag{3-62}$$

注意，根据定义，连续模的 $k$ 是实数，因此该等式仅在 $\omega_{k_2}, \omega_{k_3} \geq m$ 时成立。对此进行积分可得出在大 $t$ 时刻的介子倍增的总概率

$$P_{\text{tot}} = \frac{1}{2}\int dk_2 dk_3 P_{\text{diff}}(k_2, k_3) = \frac{\lambda\sigma\omega_{k_0}}{16\sqrt{2}\pi^{3/2}}\int dk_2 dk_3 \frac{\left|\tilde{V}_{-k_I k_2 k_3}\right|^2}{\omega_{k_2}\omega_{k_3}k_I^2}e^{-2\sigma^2(k_I-k_0)^2}. \tag{3-63}$$

由于 $\sigma \gg 1/m$，我们可以将高斯项近似为狄拉克 $\delta$ 函数，得到

$$P_{\text{diff}}(k_2, k_3) = \frac{\lambda\omega_{k_I}\left|\tilde{V}_{-k_I k_2 k_3}\right|^2}{16\pi\omega_{k_2}\omega_{k_3}k_I^2}\delta(k_I - k_0) \tag{3-64}$$

$$P_{\text{tot}} = \frac{\lambda\omega_{k_0}}{32\pi k_0^2}\int dk_2 dk_3 \frac{\left|\tilde{V}_{-k_I k_2 k_3}\right|^2}{\omega_{k_2}\omega_{k_3}}\delta(k_I - k_0)$$

$$= \frac{\lambda}{32\pi k_0}\int_{-\sqrt{(\omega_{k_0}-m)^2-m^2}}^{\sqrt{(\omega_{k_0}-m)^2-m^2}} dk_2 \frac{\left|\tilde{V}_{-k_0, k_2, \sqrt{(\omega_{k_0}-\omega_{k_2})^2-m^2}}\right|^2 + \left|\tilde{V}_{-k_0, k_2, -\sqrt{(\omega_{k_0}-\omega_{k_2})^2-m^2}}\right|^2}{\omega_{k_2}\sqrt{(\omega_{k_0}-\omega_{k_2})^2-m^2}},$$

其中我们使用了

$$\frac{\partial k_I}{\partial k_3} = \frac{\omega_{k_0}k_3}{k_0\omega_{k_3}} = \frac{\omega_{k_0}\sqrt{(\omega_{k_0}-\omega_{k_2})^2-m^2}}{k_0(\omega_{k_0}-\omega_{k_2})}. \tag{3-65}$$

## 3.4 例子：Sine-Gordon 孤子和 $\phi^4$ 扭结

### 3.4.1 Sine-Gordon 孤子

在由

$$V(\sqrt{\lambda}\phi(x)) = m^2 \left(1 - \cos(\sqrt{\lambda}\phi(x))\right) \tag{3-66}$$

定义的 Sine-Gordon 理论中，$V_{k_1 k_2 k_3}$ 在文献 [18] 中给出

$$\begin{aligned}V_{k_1 k_2 k_3} &= \frac{\pi i \sqrt{\lambda}}{4}\mathrm{sign}(k_1 k_2 k_3)\mathrm{sech}\left(\frac{\pi(k_1+k_2+k_3)}{2m}\right) \\ &\times \frac{(\omega_{k_1}+\omega_{k_2}+\omega_{k_3})(\omega_{k_1}+\omega_{k_2}-\omega_{k_3})(\omega_{k_1}+\omega_{k_3}-\omega_{k_2})(\omega_{k_2}+\omega_{k_3}-\omega_{k_1})}{\omega_{k_1}\omega_{k_2}\omega_{k_3}}.\end{aligned} \tag{3-67}$$

于是

$$V_{\pm k_I k_2 k_3} = 0. \tag{3-68}$$

原因是它正比于 $\omega_{k_2} + \omega_{k_3} - \omega_{k_I} = 0$，然后这导致

$$\tilde{V}_{-k_I k_2 k_3} = 0, \tag{3-69}$$

因为它是 $V_{\pm k_I k_2 k_3}$ 的线性组合 (3-45)。然后通过 (3-62) 我们可以得知，对于所有 $k_2$ 和 $k_3$，微分概率都为零。

这是意料之中的，Sine-Gordon 模型的可积性意味着介子的数目是守恒的，因此介子倍增不会出现在 $S$ 矩阵中。

### 3.4.2 $\phi^4$ 扭结

#### 3.4.2.1 回顾

考虑 $\phi^4$ 双势阱模型，其势为

$$V(\sqrt{\lambda}\phi(x)) = \frac{\lambda\phi^2(x)}{4}\left(\sqrt{\lambda}\phi(x) - \sqrt{2}m\right)^2. \tag{3-70}$$

为了得到 $\tilde{V}_{-k_1 k_2 k_3}$ 的表达式，我们需要知道 $\mathcal{B}_k$、$C_k$ 和 $V_{k_1 k_2 k_3}$。前两个很容易从正规模中读出

$$\mathfrak{g}_k(x) = \frac{e^{-ikx}}{\omega_k\sqrt{k^2+\beta^2}}\left[k^2 - 2\beta^2 + 3\beta^2\mathrm{sech}^2(\beta x) - 3i\beta k\tanh(\beta x)\right], \qquad \beta = \frac{m}{2}. \tag{3-71}$$

在 $x \ll -1/\beta$ 处，这变成了一个平面波，其相位是

$$\mathcal{B}_k = \frac{k^2 - 2\beta^2 + 3i\beta k}{\omega_k\sqrt{k^2+\beta^2}}, \qquad C_k = 0. \tag{3-72}$$

由于 $\phi^4$ 扭结是无反射的，乘积 $\mathcal{B}_k \mathcal{C}_k^*$ 为零。

使用式 (3-45) 和 $|\mathcal{B}_k| = 1$，无反射条件会导致简化

$$\left|\tilde{V}_{-k_0 k_2 k_3}\right| = \left|V_{-k_0 k_2 k_3}\right|. \tag{3-73}$$

因此我们只需计算 $V_{k_1 k_2 k_3}$。在文献 [25] 中，这是根据对 $x$ 的积分之和计算的，但是这些积分没有被计算，因为那篇文献涉及需要对被积函数进行微妙处理的红外发散。我们将在这里看到类似的红外发散，这是由于对应于介子倍增的三点相互作用具有非零常数模（即使在远离扭结处）。远离扭结的介子倍增被压低只是因为相应的矩阵元快速振荡，这导致即使初始动量在非常小的间隔内积分也会出现相消干涉。

让我们首先回顾文献 [25] 中 $V_{k_1 k_2 k_3}$ 的表达式。首先，势的三阶导数是

$$V^{(3)}(\sqrt{\lambda} f(x)) = 6\sqrt{2}\beta \tanh(\beta x). \tag{3-74}$$

注意，它的阶数为 $O(\sqrt{\lambda})$，正是我们振幅的阶数。另外注意，它在大的 $x$ 和 $-x$ 处趋于非零常数。

我们将使用如下等式进行 $x$-积分

$$\int dx e^{-ikx} \operatorname{sech}^{2n}(\beta x) = \begin{cases} 2\pi \delta(k) & \text{if } n = 0 \\ \frac{\pi}{(2n-1)!k} \left[\prod_{j=0}^{n-1}\left(\frac{k^2}{\beta^2} + (2j)^2\right)\right] \operatorname{csch}\left(\frac{\pi k}{2\beta}\right) & \text{if } n > 0 \end{cases}$$

$$\int dx e^{-ikx} \operatorname{sech}^{2n}(\beta x)\tanh(\beta x) = -i\frac{\pi}{(2n)!\beta} \left[\prod_{j=0}^{n-1}\left(\frac{k^2}{\beta^2} + (2j)^2\right)\right] \operatorname{csch}\left(\frac{\pi k}{2\beta}\right). \tag{3-75}$$

注意，在两个积分的 $n = 0$ 情况下，被积函数在 $|x|$ 处不会变小。这些公式对应于计算积分的一种主值方法。我们已经检验过使用这样的主值积分确实是正确的，由它得到的结果与通过光滑权重函数对 $k_1$ 中的一个小区域进行积分所得到的结果相同。在我们对振幅计算的主要公式 (3-30) 中确实存在这种相干积分，它是初始波包中动量的积分。在本章脚注 a 中解释了为何 $k$ 积分应当在 $x$ 积分之前执行。

$V_{k_1 k_2 k_3}$ 由一系列项之和组成，其中每一项都是 $\operatorname{sech}^{2I}(\beta x)\tanh^J(\beta x)$ 的对 $x$ 的积分，其中 $I \in \{0, 1, 2, 3\}$，$J \in \{0, 1\}$ 中。$I = J = 0$ 的情况产生 $\delta(k)$，其中

$$k = k_1 + k_2 + k_3. \tag{3-76}$$

由于 $\omega_{k_I} = \omega_{k_2} + \omega_{k_3}$，$k$ 不为零，因此该项消失。不过我们将保留这一项，因为在未来的问题中，我们可能还会用到 $V_{k_1 k_2 k_3}$ 的表达式。不过我们会把它单独分

开，因为它不会在树图级别对介子倍增产生贡献。我们进行如下分解

$$V_{k_1k_2k_3} = V^{00}_{k_1k_2k_3} + \hat{V}_{k_1k_2k_3}, \qquad V^{00}_{k_1k_2k_3} = -\frac{9\sqrt{2}i\beta^2 k_1k_2k_3\left(6\beta^2+k_1^2+k_2^2+k_3^2\right)2\pi\delta(k)}{\omega_{k_1}\omega_{k_2}\omega_{k_3}\sqrt{\beta^2+k_1^2}\sqrt{\beta^2+k_2^2}\sqrt{\beta^2+k_3^2}}$$
(3-77)

其中 $V^{00}$ 包含所有 $\delta(k)$ 项，而只有 $\hat{V}$ 和下面的计算相关。

让我们通过以下方式定义符号 $u$

$$\hat{V}_{k_1k_2k_3} = \frac{6\sqrt{2}\pi\beta\,\text{csch}\left(\frac{\pi k}{2\beta}\right)}{\omega_{k_1}\omega_{k_2}\omega_{k_3}\sqrt{\beta^2+k_1^2}\sqrt{\beta^2+k_2^2}\sqrt{\beta^2+k_3^2}}\sum_{J=0}^{1}\sum_{I=1-J}^{3}u^{IJ}_{k_1k_2k_3},$$
(3-78)

其中求和不包括 $I=J=0$，因为那一项在 $V^{00}$ 中。

每个 $u^{IJ}$ 被定义为 $V_{k_1k_2k_3}$ 中的一个项，其为 $e^{ixk}\text{sech}^{2I}(\beta x)\tanh^J(bx)$ 对 $x$ 的积分。我们定义符号 $\Phi$ 来对这些系数进行总结

$$u^{IJ}_{k_1k_2k_3} = \frac{\sinh\left(\frac{\pi k}{2\beta}\right)}{\pi}\Phi^{IJ}_{k_1k_2k_3}\int dx\, e^{-ixk}\text{sech}^{2I}(\beta x)\tanh^J(\beta x).$$
(3-79)

文献 [25] 提供了 $\Phi$ 的分量

$$\Phi^{10}_{k_1k_2k_3} = 3i\beta\left[16\beta^4 S^1_1 + \beta^2\left(-5S^{21}_2 - 18S^1_3\right) + S^1_3 S^1_2\right]$$
(3-80)

$$\Phi^{20}_{k_1k_2k_3} = 9i\beta^3\left[-7\beta^2 S^1_1 + S^{21}_2 + 3S^1_3\right], \qquad \Phi^{30}_{k_1k_2k_3} = 27i\beta^5 S^1_1$$

$$\Phi^{01}_{k_1k_2k_3} = -8\beta^6 + \beta^4(18S^1_2 + 4S^2_1) + \beta^2(-2S^2_2 - 9S^1_3 S^1_1) + S^2_3$$

$$\Phi^{11}_{k_1k_2k_3} = 3\beta^2\left[12\beta^4 + \beta^2(-15S^1_2 - 4S^2_1) + (S^2_2 + 3S^1_3 S^1_1)\right]$$

$$\Phi^{21}_{k_1k_2k_3} = 9\beta^4\left[-6\beta^2 + (3S^1_2 + S^2_1)\right], \qquad \Phi^{31}_{k_1k_2k_3} = 27\beta^6$$

其中我们使用了如下的 $k$ 的对称组合

$$S^n_1 = k_1^n + k_2^n + k_3^n, \qquad S^n_2 = (k_1 k_2)^n + (k_1 k_3)^n + (k_2 k_3)^n, \qquad S^n_3 = (k_1 k_2 k_3)^n$$

$$S^{mn}_2 = k_1^m k_2^n + k_1^m k_3^n + k_2^m k_3^n + k_1^n k_2^m + k_1^n k_3^m + k_2^n k_3^m.$$
(3-81)

### 3.4.2.2 计算介子倍增的概率密度和总概率

我们现在可以使用 (3-75) 对 $x$ 进行积分

$$u^{I0}_{k_1k_2k_3} = \Phi^{I0}_{k_1k_2k_3}\frac{1}{(2I-1)!k}\left[\prod_{j=0}^{I-1}\left(\frac{k^2}{\beta^2}+(2j)^2\right)\right]$$
(3-82)

$$u^{I1}_{k_1k_2k_3} = \Phi^{I1}_{k_1k_2k_3}\frac{-i}{(2I)!\beta}\left[\prod_{j=0}^{I-1}\left(\frac{k^2}{\beta^2}+(2j)^2\right)\right].$$

特别地，我们发现

$$u^{10}_{k_1 k_2 k_3} = 3ik\left[16\beta^3 S^1_1 + \beta\left(-5S^{21}_2 - 18S^1_3\right) + \frac{1}{\beta}S^1_3 S^2_2\right] \quad (3\text{-}83)$$

$$u^{20}_{k_1 k_2 k_3} = \frac{3ik}{2}\left(\frac{k^2}{\beta^2} + 4\right)\left[-7\beta^3 S^1_1 + \beta S^{21}_2 + 3\beta S^1_3\right]$$

$$u^{30}_{k_1 k_2 k_3} = \frac{9ik}{40}\left(\frac{k^4}{\beta^4} + 20\frac{k^2}{\beta^2} + 64\right)\left[\beta^3 S^1_1\right]$$

$$u^{01}_{k_1 k_2 k_3} = i\left[8\beta^5 + \beta^3(-18S^1_2 - 4S^2_1) + \beta^1(2S^2_2 + 9S^1_3 S^1_1) - \frac{S^2_3}{\beta}\right]$$

$$u^{11}_{k_1 k_2 k_3} = \frac{3ik^2}{2}\left[-12\beta^3 + \beta(15S^1_2 + 4S^2_1) + \frac{1}{\beta}(-S^2_2 - 3S^1_3 S^1_1)\right]$$

$$u^{21}_{k_1 k_2 k_3} = \frac{3ik^2}{8}\left(\frac{k^2}{\beta^2} + 4\right)\left[6\beta^3 + \beta(-3S^1_2 - S^2_1)\right]$$

$$u^{31}_{k_1 k_2 k_3} = -\frac{3ik^2}{80}\left(\frac{k^4}{\beta^4} + 20\frac{k^2}{\beta^2} + 64\right)\beta^3.$$

对这些分量进行组合，最终我们得到

$$\begin{aligned}
\hat{V}_{k_1 k_2 k_3} &= \frac{6\sqrt{2}\pi\,\mathrm{csch}\left(\frac{\pi(k_1+k_2+k_3)}{2\beta}\right)}{\omega_{k_1}\omega_{k_2}\omega_{k_3}\sqrt{\beta^2+k_1^2}\sqrt{\beta^2+k_2^2}\sqrt{\beta^2+k_3^2}} \\
&\times \Bigg\{ 8i\beta^6 + 5i\beta^4(k_1^2+k_2^2+k_3^2) + 2i\beta^2(k_1^2 k_2^2 + k_1^2 k_3^2 + k_2^2 k_3^2) \\
&\quad + i\left[\frac{3}{16}(-k_1^6 - k_2^6 - k_3^6 + k_1^4 k_2^2 + k_1^4 k_3^2 + k_2^4 k_3^2\right. \\
&\quad \left. + k_2^4 k_1^2 + k_3^4 k_1^2 + k_3^4 k_2^2) + \frac{1}{8}k_1^2 k_2^2 k_3^2\right]\Bigg\}.
\end{aligned} \quad (3\text{-}84)$$

回想一下，介子倍增概率密度 (3-62) 和总概率 (3-64) 只需要用到特殊情况 $k_1 = -k_I$。在这种情况下，系数简化为

$$V_{-k_I k_2 k_3} = -\frac{48\sqrt{2}\pi i \omega_{k_2}\omega_{k_3}\omega_{k_I}\,\mathrm{csch}\left(\frac{\pi(k_2+k_3-k_I)}{m}\right)}{\sqrt{4k_2^2+m^2}\sqrt{4k_3^2+m^2}\sqrt{4k_I^2+m^2}}. \quad (3\text{-}85)$$

为了完整性，我们同时提供 $\tilde{V}$

$$\begin{aligned}\tilde{V}_{-k_I k_2 k_3} &= \mathcal{B}_{k_I} V_{-k_I k_2 k_3} + \mathcal{C}_{-k_I} V_{k_I k_2 k_3} = \frac{k_I^2 - 2\beta^2 + 3i\beta k_I}{\omega_{k_I}\sqrt{k_I^2 + \beta^2}} V_{-k_I k_2 k_3} \\ &= \frac{48\sqrt{2}\pi \omega_{k_2}\omega_{k_3}\left(i\left(3m^2 - 2\omega_{k_I}^2\right) + 3mk_I\right)\operatorname{csch}\left(\frac{\pi(k_2+k_3-k_I)}{m}\right)}{\sqrt{4k_2^2 + m^2}\sqrt{4k_3^2 + m^2}\left(4k_I^2 + m^2\right)},\end{aligned}$$
(3-86)

其中我们用到了 (3-72) 和 (3-50)。然而，作为 (3-45) 的结果，在树图级别我们只需要绝对值 $|\tilde{V}|$，对于一个无反射性扭结它等于 $|\hat{V}|$，并且在 $k_1 \sim -k_I$ 时等于 $|V|$。

将式 (3-86) 代入式 (3-62)，我们得到介子倍增的概率密度和总概率。我们的主要结果是以下的概率密度的解析表达式

$$P_{\text{diff}}(k_2, k_3) = \frac{288\sqrt{2\pi}\lambda\sigma\omega_{k_0}\omega_{k_2}\omega_{k_3}\omega_{k_I}^2 \operatorname{csch}^2\left(\frac{\pi(k_2+k_3-k_I)}{m}\right)}{k_I^2(4k_2^2+m^2)(4k_3^2+m^2)(4k_I^2+m^2)} e^{-2\sigma^2(k_I-k_0)^2}. \quad (3\text{-}87)$$

在初始介子的单色极限 $\sigma \to \infty$ 下，这即是

$$\begin{aligned}P_{\text{diff}}(k_2, k_3) &= \frac{\lambda\omega_{k_I}\left|\tilde{V}_{-k_I k_2 k_3}\right|^2}{16\pi\omega_{k_2}\omega_{k_3}k_I^2}\delta(k_I - k_0) \\ &= \frac{288\pi\lambda\omega_{k_2}\omega_{k_3}\omega_{k_I}^3 \operatorname{csch}^2\left(\frac{\pi(k_2+k_3-k_I)}{m}\right)}{k_I^2(4k_2^2+m^2)(4k_3^2+m^2)(4k_I^2+m^2)}\delta(k_I - k_0).\end{aligned}$$
(3-88)

正如预期的那样，它是 $O(\lambda)$ 阶的。狄拉克 $\delta$ 函数要求精确的能量守恒。另一方面，介子之间的动量守恒由 csch 保证。它不是 $\delta$ 函数，因此动量可以在介子和扭结之间传递。注意，$k_2$ 和 $k_3$ 为实数的条件意味着此等式仅在以下情况下有效

$$m \leq \omega_{k_2}, \omega_{k_3} \leq \omega_{k_0} - m. \quad (3\text{-}89)$$

对 $k_3$ 进行积分，我们得到对于单个介子动量的概率密度

$$P_{\text{diff}}(k_2) = \int dk_3 P_{\text{diff}}(k_2,k_3) \tag{3-90}$$

$$= \frac{288\pi\lambda\omega_{k_2}\omega_{k_0}^2(\omega_{k_0}-\omega_{k_2})^2}{k_0(4k_2^2+m^2)(4(\omega_{k_0}-\omega_{k_2})^2-3m^2)(4k_0^2+m^2)\sqrt{(\omega_{k_0}-\omega_{k_2})^2-m^2}}$$

$$\times\left[\text{csch}^2\left(\frac{\pi\left(k_2+\sqrt{(\omega_{k_0}-\omega_{k_2})^2-m^2}-k_0\right)}{m}\right)\right.$$

$$\left.+\text{csch}^2\left(\frac{\pi\left(k_2-\sqrt{(\omega_{k_0}-\omega_{k_2})^2-m^2}-k_0\right)}{m}\right)\right].$$

分母中的最后一项导致阈值 $k_3 = 0$ 处的极点，其对应于雅可比因子 $dk_3/dk_2$ 发散这一事实。在有限的 $\sigma$ 处，这个极点被抹掉了。这两个 csch 项对应于 $k_3$ 沿原始介子方向行进或反弹，csch 项中的参数是介子和扭结之间的动量传递。

在相对论极限 $k_0 \gg m$ 中，式 (3-88) 变为

$$P_{\text{diff}}(k_2,k_3) = \frac{9\pi\lambda\text{csch}^2\left(\frac{\pi m}{2k_2k_3k_I}\left(k_I^2-k_2k_3\right)\right)}{2k_2k_3k_I}\delta(k_I-k_0) \tag{3-91}$$

$$= \frac{18\lambda k_2k_3k_0}{\pi m^2\left(k_0^2-k_2k_3\right)^2}\delta(k_2+k_3-k_0).$$

它主要分布在当 $k_2$、$k_3$ 和 $k_I$ 都是 $k_0$ 阶的情形，因此它与 $1/k_0$ 成正比。要获得总概率，需要在 $k_2 - k_3$ 平面上积分，或者更准确地说是 $k_2 + k_3 = k_0$ 线，其中 $k_2, k_3 > 0$。这条线的长度是 $O(k_0)$ 阶的，因此在大 $k_0$ 时总概率渐近到一个常数。令 $k_2 = k_0 x$ 我们得到在相对论极限下

$$P_{\text{tot}} = \frac{9\lambda}{\pi m^2}\int_0^1 dx\frac{x(1-x)}{(1-x+x^2)^2} = \frac{\lambda}{m^2}\left(\frac{6}{\pi}-\frac{2}{\sqrt{3}}\right) \sim 0.755\frac{\lambda}{m^2}. \tag{3-92}$$

## 3.5　$\phi^4$ 扭结的数值结果

在本节中，我们对刚刚得到的 $\phi^4$ 双势阱模型的一些解析计算的概率结果进行相应的数值计算。

在 $O(\lambda)$ 阶，概率密度 $P_{\text{diff}}$ 和总概率 $P_{\text{tot}}$ 与 $\lambda$ 成正比，因此在图中我们将它们除以 $\lambda$。我们使用参数 $m = 1, \sigma = 20$。我们已经通过数值验证，只要 $\sigma$ 的值满足 $1/m \ll \sigma$，$\sigma$ 的值就不会影响数值结果。

我们从图 3-1 开始，绘制概率密度 $P_\text{diff}(k_2) = \int dk_3 P_\text{diff}(k_2, k_3)$，其中 $P_\text{diff}(k_2, k_3)$ 取自式 (3-87)，两个末态介子之一将具有动量 $k_2$。每条曲线右侧的轻微抬升并非数值伪像。它们是由于，对于固定的 $k_0$，$k_3$ 积分中的雅可比因子在产生相应介子的阈值处发散。这导致在极限 $\sigma \to \infty$ 中出现一个极点，但在这里这个极点被初始波包的动量宽度抹掉了。

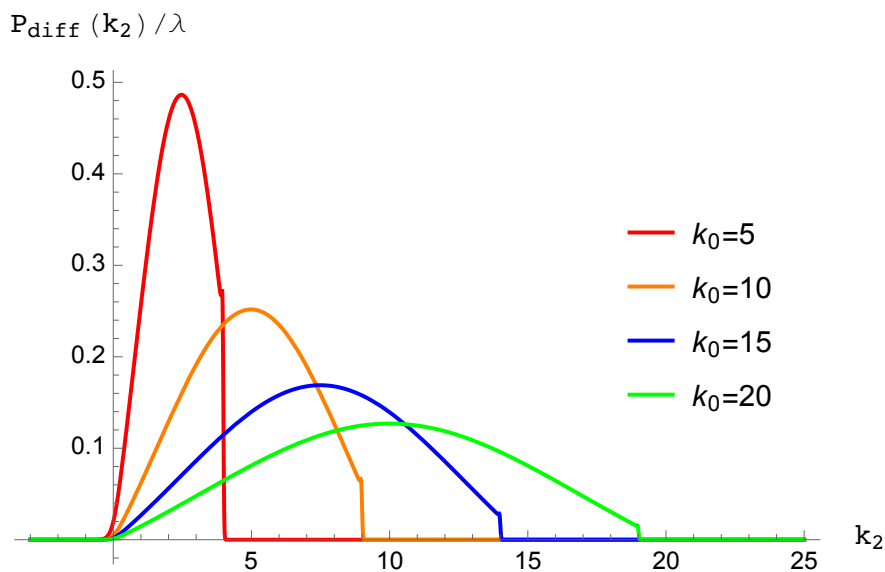

**图 3-1** 概率密度 $P_\text{diff}(k_2)$，其中一个末态介子具有动量 $k_2$，图中展示了不同 $k_0$ 值的曲线。$\lambda$ 因子已被商去。

**Figure 3-1** The probability density, $P_\text{diff}(k_2)$, that one of the final mesons has momentum $k_2$, plotted for various values of $k_0$. The factor of $\lambda$ has been divided out.

接着，在图 3-2 中，我们绘制了介子倍增的总概率，它是作为初始介子动量 $k_0$ 的函数。注意，在大 $k_0$ 时，概率渐近到式 (3-92) 中的值。

最后，在图 3-3 中，我们绘制了概率 $P_n$，其中有 $n$ 个末态介子具有动量 $k < 0$（$n = 0, 1, 2$，且 $k < 0$ 意味着它们向扭结后方移动）。该图表明，在 $O(\lambda)$ 阶，即使是无反射性扭结也会导致一些反射。然而，正如所料，当初始介子的动量 $k_0$ 远大于介子质量 $m$ 时，这种情况非常罕见。

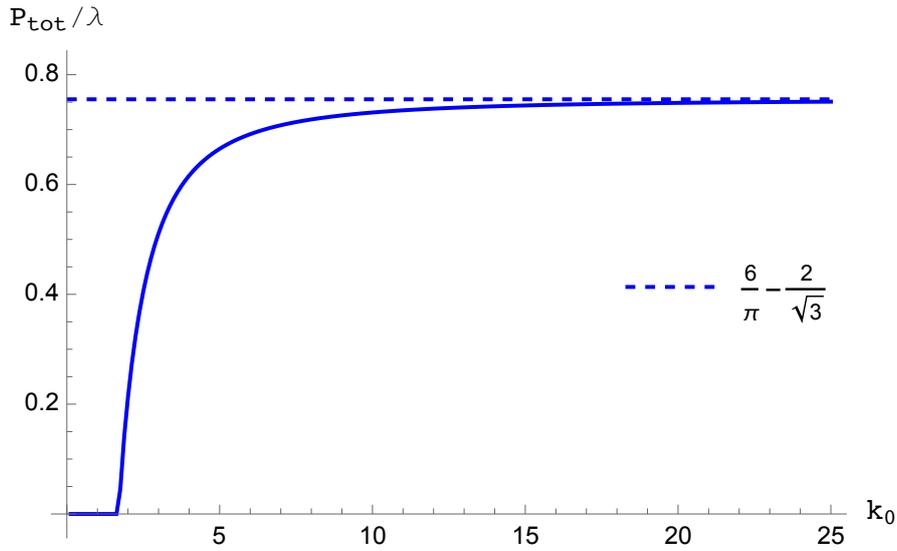

**图 3-2** 总介子倍增概率 $P_{\text{tot}}$ 作为 $k_0$ 的函数，其按 $1/\lambda$ 缩放。虚线是式 (3-92) 中导出的渐近值。

**Figure 3-2** The total meson multiplication probability $P_{\text{tot}}$ as a function of $k_0$, rescaled by $1/\lambda$. The dashed line is the asymptotic value derived in Eq. (3-92).

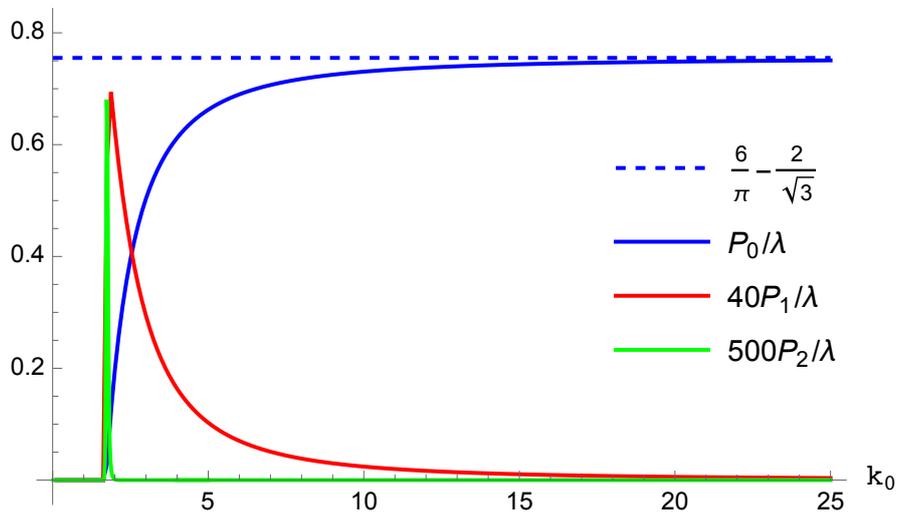

**图 3-3** $n$ 个出射介子的动量为负的概率 $P_n$。为使它们在图中可见，我们对它们按 $1/\lambda$ 以及图例中给出的其他比例因子缩放。虚线同样是式 (3-92) 中的渐近值。

**Figure 3-3** The probability $P_n$ that $n$ of the momenta of the outgoing mesons are negative. These are all rescaled by $1/\lambda$ and also by other factors, given in the legend, to make them visible in the plot. The dashed line is again the asymptotic value in Eq. (3-92).

## 3.6 本章小结

在 $\phi^4$ 双势阱模型的势的最小值之一展开，我们得到立方的相互作用项。原则上，这种相互作用允许一个介子倍增成两个介子。然而，这个过程在真空中是被禁止的，因为不可能同时满足能量和动量守恒。

另一方面，如果出现扭结，情况则会发生变化。在微扰理论的领头阶，介子仍然不能将能量转移到扭结。然而，如果介子倍增发生在离扭结足够近的地方，则动量可以转移。这种转移以一个 $\text{csch}^2$ 项出现在概率密度 (3-88)，它强制介子之间满足近似的动量守恒。

尽管如此，远距离的动量转移使我们的计算复杂化，因为介子倍增可以发生在任何位置，所有这些位置都需要被积分，朴素地看这将导致发散。我们找到了三种处理这些发散的方法。首先，初始介子波包动量的相干积分导致大 |x| 处的快速振荡而振幅被抑制。其次，将指数阻尼项添加到振幅，然后在阻尼消失时取极限同时也消除了发散。最后，本章使用的 tanh 的 $x$ 积分的主值使得结果有限。我们已经验证了消除发散的所有三种方法都会产生相同的结果。而只有第一个在物理上是合理的，因为它是固有的波包传播的结果，而不是一种特设的修改。然而，后两种方法由于其简易性在计算中更容易被实现。

在具有单个介子的扭结散射中，在 $O(\lambda)$ 阶概率下只有三个非弹性过程可能发生。一种是本章处理的介子倍增，还有两种是介子透射或反射时形模的激发和退激，我们分别称之为斯托克斯散射和反斯托克斯散射，这将是第 4 章的主题。

# 第 4 章 （反）斯托克斯散射

## 4.1 引言

1+1 维的标量理论提供了一些最简单的量子场论模型。如果标量场受到简并势的影响，除了基本的介子激发之外，还会存在非微扰扭结。

这种模型的经典场论已经非常丰富了，人们的大多数注意力集中在扭结-反扭结散射 [70–73]。人们早就知道 [1] 导致不同结果的初始相对速度的范围具有分数结构的共振窗口。人们曾经认为这是由扭结内部的激发谱引起的。当然，扭结内部的激发谱有它们的作用 [56, 74, 75]。然而，后来发现 [2, 4] 这种窗口甚至出现在扭结没有内部激发的模型中。因此，很明显扭结与整体动力学的相互作用很重要 [76]。

这种整体动力学本身就相当丰富。存在谱阱 [77, 78]，在这些谱阱之上，扭结的内部激发逃逸成为连续型。虽然它们有着显著的经典结果，但在量子理论中它们相当平滑 [23]。此外，在扭结-反扭结碰撞后，会激发多种模。而最后只有寿命最长的留下。其中，振子 [79, 80] 可以存活相当长的时间。然而，量子理论似乎又有所不同。在量子理论中，有了新的衰变道，大大缩短了振子的寿命 [62]。

总之，拼图中缺少两片关键部分。首先是对扭结和整体自由度之间相互作用的系统理解。第二个是理解量子修正所起的作用，以及它们是否在经典极限下消失，或者像振子情况和可能的谱阱情况似乎暗示的那样，从根本上影响物理。

这激发了对量子理论中扭结和基本介子之间相互作用的理解。这种相互作用比扭结-反扭结相互作用简单得多。然而，关于量子扭结-介子散射的经典文献主要局限于寻找扭结和介子之间的有效汤川耦合 [43, 44]。

最近，在文献 [17] 和文献 [18] 中，此类模型的线性化的微扰理论被单圈和高圈地开发。相比于文献 [15, 16] 中传统的集体坐标方法，它极大地简化了单扭结空间的计算，该空间由具有一个扭结和任意有限数量的介子的态组成。

使用这种方法，我们很快意识到，在领头阶恰好存在三个非弹性散射过程，这些过程以单个扭结和单个介子开始。第一个是介子倍增，其中介子被扭结吸收并发射出两个介子。我们在第 3 章中对此进行了研究。另外两个是斯托克斯散射和反斯托克斯散射。斯托克斯散射是介子在和基态扭结散射过程中激发扭结的形模的过程。反斯托克斯散射是介子从激发的扭结中散射出去并退激发其形模的过程。

在本章中，我们首次对这两个过程进行了处理。我们研究了斯托克斯散射和反斯托克斯散射的概率作为入射介子动量的函数，并将我们得到的一般性结果应用到了 $\phi^4$ 扭结模型。

## 4.2 斯托克斯散射

在斯托克斯散射中，一个介子被基态扭结吸收，另一个介子被发射并且一个形模被激发。初始条件是单介子态

$$|k_1\rangle_0 = B_{k_1}^{\ddagger}|0\rangle_0 \tag{4-1}$$

在扭结空间中的叠加

$$|\Phi\rangle_0 = \int \frac{dk_1}{2\pi} \alpha_{k_1}|k_1\rangle_0, \qquad \alpha_k = \int dx \Phi(x)\mathfrak{g}_k(x) \tag{4-2}$$

$$\Phi(x) = \mathrm{Exp}\left[-\frac{(x-x_0)^2}{4\sigma^2} + ixk_0\right], \quad x_0 \ll -\frac{1}{m}, \quad \frac{1}{k_0}, \frac{1}{m} \ll \sigma \ll |x_0|.$$

这里介子波包的初始位置是 $x = x_0$，它在扭结左侧并距其很远，扭结的中心位于 $x = 0$。介子以大约 $k_0$ 的动量向右移动。末态由一个介子和一个形模被激发的扭结组成。它是如下态的叠加

$$|Sk_2\rangle_0 = B_S^{\ddagger} B_{k_2}^{\ddagger}|0\rangle_0. \tag{4-3}$$

在最低阶 $O(\sqrt{\lambda})$ 阶，扭结哈密顿量中唯一可以在这些态之间产生我们所需要的贡献的项是

$$H_I = \frac{\sqrt{\lambda}}{2} \int \frac{dk_1}{2\pi} \frac{dk_2}{2\pi} \frac{V_{S,k_2,-k_1}}{\omega_{k_1}} B_S^{\ddagger} B_{k_2}^{\ddagger} B_{k_1} \tag{4-4}$$

$$V_{S,k_2,-k_1} = \int dx V^{(3)}(\sqrt{\lambda}f(x))\mathfrak{g}_S(x)\mathfrak{g}_{k_2}(x)\mathfrak{g}_{-k_1}(x).$$

在 $O(\sqrt{\lambda})$ 阶，时间演化算符中对应的项为

$$e^{-it(H_2' + H_I)} = e^{-itH_2'} - i \int_0^t dt_1 e^{-i(t-t_1)H_2'} H_I e^{-it_1 H_2'} + O(\lambda). \tag{4-5}$$

我们将丢掉第一项，因为它对下面的矩阵元没有贡献。把它作用于单扭结单介子态，我们得到态的斯托克斯散射

$$e^{-iH't}|k_1\rangle_0\Big|_{O(\sqrt{\lambda})} = \frac{-i\sqrt{\lambda}}{2\omega_{k_1}} \int \frac{dk_2}{2\pi} V_{S,k_2,-k_1} e^{-\frac{it}{2}(\omega_{k_1}+\omega_S+\omega_{k_2})} \frac{\sin\left[\left(\frac{\omega_S+\omega_{k_2}-\omega_{k_1}}{2}\right)t\right]}{(\omega_S+\omega_{k_2}-\omega_{k_1})/2}|Sk_2\rangle_0. \tag{4-6}$$

第 4 章　（反）斯托克斯散射当 $k_1 = \pm k_I^S$ 时这个过程是在壳的，这里我们定义了

$$\omega_{k_I^S} = \omega_{k_2} + \omega_S, \qquad k_I^S > 0. \tag{4-7}$$

在大时间极限下，我们可以使用如下等式

$$\lim_{t \to \infty} \frac{\sin\left[\left(\frac{\omega_S+\omega_{k_2}-\omega_{k_1}}{2}\right)t\right]}{(\omega_S + \omega_{k_2} - \omega_{k_1})/2} = 2\pi\delta(\omega_S+\omega_{k_2}-\omega_{k_1}) = \left(\frac{\omega_{k_I^S}}{k_I^S}\right)\left(2\pi\delta(k_1 - k_I^S) + 2\pi\delta(k_1 + k_I^S)\right) \tag{4-8}$$

来计算对 $k_2$ 的积分。将此结果应用到波包 (4-2) 中，可以得到在大 $t$ 时刻态的斯托克斯散射部分

$$\begin{aligned} & \left. e^{-iH't}|\Phi\rangle_0 \right|_{O(\sqrt{\lambda})} \\ &= -i\sqrt{\lambda}\int \frac{dk_1}{2\pi}\frac{\alpha_{k_1}}{2\omega_{k_1}}\int \frac{dk_2}{2\pi}V_{S,k_2,-k_1}e^{-\frac{it}{2}(\omega_{k_1}+\omega_S+\omega_{k_2})}\frac{\sin\left[\left(\frac{\omega_S+\omega_{k_2}-\omega_{k_1}}{2}\right)t\right]}{(\omega_S+\omega_{k_2}-\omega_{k_1})/2}|Sk_2\rangle_0 \\ &= \frac{-i\sqrt{\lambda}}{2}\int \frac{dk_2}{2\pi}\frac{e^{-i\omega_{k_I^S}t}}{k_I^S}\left(\alpha_{k_I^S}V_{S,k_2,-k_I^S} + \alpha_{-k_I^S}V_{S,k_2,k_I^S}\right)|Sk_2\rangle_0. \end{aligned} \tag{4-9}$$

介子波包从远离扭结的地方出发，在那里可以使用正规模的渐近形式

$$\begin{aligned} \mathfrak{g}_k(x) &= \begin{cases} \mathcal{B}_k e^{-ikx} + \mathcal{C}_k e^{ikx} & \text{若} \quad x \ll -1/m \\ \mathcal{D}_k e^{-ikx} + \mathcal{E}_k e^{ikx} & \text{若} \quad x \gg 1/m \end{cases} \\ & |\mathcal{B}_k|^2 + |\mathcal{C}_k|^2 = |\mathcal{D}_k|^2 + |\mathcal{E}_k|^2 = 1 \\ \mathcal{B}_k^* &= \mathcal{B}_{-k}, \qquad \mathcal{C}_k^* = \mathcal{C}_{-k}, \qquad \mathcal{D}_k^* = \mathcal{D}_{-k}, \qquad \mathcal{E}_k^* = \mathcal{E}_{-k} \end{aligned} \tag{4-10}$$

来计算波包的系数 $\alpha_k$。由于 $k_I^S$ 被定义为正数并且 $k_0$ 被设定为正数，在式 (4-9) 中只有如下两种情况

$$\begin{aligned} \alpha_{k_I^S} &= 2\sigma\sqrt{\pi}\left[\mathcal{B}_{k_I^S}e^{ix_0(k_0-k_I^S)}e^{-\sigma^2(k_0-k_I^S)^2} + \mathcal{C}_{k_I^S}e^{ix_0(k_0+k_I^S)}e^{-\sigma^2(k_0+k_I^S)^2}\right] \\ &= 2\sigma\sqrt{\pi}\mathcal{B}_{k_I^S}e^{ix_0(k_0-k_I^S)}e^{-\sigma^2(k_0-k_I^S)^2} \end{aligned} \tag{4-11}$$

和

$$\begin{aligned} \alpha_{-k_I^S} &= 2\sigma\sqrt{\pi}\left[\mathcal{B}_{k_I^S}^* e^{ix_0(k_0+k_I^S)}e^{-\sigma^2(k_0+k_I^S)^2} + \mathcal{C}_{k_I^S}^* e^{ix_0(k_0-k_I^S)}e^{-\sigma^2(k_0-k_I^S)^2}\right] \\ &= 2\sigma\sqrt{\pi}\mathcal{C}_{k_I^S}^* e^{ix_0(k_0-k_I^S)}e^{-\sigma^2(k_0-k_I^S)^2}. \end{aligned} \tag{4-12}$$

将这些代回式 (4-9)，我们得到大 $t$ 时刻态的相关部分

$$e^{-iH't}|\Phi\rangle_0\Big|_{O(\sqrt{\lambda})} = -i\sigma\sqrt{\pi\lambda}\int\frac{dk_2}{2\pi}e^{ix_0(k_0-k_I^S)}e^{-\sigma^2(k_0-k_I^S)^2}e^{-i\omega_{k_I^S}t}\left(\frac{\tilde{V}_{S,k_2,-k_I^S}}{k_I^S}\right)|Sk_2\rangle_0$$

$$\tilde{V}_{S,k_2,-k_I^S} = \mathcal{B}_{k_I^S}V_{S,k_2,-k_I^S} + \mathcal{C}^*_{k_I^S}V_{S,k_2,k_I^S}. \tag{4-13}$$

注意在无反射性扭结的情况下，$\mathcal{C} = 0$，因而 $|\tilde{V}| = |V|$。

使用内积

$$_0\langle Sk_1|Sk_2\rangle_0 = \frac{2\pi\delta(k_1-k_2)}{4\omega_S\omega_{k_2}}{}_0\langle 0|0\rangle_0 \tag{4-14}$$

我们得到矩阵元

$$\frac{_0\langle Sk_2|e^{-iH't}|\Phi\rangle_0}{_0\langle 0|0\rangle_0} = \frac{-i\sigma\sqrt{\pi\lambda}}{4\omega_S\omega_{k_2}k_I^S}e^{ix_0(k_0-k_I^S)}e^{-\sigma^2(k_0-k_I^S)^2}e^{-i\omega_{k_I^S}t}\tilde{V}_{S,k_2,-k_I^S} \tag{4-15}$$

平方之后得到

$$\left|\frac{_0\langle Sk_2|e^{-iH't}|\Phi\rangle_0}{_0\langle 0|0\rangle_0}\right|^2 = \frac{\sigma^2\pi\lambda}{16\omega_S^2\omega_{k_2}^2 k_I^{S2}}\left|\tilde{V}_{S,k_2,-k_I^S}\right|^2 e^{-2\sigma^2(k_0-k_I^S)^2}$$

$$= \frac{\sigma\pi^{3/2}\lambda}{16\sqrt{2}\omega_S^2\omega_{k_2}^2 k_I^{S2}}\left|\tilde{V}_{S,k_2,-k_I^S}\right|^2 \delta(k_I^S-k_0). \tag{4-16}$$

最后一个等式在 $\sigma \to \infty$ 极限下成立。

为了计算斯托克斯散射概率，我们需要投影算符 $\mathcal{P}$ 来投影到具有激发扭结和单个介子的末态

$$\mathcal{P} = \int dk_2 \mathcal{P}_{\text{diff}}(k_2), \qquad \mathcal{P}_{\text{diff}}(k_2) = \frac{4\omega_S\omega_{k_2}}{2\pi}\frac{|Sk_2\rangle_{00}\langle Sk_2|}{_0\langle 0|0\rangle_0}. \tag{4-17}$$

使用内积

$$\frac{_0\langle k_1|k_2\rangle_0}{_0\langle 0|0\rangle_0} = \frac{2\pi\delta(k_1-k_2)}{2\omega_{k_1}} \tag{4-18}$$

我们得到初态的归一化

$$\frac{_0\langle\Phi|\Phi\rangle_0}{_0\langle 0|0\rangle_0} = \int\frac{d^2k}{(2\pi)^2}\alpha_{k_1}\alpha^*_{k_2}\frac{_0\langle k_2|k_1\rangle_0}{_0\langle 0|0\rangle_0} = \int\frac{dk}{2\pi}\frac{|\alpha_k|^2}{2\omega_k} = \frac{1}{2\omega_{k_0}}\int\frac{dk}{2\pi}|\alpha_k|^2$$

$$= \frac{1}{2\omega_{k_0}}\int\frac{dk}{2\pi}\int dx\int dy\, g_k(x)g_k^*(y)\Phi(x)\Phi^*(y)$$

$$= \frac{1}{2\omega_{k_0}}\int dx|\Phi(x)|^2 = \frac{\sigma\sqrt{\pi}}{\sqrt{2}\omega_{k_0}} \tag{4-19}$$

其中我们在第一行的最后一步使用了 $\omega_k \sim \omega_{k_0}$。

${}_0\langle k_1|k_2\rangle_0$ 和 ${}_0\langle 0|0\rangle_0$ 都是无限的，所以严格来说前面的表达式没有定义。在第 2 章中，我们描述了如何通过将分子和分母除以平移群来系统地计算此类内积。扭结标架中平移算符的非对角作用对上述朴素操作进行了修正。然而，这些修正并不会影响我们在 $O(\lambda)$ 阶得到的概率。

最后，我们可以将所有这些成分组合起来，写出 $O(\lambda)$ 阶的斯托克斯散射总概率

$$
\begin{aligned}
P_S &= \frac{{}_0\langle \Phi|e^{iH't}\mathcal{P}e^{-iH't}|\Phi\rangle_0}{{}_0\langle \Phi|\Phi\rangle_0} = \int \frac{dk_2}{2\pi} 4\omega_S \omega_{k_2} \frac{\left|{}_0\langle Sk_2|e^{-iH't}|\Phi\rangle_0\right|^2}{{}_0\langle 0|0\rangle_0 \, {}_0\langle \Phi|\Phi\rangle_0/{}_0\langle 0|0\rangle_0} \frac{1}{{}_0\langle 0|0\rangle_0} \\
&= \int \frac{dk_2}{2\pi} 4\omega_S \omega_{k_2} \frac{\frac{\sigma \pi^{3/2}\lambda}{16\sqrt{2}\omega_S^2\omega_{k_2}^2 k_I^{S\,2}}\left|\tilde{V}_{S,k_2,-k_I^S}\right|^2 \delta(k_I^S - k_0)}{\left(\frac{\sigma\sqrt{\pi}}{\sqrt{2}\omega_{k_0}}\right)} \\
&= \frac{\pi\lambda\omega_{k_0}}{4\omega_S(\omega_{k_0}-\omega_S)k_0^2} \int \frac{dk_2}{2\pi}\left|\tilde{V}_{S,k_2,-k_I^S}\right|^2 \delta(k_I^S - k_0) \\
&= \lambda \frac{\left|\tilde{V}_{S,\sqrt{(\omega_{k_0}-\omega_S)^2-m^2},-k_0}\right|^2 + \left|\tilde{V}_{S,-\sqrt{(\omega_{k_0}-\omega_S)^2-m^2},-k_0}\right|^2}{8\omega_S k_0 \sqrt{(\omega_{k_0}-\omega_S)^2 - m^2}}.
\end{aligned}
\tag{4-20}
$$

我们看到概率是两项的总和。第一个是出射介子与入射介子沿相同方向传播的概率，第二个则是出射介子与入射介子沿相反方向传播的概率。我们将在下面的例子中看到，即使在无反射性扭结的情况下也会发生这种反射。

在初态和末态 (4-2) 和 (4-3) 中，介子在远离扭结时以恒定速度 $k_0/\omega_{k_0}$ 行进。然而，当用 $H_2'$ 演化时，这些态的 $O(\sqrt{\lambda})$ 阶量子修正，原则上有可能和态的领头阶用 $e^{-it(H_2'+H_I)}$ 演化产生同阶的贡献。即便我们的初态和末态 (4-2) 和 (4-3) 中不包含这样的 $O(\sqrt{\lambda})$ 阶修正，假如使用 $e^{-itH'}$ 进行演化，在介子从远处向扭结运动时，这样的修正也会产生。

正如第 3 章中所述，可以对初态和末态包含 $O(\sqrt{\lambda})$ 阶量子修正，使得它们在远离扭结处移动时不发生形变。这些态不是扭结哈密顿量 $H'$ 的本征态，而是左右真空哈密顿量的本征态，它们是通过将关于真空的定义哈密顿量扩展到扭结的左侧和右侧来定义的。这些量子修正来自远离扭结的三介子顶点。远离扭结处，介子各自保持动量守恒。结果是，这些过程相当离壳，导致在初态和末态介子周围形成一团相当离壳的介子云。因此，我们预计这种云不会影响介子倍增

或斯托克斯散射的渐近概率。第 3 章中表明，在介子倍增的情况下，情况确实如此。在这里我们可以进行相同的论证，因为斯托克斯散射只是介子倍增中把其中一个产生的介子换成束缚态，而反斯托克斯散射情形的论证也很类似。因此我们得出结论，在本章初态和末态修正也不会对结果产生影响。

## 4.3  反斯托克斯散射

在反斯托克斯散射中，初始扭结有一个激发的形模，介子波包从远处接近。因此初态是

$$|\Phi\rangle_0 = \int \frac{dk_1}{2\pi} \alpha_{k_1} |Sk_1\rangle_0, \qquad |Sk_1\rangle_0 = B_S^{\ddagger} B_{k_1}^{\ddagger} |0\rangle_0. \qquad (4\text{-}21)$$

末态由介子波包和退激扭结组成，故而它处在由 $|k_2\rangle_0$ 张成的态空间。

在 $O(\sqrt{\lambda})$ 阶，在这两个态之间进行起作用的唯一项是

$$H_I = \frac{\sqrt{\lambda}}{4\omega_S} \int \frac{dk_1}{2\pi} \frac{dk_2}{2\pi} \frac{V_{S,k_2,-k_1}}{\omega_{k_1}} B_{k_2}^{\ddagger} B_S B_{k_1}, \qquad H_I|Sk_1\rangle_0 = \frac{\sqrt{\lambda}}{4\omega_S} \int \frac{dk_2}{2\pi} \frac{V_{S,k_2,-k_1}}{\omega_{k_1}} |k_2\rangle_0. \qquad (4\text{-}22)$$

在领头阶，有限时间的演化导致

$$e^{-iH't}|Sk_1\rangle_0\Big|_{O(\sqrt{\lambda})} = \frac{-i\sqrt{\lambda}}{4\omega_S \omega_{k_1}} \int \frac{dk_2}{2\pi} V_{S,k_2,-k_1} e^{-\frac{it}{2}(\omega_{k_1}+\omega_S+\omega_{k_2})} \frac{\sin\left[\left(\frac{\omega_{k_1}+\omega_S-\omega_{k_2}}{2}\right)t\right]}{(\omega_{k_1}+\omega_S-\omega_{k_2})/2} |k_2\rangle_0. \qquad (4\text{-}23)$$

这个过程只有当 $k_2 = \pm k_I^{\mathrm{aS}}$ 时才是在壳的，不同于第 4.2 节中斯托克斯散射的情况，我们现在定义 $k_I^{\mathrm{aS}}$

$$\omega_{k_I^{\mathrm{aS}}} = \omega_{k_2} - \omega_S, \qquad k_I^{\mathrm{aS}} > 0. \qquad (4\text{-}24)$$

在大时间极限下，只有在壳的 $k_2$ 值有贡献

$$\lim_{t\to\infty} \frac{\sin\left[\left(\frac{\omega_{k_1}+\omega_S-\omega_{k_2}}{2}\right)t\right]}{(\omega_{k_1}+\omega_S-\omega_{k_2})/2} = \left(\frac{\omega_{k_I^{\mathrm{aS}}}}{k_I^{\mathrm{aS}}}\right) \left(2\pi\delta(k_1 - k_I^{\mathrm{aS}}) + 2\pi\delta(k_1 + k_I^{\mathrm{aS}})\right). \qquad (4\text{-}25)$$

将此极限代入式 (4-23) 并将结果代入到我们的初始波包 (4-21) 中，我们得到 $t$ 时

刻态的反斯托克斯散射部分

$$e^{-iH't}|\Phi\rangle_0\Big|_{O(\sqrt{\lambda})}$$

$$= \frac{-i\sqrt{\lambda}}{4\omega_S}\int \frac{dk_1}{2\pi}\frac{\alpha_{k_1}}{\omega_{k_1}}\int \frac{dk_2}{2\pi}V_{S,k_2,-k_1}e^{-\frac{it}{2}(\omega_{k_1}+\omega_S+\omega_{k_2})}\frac{\sin\left[\left(\frac{\omega_{k_1}+\omega_S-\omega_{k_2}}{2}\right)t\right]}{(\omega_{k_1}+\omega_S-\omega_{k_2})/2}|k_2\rangle_0$$

$$= \frac{-i\sqrt{\lambda}}{4\omega_S}\int \frac{dk_2}{2\pi}e^{-i\omega_{k_2}t}\left(\frac{1}{k_I^{\text{aS}}}\right)\left(\alpha_{k_I^{\text{aS}}}V_{S,k_2,-k_I^{\text{aS}}}+\alpha_{-k_I^{\text{aS}}}V_{S,k_2,k_I^{\text{aS}}}\right)|k_2\rangle_0$$

$$= \frac{-i\sigma\sqrt{\pi\lambda}}{2\omega_S}\int \frac{dk_2}{2\pi}e^{ix_0(k_0-k_I^{\text{aS}})}e^{-\sigma^2(k_0-k_I^{\text{aS}})^2}e^{-i\omega_{k_2}t}\left(\frac{\tilde{V}_{S,k_2,-k_I^{\text{aS}}}}{k_I^{\text{aS}}}\right)|k_2\rangle_0. \quad (4\text{-}26)$$

于是我们得到如下矩阵元

$$\frac{{}_0\langle k_2|e^{-iH't}|\Phi\rangle_0}{{}_0\langle 0|0\rangle_0} = \frac{-i\sigma\sqrt{\pi\lambda}}{4\omega_S\omega_{k_2}k_I^{\text{aS}}}e^{ix_0(k_0-k_I^{\text{aS}})}e^{-\sigma^2(k_0-k_I^{\text{aS}})^2}e^{-i\omega_{k_2}t}\tilde{V}_{S,k_2,-k_I^{\text{aS}}}. \quad (4\text{-}27)$$

在极限 $\sigma \to \infty$ 下，初始介子波包是单色的，结果简化为

$$\left|\frac{{}_0\langle k_2|e^{-iH't}|\Phi\rangle_0}{{}_0\langle 0|0\rangle_0}\right|^2 = \frac{\sigma^2\pi\lambda}{16\omega_S^2\omega_{k_2}^2 k_I^{\text{aS}\,2}}\left|\tilde{V}_{S,k_2,-k_I^{\text{aS}}}\right|^2 e^{-2\sigma^2(k_0-k_I^{\text{aS}})^2} \quad (4\text{-}28)$$

$$= \frac{\sigma\pi^{3/2}\lambda}{16\sqrt{2}\omega_S^2\omega_{k_2}^2 k_I^{\text{aS}\,2}}\left|\tilde{V}_{S,k_2,-k_I^{\text{aS}}}\right|^2 \delta(k_I^{\text{aS}}-k_0).$$

我们将要计算末态有一个基态扭结和一个介子的概率。这需要用到投影算符

$$\mathcal{P} = \int dk_2 \mathcal{P}_{\text{diff}}(k_2), \qquad \mathcal{P}_{\text{diff}}(k_2) = \frac{1}{2\pi}\frac{2\omega_{k_2}}{{}_0\langle 0|0\rangle_0}|k_2\rangle_{0\,0}\langle k_2|. \quad (4\text{-}29)$$

考虑初态的归一化

$$\frac{{}_0\langle\Phi|\Phi\rangle_0}{{}_0\langle 0|0\rangle_0} = \int \frac{d^2k}{(2\pi)^2}\alpha_{k_1}\alpha_{k_2}^*\frac{{}_0\langle Sk_2|Sk_1\rangle_0}{{}_0\langle 0|0\rangle_0} = \int \frac{dk}{2\pi}\frac{|\alpha_k|^2}{4\omega_S\omega_k} = \frac{1}{4\omega_S\omega_{k_0}}\int \frac{dk}{2\pi}|\alpha_k|^2$$

$$= \frac{1}{4\omega_S\omega_{k_0}}\int \frac{dk}{2\pi}\int dx\int dy g_k(x)g_k^*(y)\Phi(x)\Phi^*(y)$$

$$= \frac{1}{4\omega_S\omega_{k_0}}\int dx|\Phi(x)|^2 = \frac{\sigma\sqrt{\pi}}{2\sqrt{2}\omega_S\omega_{k_0}} \quad (4\text{-}30)$$

其中我们在第一行的最后一步再次使用 $\omega_k \sim \omega_{k_0}$，我们发现反斯托克斯散射的

总概率为

$$\begin{aligned}
P_{\text{aS}} &= \frac{{}_0\langle\Phi|e^{iH't}\mathcal{P}e^{-iH't}|\Phi\rangle_0}{{}_0\langle\Phi|\Phi\rangle_0} = \int \frac{dk_2}{2\pi} \frac{2\omega_{k_2}}{{}_0\langle 0|0\rangle_0} \frac{\left|{}_0\langle k_2|e^{-iH't}|\Phi\rangle_0\right|^2}{{}_0\langle\Phi|\Phi\rangle_0/{}_0\langle 0|0\rangle_0} \frac{1}{{}_0\langle 0|0\rangle_0} \\
&= \int \frac{dk_2}{2\pi} 2\omega_{k_2} \frac{\frac{\sigma\pi^{3/2}\lambda}{16\sqrt{2}\omega_S^2\omega_{k_2}^2 k_I^{\text{aS}\,2}}\left|\tilde{V}_{S,k_2,-k_I^{\text{aS}}}\right|^2 \delta(k_I^{\text{aS}} - k_0)}{\left(\frac{\sigma\sqrt{\pi}}{2\sqrt{2}\omega_S\omega_{k_0}}\right)} \\
&= \frac{\pi\lambda\omega_{k_0}}{4\omega_S(\omega_{k_0}+\omega_S)k_0^2}\int \frac{dk_2}{2\pi}\left|\tilde{V}_{S,k_2,-k_I^{\text{aS}}}\right|^2 \delta(k_I^{\text{aS}}-k_0) \\
&= \lambda\frac{\left|\tilde{V}_{S,\sqrt{(\omega_{k_0}+\omega_S)^2-m^2},-k_0}\right|^2 + \left|\tilde{V}_{S,-\sqrt{(\omega_{k_0}+\omega_S)^2-m^2},-k_0}\right|^2}{8\omega_S k_0 \sqrt{(\omega_{k_0}+\omega_S)^2-m^2}}.
\end{aligned} \quad (4\text{-}31)$$

同样地，第一项是出射介子与入射介子沿相同方向传播的概率。

## 4.4 例子：$\phi^4$ 扭结

### 4.4.1 解析结果

考虑 $\phi^4$ 双势阱模型，它由以下的势定义

$$V(\sqrt{\lambda}\phi(x)) = \frac{\lambda\phi^2(x)}{4}\left(\sqrt{\lambda}\phi(x) - \sqrt{2}m\right)^2. \quad (4\text{-}32)$$

它只有一个形模，其频率为

$$\omega_S = \sqrt{3}\beta, \qquad \beta = \frac{m}{2}. \quad (4\text{-}33)$$

由正规模

$$\begin{aligned}
\mathfrak{g}_k(x) &= \frac{e^{-ikx}}{\omega_k\sqrt{k^2+\beta^2}}\left[k^2 - 2\beta^2 + 3\beta^2\text{sech}^2(\beta x) - 3i\beta k\tanh(\beta x)\right] \quad (4\text{-}34) \\
\mathfrak{g}_S(x) &= \sqrt{\frac{3\beta}{2}}\tanh(\beta x)\text{sech}(\beta x), \qquad \mathfrak{g}_B(x) = \frac{\sqrt{3\beta}}{2}\text{sech}^2(\beta x)
\end{aligned}$$

得出

$$V_{Sk_1k_2} = \pi\frac{3\sqrt{3}}{8}\frac{\left(17\beta^4 - (\omega_{k_1}^2-\omega_{k_2}^2)^2\right)(\beta^2+k_1^2+k_2^2) + 8\beta^2 k_1^2 k_2^2}{\beta^{3/2}\omega_{k_1}\omega_{k_2}\sqrt{\beta^2+k_1^2}\sqrt{\beta^2+k_2^2}}\text{sech}\left(\frac{\pi(k_1+k_2)}{2\beta}\right), \quad (4\text{-}35)$$

然后

$$\left| \tilde{V}_{S,\pm\sqrt{(\omega_{k_0}-\omega_S)^2-m^2},-k_0} \right| = \left| V_{S,\pm\sqrt{(\omega_{k_0}-\omega_S)^2-m^2},-k_0} \right|$$

$$= \frac{\left| \left(-10\beta^2 + 3\omega_S\omega_{k_0} - 3k_0^2\right)(k_0^2 - \omega_S\omega_{k_0} + 2\beta^2) + \left(k_0^2 - 2\omega_S\omega_{k_0} + 3\beta^2\right)k_0^2 \right|}{(\omega_{k_0}-\omega_S)\omega_{k_0}\sqrt{\omega_{k_0}^2 - 2\omega_S\omega_{k_0}}\sqrt{\beta^2 + k_0^2}}$$

$$\times 3\sqrt{3}\beta\pi \operatorname{sech}\left(\frac{\pi(\pm\sqrt{k_0^2 - 2\omega_S\omega_{k_0} + 3\beta^2} - k_0)}{2\beta}\right)$$

$$= 6\sqrt{3}\beta\pi \frac{\sqrt{\omega_{k_0}}(\omega_{k_0}-\omega_S)}{\sqrt{\omega_{k_0}-2\omega_S}\sqrt{\beta^2+k_0^2}} \operatorname{sech}\left(\frac{\pi(\pm\sqrt{k_0^2 - 2\omega_S\omega_{k_0} + 3\beta^2} - k_0)}{2\beta}\right).$$

(4-36)

斯托克斯散射的概率则是

$$P_S = \frac{\omega_{k_0}(\omega_{k_0}-\omega_S)^2}{(\omega_{k_0}-2\omega_S)(\beta^2+k_0^2)k_0\sqrt{k_0^2-2\omega_S\omega_{k_0}+3\beta^2}}$$

$$\times \frac{9\sqrt{3}\pi^2\lambda}{2}\left[\operatorname{sech}^2\left(\frac{\pi(\sqrt{k_0^2-2\omega_S\omega_{k_0}+3\beta^2}-k_0)}{2\beta}\right)\right.$$

$$\left. + \operatorname{sech}^2\left(\frac{\pi(\sqrt{k_0^2-2\omega_S\omega_{k_0}+3\beta^2}+k_0)}{2\beta}\right)\right].$$

(4-37)

第一个 sech 项是出射介子继续沿与入射介子相同方向运动的概率，而第二项是出射介子沿相反方向运动的概率。这两种情况概率的比值就是两个 sech 项的比值。

在这一阶，由介子携带的总能量的转移正好是 $\omega_S$。然而介子的总动量并不守恒。远离扭结处，初始介子动量为 $k_0$，而出射介子的动量为 $\sqrt{k_0^2-2\omega_S\omega_{k_0}+3\beta^2}$。sech 项中的参数是介子在向前散射和向后散射情况下和扭结之间的动量传递。

同样，在反斯托克斯散射的情况下

$$\begin{aligned}\left|\tilde{V}_{S,\pm\sqrt{(\omega_{k_0}+\omega_S)^2-m^2},-k_0}\right| &= \left|V_{S,\pm\sqrt{(\omega_{k_0}+\omega_S)^2-m^2},-k_0}\right| \\ &= \frac{\left|\left(-3k_0^2 - 3\omega_S\omega_{k_0} - 10\beta^2\right)(k_0^2 + \omega_S\omega_{k_0} + 2\beta^2) + (k_0^2 + 2\omega_S\omega_{k_0} + 3\beta^2)k_0^2\right|}{\omega_{k_0}(\omega_{k_0}+\omega_S)\sqrt{\omega_{k_0}^2 + 2\omega_S\omega_{k_0}}\sqrt{\beta^2+k_0^2}} \\ &\quad \times 3\sqrt{3}\beta\pi\,\mathrm{sech}\left(\frac{\pi\left(\sqrt{k_0^2 + 2\omega_S\omega_{k_0} + 3\beta^2} \pm k_0\right)}{2\beta}\right) \\ &= 6\sqrt{3}\beta\pi\frac{\sqrt{\omega_{k_0}}(\omega_{k_0}+\omega_S)}{\sqrt{\omega_{k_0}+2\omega_S}\sqrt{\beta^2+k_0^2}}\,\mathrm{sech}\left(\frac{\pi(\pm\sqrt{k_0^2 + 2\omega_S\omega_{k_0} + 3\beta^2} - k_0)}{2\beta}\right)\end{aligned}$$
(4-38)

导出概率

$$\begin{aligned}P_{\mathrm{aS}} &= \frac{\omega_{k_0}(\omega_{k_0}+\omega_S)^2}{(\omega_{k_0}+2\omega_S)(\beta^2+k_0^2)k_0\sqrt{k_0^2+2\omega_S\omega_{k_0}+3\beta^2}} \\ &\quad \times \frac{9\sqrt{3}\pi^2\lambda}{2}\left[\mathrm{sech}^2\left(\frac{\pi(\sqrt{k_0^2+2\omega_S\omega_{k_0}+3\beta^2}-k_0)}{2\beta}\right)\right. \\ &\quad \left. + \mathrm{sech}^2\left(\frac{\pi(\sqrt{k_0^2+2\omega_S\omega_{k_0}+3\beta^2}+k_0)}{2\beta}\right)\right].\end{aligned}$$
(4-39)

### 4.4.2 数值结果

概率取决于无量纲耦合 $\lambda/m^2$ 以及无量纲动量 $k_0/m$。我们已经固定了我们的单位，使得远离扭结的介子质量为 $m=1$。对于基态扭结的斯托克斯散射的概率和对于激发态扭结的反斯托克斯散射的概率分别绘制在图 4-1 和 4-2 中。

注意，只有足够高的初始动量才在能量上允许斯托克斯散射，而反斯托克斯散射的概率在小动量时发散。无论两种情况的何种，接近阈值时，向后和向前散射的可能性变得相等。

在图 4-3 中，我们将这些过程的总概率与此阶唯一允许的另一个非弹性散射过程——介子倍增的总概率进行比较。对于较大的初始动量，斯托克斯散射和反斯托克斯散射的概率趋于零，而介子倍增的概率趋于恒定。特别地，我们看到斯

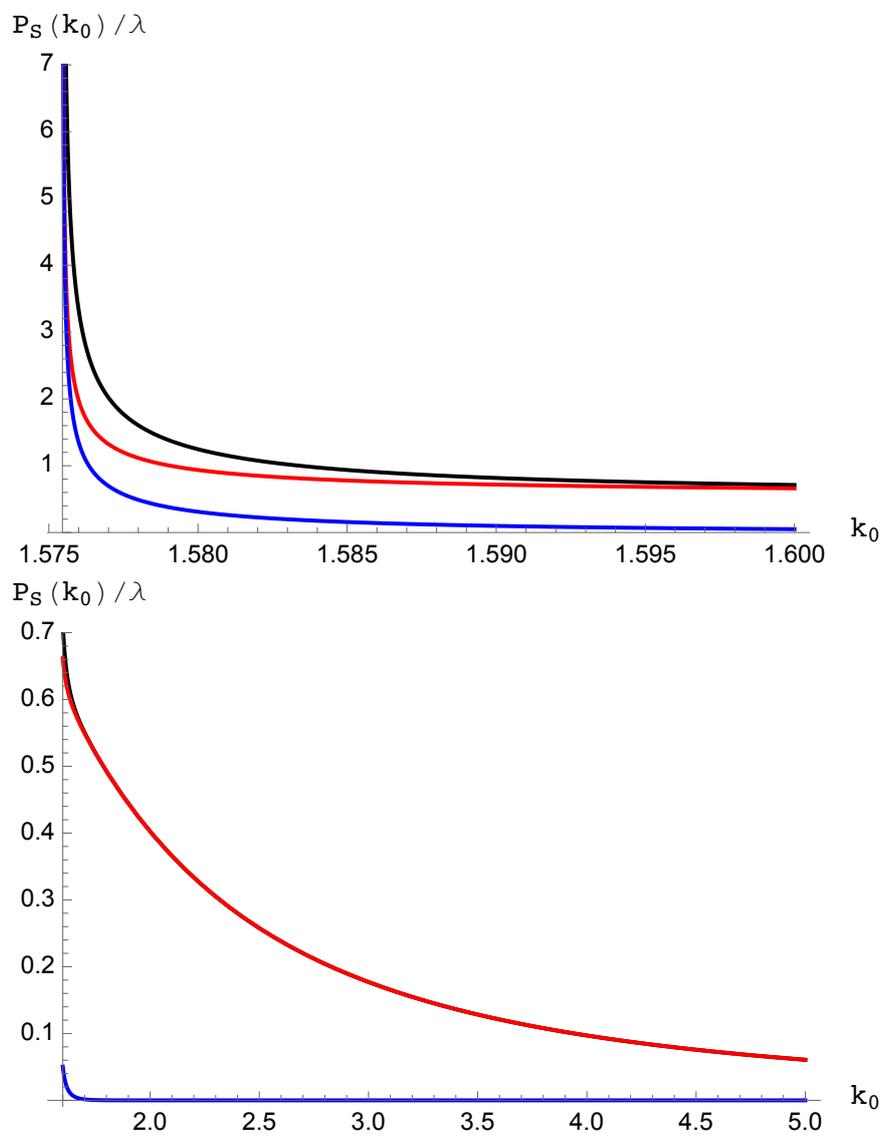

**图 4-1** 向前斯托克斯散射（红色）、向后斯托克斯散射（蓝色）和总的斯托克斯散射（黑色）概率 $P_S(k_0)$，$m=1$。

**Figure 4-1** The forward (red), backward (blue) and total (black) probabilities $P_S(k_0)$ of Stokes scattering, with $m=1$.

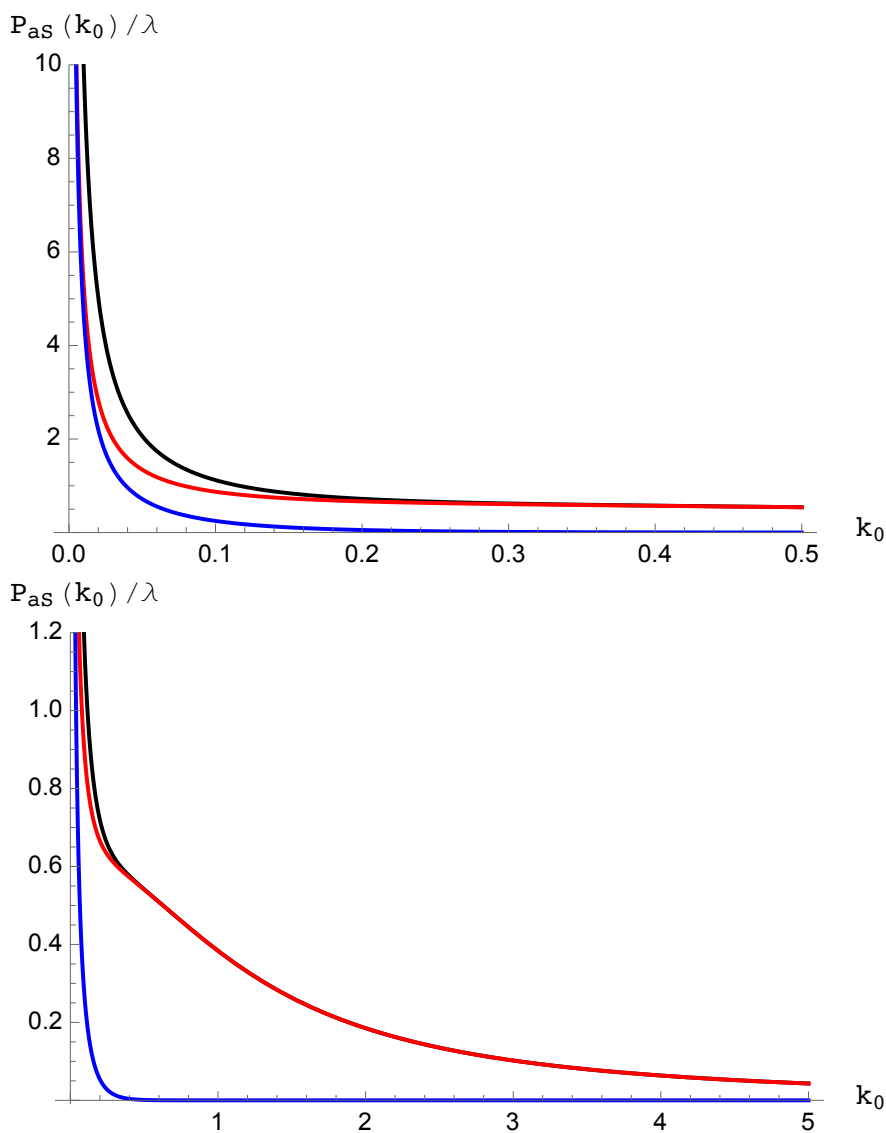

**图 4-2 向前反斯托克斯散射（红色）、向后反斯托克斯散射（蓝色）和总的反斯托克斯散射（黑色）概率 $P_{aS}(k_0)$，$m=1$。**

**Figure 4-2 The forward (red), backward (blue) and total (black) probabilities $P_{aS}(k_0)$ of anti-Stokes scattering, with $m=1$.**

托克斯散射和反斯托克斯散射在初始介子动量较低时占主导地位，而介子倍增在初始介子动量较高时占主导地位，当初始介子的动量约为介子质量的两倍时图中曲线出现交叉。

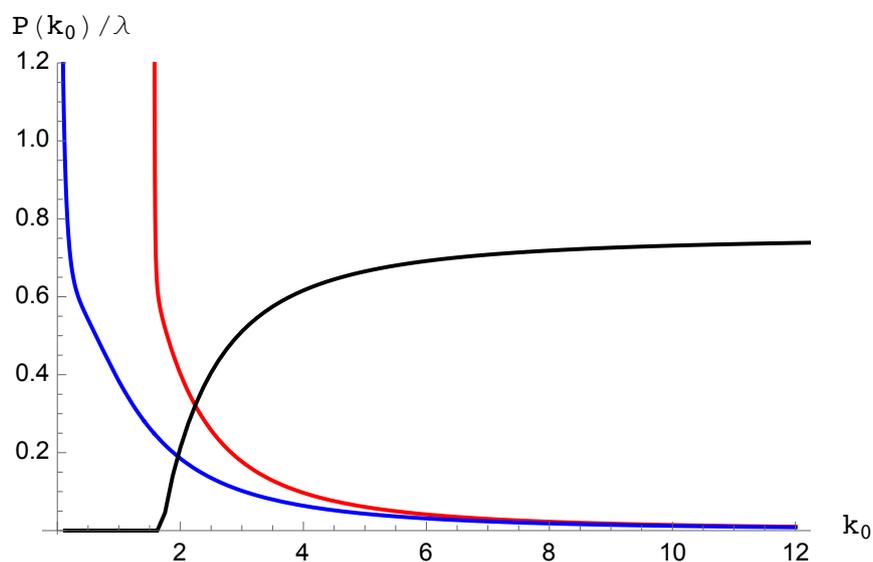

**图 4-3 第 3 章中介子倍增（黑色），斯托克斯散射（红色）和反斯托克斯散射（蓝色）的总概率比较。**

**Figure 4-3 The total probability of meson multiplication (black) from Chapter 3, plotted against the probability of Stokes (red) and anti-Stokes (blue) scattering.**

# 第 5 章　总结

二维标量模型为开发工具以处理现实世界中的孤子提供了理想的沙盒。如果标量场受到具有简并最小值的势的影响，那么该理论将有扭结和反扭结解。通常，在弱耦合下，可以将一个给定构型分解为扭结和标量场的微扰基本量子，称为介子。在弱耦合下对这些理论的理解就被简化为理解：介子之间的相互作用、扭结与（反）扭结之间的相互作用、以及扭结与介子的相互作用。

在 1+1 维情形，无质量标量场是量子化的阻碍 [81]。因此本文只考虑了 $m > 0$ 的模型。其结果是，扭结施加在介子上的力被距离乘以 $m$ 这个无量纲数指数地压低。这意味着在两者间隔的距离远大于 $1/m$ 时，介子和扭结对 $P'$ 的贡献基本上是分别守恒的。此外，在如此大的间隔下，直到 $O(\lambda)$ 阶的修正，每个介子对动量的贡献仅由 $k$ 给出。这并非是说扭结不会影响非常远的介子，而是说距离介子非常远的扭结只会改变一些平移不变的介子自耦合的值，而不会导致介子加速。

在 $O(\lambda)$ 阶，量子扭结和基本介子的非弹性散射现在已被完全理解。允许的过程有三种。首先是介子倍增。其次，如果扭结处于基态，那么当它和介子发生相互作用时，可能会激发其形模。最后，如果一个形模最初被激发，那么当介子和扭结相互作用时它可能会退激发形模。第一种情况在高能量下占主导地位，而另外两种情况发生的概率在接近其低能量阈值时变得非常大。

对这三种过程的计算，我们总是从自由哈密顿量的本征态开始，并在自由哈密顿量的本征态中对态进行测量。由于我们必须对紧空间中质心的所有可能位置进行积分，这将涉及形式上无限的矩阵元。相同的矩阵元出现在分子和分母中，因而它们互相抵消。在第 2 章中，我们更仔细地处理了此类比值，在那里我们除以了平移对称群，使得分子和分母都有限。我们发现相比于朴素的抵消，确实存在对结果的修正。然而，这些修正被 $\lambda$ 的幂压低了，因此在我们关心的领头阶不影响结果。但是，如果我们希望计算高阶修正，则必须包括第 2 章中我们研究的修正，因为在考虑高阶修正时它们会开始产生贡献。

本文对三种扭结-介子非弹性散射的研究，均为树图阶的计算。更丰富的非弹性散射现象（比如初态一个介子和扭结发生散射后末态为三个介子；或者斯托克斯散射和反斯托克斯散射情形的激发态扭结含有两个形模，等等），以及我们未来要研究的扭结-介子弹性散射，则会涉及到圈图计算。在线性化的孤子微扰理论的框架下，使用约化内积可以正确计算更高圈情形，与本文的计算相比并无

本质困难，只是计算过程更为繁琐。

## 参考文献

- [1] Campbell D K, Schonfeld J F, Wingate C A. Resonance Structure in Kink - Antikink Interactions in $\phi^4$ Theory [J]. Physica D, 1983, 9: 1.

- [2] Dorey P, Mersh K, Romanczukiewicz T, et al. Kink-antikink collisions in the $\phi^6$ model [J/OL]. Phys. Rev. Lett., 2011, 107: 091602. DOI: 10.1103/PhysRevLett.107.091602.

- [3] Dorey P, Gorina A, Perapechka I, et al. Resonance structures in kink-antikink collisions in a deformed sine-Gordon model [J/OL]. JHEP, 2021, 09: 145. DOI: 10.1007/JHEP09(2021)145.

- [4] Adam C, Dorey P, Garcia Martin-Caro A, et al. Multikink scattering in the $\phi^6$ model revisited [J/OL]. Phys. Rev. D, 2022, 106(12): 125003. DOI: 10.1103/PhysRevD.106.125003.

- [5] Faddeev L D, Korepin V E. Quantum Theory of Solitons: Preliminary Version [J/OL]. Phys. Rept., 1978, 42: 1-87. DOI: 10.1016/0370-1573(78)90058-3.

- [6] Lowe M. BOSON - SOLITON SCATTERING IN THE SINE-GORDON MODEL [J/OL]. Nucl. Phys. B, 1979, 159: 349-362. DOI: 10.1016/0550-3213(79)90339-0.

- [7] Parmentola J A, Zahed I. Meson-soliton scattering with soliton recoil [J/OL]. Chiral Solitons, 1987: 537-564. DOI: 10.1142/9789814415668_0014.

- [8] Swanson M S. SOLITON-PARTICLE SCATTERING AND BERRY'S PHASE [J/OL]. Phys. Rev. D, 1988, 38: 3122-3127. DOI: 10.1103/PhysRevD.38.3122.

- [9] Uehara M, Hayashi A, Saito S. Meson - soliton scattering with full recoil in standard collective coordinate quantization [J/OL]. Nucl. Phys. A, 1991, 534: 680-696. DOI: 10.1016/0375-9474(91)90466-J.

- [10] Abdelhady A M H H, Weigel H. Wave-Packet Scattering off the Kink-Solution [J/OL]. Int. J. Mod. Phys. A, 2011, 26: 3625-3640. DOI: 10.1142/S0217751X11054012.

- [11] Romanczukiewicz T. Interaction between kink and radiation in phi**4 model [J]. Acta Phys. Polon. B, 2004, 35: 523-540.

- [12] Romanczukiewicz T. Interaction between topological defects and radiation [J]. Acta Phys. Polon. B, 2005, 36: 3877-3887.

- [13] Forgacs P, Lukacs A, Romanczukiewicz T. Negative radiation pressure exerted on kinks [J/OL]. Phys. Rev. D, 2008, 77: 125012. DOI: 10.1103/PhysRevD.77.125012.

- [14] Dashen R F, Hasslacher B, Neveu A. Nonperturbative Methods and Extended Hadron Models in Field Theory 2. Two-Dimensional Models and Extended Hadrons [J/OL]. Phys. Rev. D, 1974, 10: 4130-4138. DOI: 10.1103/PhysRevD.10.4130.

- [15] Gervais J L, Jevicki A, Sakita B. Collective Coordinate Method for Quantization of Extended Systems [J/OL]. Phys. Rept., 1976, 23: 281-293. DOI: 10.1016/0370-1573(76)90049-1.

- [16] Gervais J L, Jevicki A. Point Canonical Transformations in Path Integral [J/OL]. Nucl. Phys. B, 1976, 110: 93-112. DOI: 10.1016/0550-3213(76)90422-3.

[17] Evslin J. Manifestly Finite Derivation of the Quantum Kink Mass [J/OL]. JHEP, 2019, 11: 161. DOI: 10.1007/JHEP11(2019)161.

[18] Evslin J, Guo H. Two-Loop Scalar Kinks [J/OL]. Phys. Rev. D, 2021, 103(12): 125011. DOI: 10.1103/PhysRevD.103.125011.

[19] Evslin J, Royston A B, Zhang B. Cut-off kinks [J/OL]. JHEP, 2023, 01: 073. DOI: 10.1007/JHEP01(2023)073.

[20] Evslin J, García Martín-Caro A. Spontaneous emission from excited quantum kinks [J/OL]. JHEP, 2022, 12: 111. DOI: 10.1007/JHEP12(2022)111.

[21] Guo H. Leading quantum correction to the $\Phi^4$ kink form factor [J/OL]. Phys. Rev. D, 2022, 106(9): 096001. DOI: 10.1103/PhysRevD.106.096001.

[22] Evslin J. Form factors for meson-kink scattering [J/OL]. Phys. Lett. B, 2022, 830: 137177. DOI: 10.1016/j.physletb.2022.137177.

[23] Evslin J, Halcrow C, Romanczukiewicz T, et al. Spectral walls at one loop [J/OL]. Phys. Rev. D, 2022, 105(12): 125002. DOI: 10.1103/PhysRevD.105.125002.

[24] Evslin J, Guo H. Removing tadpoles in a soliton sector [J/OL]. JHEP, 2021, 11: 128. DOI: 10.1007/JHEP11(2021)128.

[25] Evslin J. $\phi^4$ kink mass at two loops [J/OL]. Phys. Rev. D, 2021, 104(8): 085013. DOI: 10.1103/PhysRevD.104.085013.

[26] Evslin J. The two-loop $\phi^4$ kink mass [J/OL]. Phys. Lett. B, 2021, 822: 136628. DOI: 10.1016/j.physletb.2021.136628.

[27] Evslin J. Evidence for the unbinding of the $\phi^4$ kink's shape mode [J/OL]. JHEP, 2021, 09: 009. DOI: 10.1007/JHEP09(2021)009.

[28] Evslin J, Guo H. Excited Kinks as Quantum States [J/OL]. Eur. Phys. J. C, 2021, 81(10): 936. DOI: 10.1140/epjc/s10052-021-09739-9.

[29] Evslin J, Guo H. Alternative to collective coordinates [J/OL]. Phys. Rev. D, 2021, 103(4): L041701. DOI: 10.1103/PhysRevD.103.L041701.

[30] Evslin J. The Ground State of the Sine-Gordon Soliton [J/OL]. JHEP, 2020, 07: 099. DOI: 10.1007/JHEP07(2020)099.

[31] Evslin J, Liu H. Quantum Reflective Kinks [J]. 2022.

[32] Evslin J. Moving kinks and their wave packets [J/OL]. Phys. Rev. D, 2022, 105(10): 105001. DOI: 10.1103/PhysRevD.105.105001.

[33] Weigel H. Quantum Instabilities of Solitons [J/OL]. AIP Conf. Proc., 2019, 2116(1): 170002. DOI: 10.1063/1.5114153.

[34] Cahill K E, Comtet A, Glauber R J. Mass Formulas for Static Solitons [J/OL]. Phys. Lett. B, 1976, 64: 283-285. DOI: 10.1016/0370-2693(76)90202-1.

[35] Romanczukiewicz T. Could the primordial radiation be responsible for vanishing of topological defects? [J/OL]. Phys. Lett. B, 2017, 773: 295-299. DOI: 10.1016/j.physletb.2017.08.045.

[36] Evslin J, Liu H. A reduced inner product for kink states [J/OL]. JHEP, 2023, 03: 070. DOI: 10.1007/JHEP03(2023)070.

[37] Liu H, Evslin J, Zhang B. Meson production from kink-meson scattering [J/OL]. Phys. Rev. D, 2023, 107(2): 025012. DOI: 10.1103/PhysRevD.107.025012.

[38] Evslin J, Liu H. (Anti-)Stokes Scattering on Kinks [J]. 2023.

[39] Graham N, Weigel H. Quantum corrections to soliton energies [J/OL]. Int. J. Mod. Phys. A, 2022, 37(19): 2241004. DOI: 10.1142/S0217751X22410044.

[40] de Vega H J. Two-Loop Quantum Corrections to the Soliton Mass in Two-Dimensional Scalar Field Theories [J/OL]. Nucl. Phys. B, 1976, 115: 411-428. DOI: 10.1016/0550-3213(76)90497-1.

[41] Verwaest J. Higher Order Correction to the Sine-Gordon Soliton Mass [J/OL]. Nucl. Phys. B, 1977, 123: 100-108. DOI: 10.1016/0550-3213(77)90343-1.

[42] Shifman M A, Vainshtein A I, Voloshin M B. Anomaly and quantum corrections to solitons in two-dimensional theories with minimal supersymmetry [J/OL]. Phys. Rev. D, 1999, 59: 045016. DOI: 10.1103/PhysRevD.59.045016.

[43] Hayashi A, Saito S, Uehara M. Pion - nucleon scattering in the Skyrme model and the P wave Born amplitudes [J/OL]. Phys. Rev. D, 1991, 43: 1520-1531. DOI: 10.1103/PhysRevD.43.1520.

[44] Hayashi A, Saito S, Uehara M. Pion - nucleon scattering in the soliton model [J/OL]. Prog. Theor. Phys. Suppl., 1992, 109: 45-72. DOI: 10.1143/PTPS.109.45.

[45] Melnikov I V, Papageorgakis C, Royston A B. Accelerating solitons [J/OL]. Phys. Rev. D, 2020, 102(12): 125002. DOI: 10.1103/PhysRevD.102.125002.

[46] Melnikov I V, Papageorgakis C, Royston A B. Forced Soliton Equation and Semiclassical Soliton Form Factors [J/OL]. Phys. Rev. Lett., 2020, 125(23): 231601. DOI: 10.1103/PhysRevLett.125.231601.

[47] Wheater J F, Xavier P D. The Size of a Soliton [J]. 2022.

[48] Evslin J, Gudnason S B. Dwarf Galaxy Sized Monopoles as Dark Matter? [J]. 2012.

[49] Schive H Y, Chiueh T, Broadhurst T. Cosmic Structure as the Quantum Interference of a Coherent Dark Wave [J/OL]. Nature Phys., 2014, 10: 496-499. DOI: 10.1038/nphys2996.

[50] Adam C, Oles K, Queiruga J M, et al. Solvable self-dual impurity models [J/OL]. JHEP, 2019, 07: 150. DOI: 10.1007/JHEP07(2019)150.

[51] Schwesinger B, Weigel H, Holzwarth G, et al. The Skyrme Soliton in Pion, Vector and Scalar Meson Fields: $\pi N$ Scattering and Photoproduction [J/OL]. Phys. Rept., 1989, 173: 173. DOI: 10.1016/0370-1573(89)90022-7.

[52] Skyrme T H R. A Nonlinear field theory [J/OL]. Proc. Roy. Soc. Lond. A, 1961, 260: 127-138. DOI: 10.1098/rspa.1961.0018.

[53] Gudnason S B, Halcrow C. A Smörgåsbord of Skyrmions [J/OL]. JHEP, 2022, 08: 117. DOI: 10.1007/JHEP08(2022)117.

[54] Martin M A A, Schlesier R, Zahn J. The semiclassical energy density of kinks and solitons [J]. 2022.

[55] Evslin J. Normal ordering normal modes [J/OL]. Eur. Phys. J. C, 2021, 81(1): 92. DOI: 10.1140/epjc/s10052-021-08890-7.

[56] Alonso-Izquierdo A, Miguélez-Caballero D, Nieto L M, et al. Wobbling kinks in a two-component scalar field theory: Interaction between shape modes [J/OL]. Physica D: Nonlinear Phenomena, 2023, 443: 133590. DOI: 10.1016/j.physd.2022.133590.

[57] Weigel H, Graham N. Vacuum polarization energy of the Shifman–Voloshin soliton [J/OL]. Phys. Lett. B, 2018, 783: 434-439. DOI: 10.1016/j.physletb.2018.07.027.

[58] Takyi I, Matfunjwa M K, Weigel H. Quantum corrections to solitons in the $\Phi^8$ model [J/OL]. Phys. Rev. D, 2020, 102(11): 116004. DOI: 10.1103/PhysRevD.102.116004.

[59] Takyi I, Barnes B, Ackora-Prah J. Vacuum Polarization Energy of the Kinks in the Sinh-Deformed Models [J]. Turk. J. Phys., 2021, 45: 194-206.

[60] Zhong Y. Normal modes for two-dimensional gravitating kinks [J/OL]. Phys. Lett. B, 2022, 827: 136947. DOI: 10.1016/j.physletb.2022.136947.

[61] Zhong Y. Singular Pöschl-Teller II potentials and gravitating kinks [J/OL]. JHEP, 2022, 09: 165. DOI: 10.1007/JHEP09(2022)165.

[62] Hertzberg M P. Quantum Radiation of Oscillons [J/OL]. Phys. Rev. D, 2010, 82: 045022. DOI: 10.1103/PhysRevD.82.045022.

[63] Kovtun A. Analytical computation of quantum corrections to a nontopological soliton within the saddle-point approximation [J/OL]. Phys. Rev. D, 2022, 105(3): 036011. DOI: 10.1103/PhysRevD.105.036011.

[64] Manton N S, Merabet H. Phi**4 kinks: Gradient flow and dynamics [J/OL]. Nonlinearity, 1997, 10: 3. DOI: 10.1088/0951-7715/10/1/002.

[65] Dollard J D. Adiabatic switching in the schrödinger theory of scattering [J]. Journal of Mathematical Physics, 1966, 7(5): 802-810.

[66] Morchio G, Strocchi F. Dynamics of Dollard asymptotic variables. Asymptotic fields in Coulomb scattering [J/OL]. Rev. Math. Phys., 2016, 28(01): 1650001. DOI: 10.1142/S0129055X1650001X.

[67] Strominger A. Lectures on the Infrared Structure of Gravity and Gauge Theory [J]. 2017.

[68] Kulish P P, Faddeev L D. Asymptotic conditions and infrared divergences in quantum electrodynamics [J/OL]. Theor. Math. Phys., 1970, 4: 745. DOI: 10.1007/BF01066485.

[69] Prabhu K, Satishchandran G, Wald R M. Infrared finite scattering theory in quantum field theory and quantum gravity [J/OL]. Phys. Rev. D, 2022, 106(6): 066005. DOI: 10.1103/PhysRevD.106.066005.

[70] Zhong Y, Du X L, Jiang Z C, et al. Collision of two kinks with inner structure [J/OL]. JHEP, 2020, 02: 153. DOI: 10.1007/JHEP02(2020)153.

[71] Yan H, Zhong Y, Liu Y X, et al. Kink-antikink collision in a Lorentz-violating $\phi^4$ model [J/OL]. Phys. Lett. B, 2020, 807: 135542. DOI: 10.1016/j.physletb.2020.135542.

[72] Mohammadi M, Momeni E. Scattering of kinks in the B$\varphi$4 model [J/OL]. Chaos Solitons and Fractals: the interdisciplinary journal of Nonlinear Science and Nonequilibrium and Complex Phenomena, 2022, 165: 112834. DOI: 10.1016/j.chaos.2022.112834.

[73] Takyi I, Gyampoh S, Barnes B, et al. Kink collision in the noncanonical $\varphi^6$ model: A model with localized inner structures [J/OL]. Results Phys., 2023, 44: 106197. DOI: 10.1016/j.rinp.2022.106197.

[74] Manton N S, Oles K, Romanczukiewicz T, et al. Collective Coordinate Model of Kink-Antikink Collisions in $\phi$4 Theory [J/OL]. Phys. Rev. Lett., 2021, 127(7): 071601. DOI: 10.1103/PhysRevLett.127.071601.

[75] Moradi Marjaneh A, Simas F C, Bazeia D. Collisions of kinks in deformed $\varphi^4$ and $\varphi^6$ models [J/OL]. Chaos Solitons and Fractals: the interdisciplinary journal of Nonlinear Science and Nonequilibrium and Complex Phenomena, 2022, 164: 112723. DOI: 10.1016/j.chaos.2022.112723.

[76] Adam C, Ciurla D, Oles K, et al. Sphalerons and resonance phenomenon in kink-antikink collisions [J/OL]. Phys. Rev. D, 2021, 104(10): 105022. DOI: 10.1103/PhysRevD.104.105022.

[77] Adam C, Oles K, Romanczukiewicz T, et al. Spectral Walls in Soliton Collisions [J/OL]. Phys. Rev. Lett., 2019, 122(24): 241601. DOI: 10.1103/PhysRevLett.122.241601.

[78] Campos J a G F, Mohammadi A, Queiruga J M, et al. Fermionic spectral walls in kink collisions [J/OL]. JHEP, 2023, 01: 071. DOI: 10.1007/JHEP01(2023)071.

[79] Segur H, Kruskal M D. Nonexistence of Small Amplitude Breather Solutions in $\phi^4$ Theory [J/OL]. Phys. Rev. Lett., 1987, 58: 747-750. DOI: 10.1103/PhysRevLett.58.747.

[80] Fodor G, Forgacs P, Grandclement P, et al. Oscillons and Quasi-breathers in the phi**4 Klein-Gordon model [J/OL]. Phys. Rev. D, 2006, 74: 124003. DOI: 10.1103/PhysRevD.74.124003.

[81] Coleman S R. There are no Goldstone bosons in two-dimensions [J/OL]. Commun. Math. Phys., 1973, 31: 259-264. DOI: 10.1007/BF01646487.

# 致 谢

作者感谢中国科学院近代物理研究所 Jarah Evslin 研究员和河南大学张柏阳博士与作者的合作以及很多有益讨论，使作者受益匪浅。感谢 CAS-DAAD 联合培养博士生项目的资助。感谢好友杜杰、好友顾捷和好友石健伟的精神支持（按姓氏排序，排名不分先后）。最后尤其感谢女神郁可唯的温暖歌声陪伴作者度过人生中的艰难时刻。

<div style="text-align:right">2023 年 6 月</div>

# 作者简历及攻读学位期间发表的学术论文与其他相关学术成果

**作者简历：**

2011 年 9 月——2016 年 6 月，在中国科学技术大学生命科学学院获得学士学位。

2017 年 9 月——2020 年 6 月，在中国科学院近代物理研究所获得硕士学位。

2020 年 9 月——2023 年 6 月，在中国科学院理论物理研究所攻读博士学位，其中

2021 年 7 月——2023 年 4 月，在慕尼黑大学索末菲理论物理中心访问。

**攻读学位期间已发表（或正式接受）的学术论文：**

1. Nuclear Fermi Momenta of $^2$H, $^{27}$Al and $^{56}$Fe from an Analysis of CLAS Data, 2022 (Nucl. Phys. A, arXiv:2103.15609 [nucl-th])
   Hui Liu, Na-Na Ma, Rong Wang

2. Meson Production from Kink-Meson Scattering, 2023 (PRD, arXiv:2211.01794 [hep-th])
   Hui Liu, Jarah Evslin, Baiyang Zhang

3. (Anti-)Stokes Scattering on Kinks, 2023 (JHEP, arXiv:2301.04099 [hep-th])
   Jarah Evslin, Hui Liu

4. A Reduced Inner Product for Kink States, 2023 (JHEP, arXiv:2212.10344 [hep-th])
   Jarah Evslin, Hui Liu

5. Squeezing the Free Scalar Ground State, 2023 (IJMPA, arXiv:1909.13497 [hep-th])
   Hui Liu, Yao Zhou, Jarah Evslin